\documentclass[12pt]{article}

\usepackage{times}

\usepackage{graphicx}
\usepackage{caption}
\usepackage{subcaption}
\usepackage{amsmath,amssymb}
\usepackage{booktabs}
\usepackage{array}
\usepackage{makecell}
\newcolumntype{C}[1]{>{\centering\let\newline\\\arraybackslash\hspace{0pt}}m{#1}}

\newcommand{\be}{\begin{equation}}
\newcommand{\ee}{\end{equation}}

\renewcommand{\thefigure}{\arabic{figure}}
\renewcommand{\thesubfigure}{\Alph{subfigure}}
\mathchardef\mhyphen="2D 
\usepackage[colorlinks=true
,urlcolor=blue
,anchorcolor=blue
,citecolor=blue
,filecolor=blue
,linkcolor=red
,menucolor=blue
,linktocpage=true
,pdfproducer=medialab
,pdfa=true
]{hyperref}

\newenvironment{sciabstract}{%
\begin{quote} \bf}
{\end{quote}}

\title{The dark matter interpretation of the 3.5-keV line is inconsistent with blank-sky observations}

\author
{Christopher Dessert,$^{1}$ Nicholas L. Rodd,$^{2,3}$ Benjamin R. Safdi$^{1}$\\
\\
\normalsize{$^{1}$Leinweber Center for Theoretical Physics, Department of Physics,}\\
\normalsize{University of Michigan, Ann Arbor, MI 48109, USA}\\
\normalsize{$^{2}$Berkeley Center for Theoretical Physics,}\\
\normalsize{University of California, Berkeley, CA 94720, USA}\\
\normalsize{$^{3}$Theoretical Physics Group, Lawrence Berkeley National Laboratory,}\\
\normalsize{Berkeley, CA 94720, USA}
}

\date{}

\begin{document} 


\baselineskip24pt


\maketitle


\begin{sciabstract}
Observations of nearby galaxies and galaxy clusters have reported an unexpected X-ray emission line around 3.5 kilo\textendash electron volts (keV). Proposals to explain this line include decaying dark matter\textemdash in particular, that the decay of sterile neutrinos with a mass around 7 keV could match the available data. If this interpretation is correct, the 3.5 keV line should also be emitted by dark matter in the halo of the Milky Way. We used more than 30 megaseconds of XMM-Newton (X-ray Multi-Mirror Mission) blank-sky observations to test this hypothesis, finding no evidence of the 3.5-keV line emission from the Milky Way halo.  We set an upper limit on the decay rate of dark matter in this mass range, which is inconsistent with the possibility that the 3.5-keV line originates from dark matter decay.
\end{sciabstract}


A plethora of cosmological and astrophysical measurements indicate that dark matter (DM) exists and makes up $\sim$80\% of the matter in the Universe, but its microscopic nature is unknown. If DM consists of particles that can decay into ordinary matter, the decay process may produce photons that are detectable with X-ray telescopes. Some DM models, such as sterile neutrino DM, predict such X-ray emission lines~\cite{PhysRevD.25.766}. If the sterile neutrinos exist with a mass-energy of a few kilo\textendash electron volts, they may explain the observed abundance of DM~\cite{Dodelson:1993je,Shi:1998km,Kusenko:2006rh}. The detection of an unidentified X-ray line (UXL) around 3.5 keV in a stacked sample of nearby galaxy clusters~\cite{Bulbul:2014sua} and an independent detection in one of those clusters and a galaxy~\cite{Boyarsky:2014jta} have been interpreted as evidence for DM decay~\cite{Abazajian:2017tcc}.  Other less-exotic explanations have also been proposed, such as emission lines of potassium or argon, from hot gas within the clusters~\cite{Jeltema:2014qfa}, or charge-exchange lines from interactions of the hot intracluster plasmas and cold gas clouds~\cite{Gu:2015gqm,Shah:2016efh}. 

\begin{figure}[p]
\includegraphics[width = 0.99\textwidth]{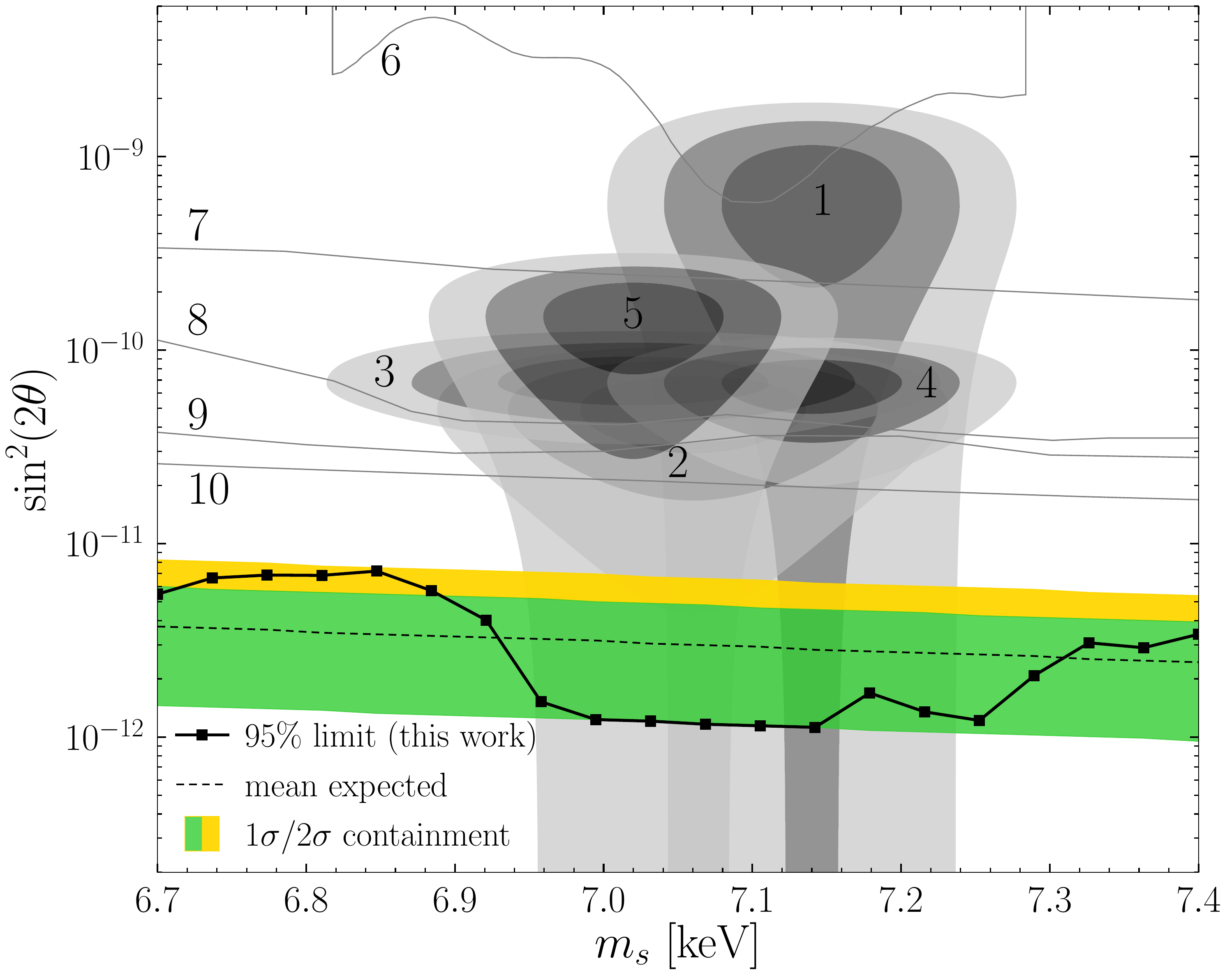}
\caption{\noindent {\bf Our upper limits on sterile neutrino decay.} The one-sided 95\% upper limit on the sterile neutrino DM mixing parameter $\sin^2 (2 \theta)$ as a function of the DM mass $m_s$ from our analysis of {\it XMM-Newton} BSOs (black squares). We compare with the expected sensitivity from the Asimov procedure (1$\sigma$ shown in green and 2$\sigma$ in yellow), and previous constraints (gray lines) and parameters required for DM decay explanations of previous UXL detections (3$\sigma$ in dark gray, 2$\sigma$ in gray, and 1$\sigma$ in light gray). We also show several existing detections (labelled 1 to 5) and constraints (6 to 10)~\cite{Abazajian:2017tcc}.
}
\label{fig: main-figure}
\end{figure}

The 3.5-keV UXL (hereafter just UXL) has been confirmed by several groups using different astrophysical targets and telescopes.  These include observations of the Perseus cluster using the Chandra~\cite{Bulbul:2014sua} and Suzaku~\cite{Urban:2014yda} telescopes, observations of the Galactic Center of the Milky Way with XMM-Newton (X-ray Multi-Mirror Mission)~\cite{Boyarsky:2014ska}, and observations of the diffuse Milky Way halo with {\it Chandra} deep-field data~\cite{Cappelluti:2017ywp}.  Several non-detections of the UXL have also been reported~\cite{Horiuchi:2013noa,Malyshev:2014xqa,Anderson:2014tza,Tamura:2014mta,Aharonian:2016gzq}. It is possible for a decaying DM model to be consistent with both the positive detections and negative results. Fig.~\ref{fig: main-figure} shows the existing detections and upper limits for the UXL, in the plane of sterile neutrino DM mass $m_s$ and sterile-active mixing parameter $\sin^2(2 \theta)$, which characterizes (and linearly scales with) the decay rate of the sterile neutrino DM state~\cite{SM}.

We seek to constrain the DM decay rate in the mass range relevant for the UXL by using {\it XMM-Newton} blank-sky observations (BSOs).
Our analysis utilizes $\sim$$10^3$ BSOs, which we define as observations away from large X-ray emitting regions, for a total of $30.6$ Ms of exposure time.  We focus on the line signal predicted from DM decay within the Milky Way, which should be present at every point in the sky.
The sensitivity of this technique can be estimated in the limit of large counts, in other words, detected photons. Then the test statistic (TS) in favor of detection of DM decay (related to the significance $\sigma \sim \sqrt{\textrm{TS}}$), scales as $\textrm{TS} \sim S^2 / B$, where $S$ is the number of signal photons from DM decay and $B$ is the number of background photons. The number of signal photons expected from a given location in the sky is proportional to the product of the decay rate of DM and the integrated column density of DM along the line of sight, which is quantified by the $D$ factor, $D = \int d s\,\rho_\textrm{DM}(s)$ where $\rho_\textrm{DM}$ is the DM density and $s$ is the line-of-sight distance.

We use these scalings to estimate the expected sensitivity of a BSO analysis, given the previous UXL observations. For example, the UXL has been detected with a 320-ks observation of the Perseus cluster using the {\it XMM-Newton} Metal Oxide Semiconductor (MOS) camera at roughly the $4\sigma$ level ($\textrm{TS} \sim 16$)~\cite{Bulbul:2014sua}. The background X-ray flux from Perseus is much higher than that for the BSOs, typically by a factor of 50. Averaged over the field of view of {\it XMM-Newton}, the $D$ factor of the Perseus cluster is $D_\textrm{Pers} \sim 3 \times 10^{28}\,\textrm{keV} \textrm{cm}^{-2}$, which is approximately the same as $D_\textrm{BSO}$, the $D$ factor within the Milky Way halo for observations $\sim$$45^\circ$ away from the Galactic Center. We calculated both $D$ factors assuming a Navarro-Frenk-White (NFW) DM profile~\cite{Navarro:1995iw}. Although the signal power should therefore be the same between Perseus and the BSO, we expect the same sensitivity to the UXL with a 6 ks BSO observation\textemdash assuming a DM origin\textemdash because the BSO background is expected to be lower than that of Perseus. Our analysis below uses $\sim30$ Ms of BSO exposure time, which implies that the UXL would be seen with a $\textrm{TS} \sim 10^5$, corresponding to a detection significance of $>100 \sigma$, if it is caused by decaying DM with the same properties as that in the Perseus cluster.  

We analyzed all publicly-available archival {\it XMM-Newton} observations that pass a set of quality cuts.  For our fiducial analysis, we first restrict to the observations used to those between $5^\circ$ and $45^\circ$ of the Galactic Center.  Within this region there are 1492~observations, with 4303~total exposures, for $\sim$$86$~Ms of exposure time.  These observations are distributed quite uniformly through our fiducial region, although there is a bias towards the Galactic plane. There are more exposures than observations because each of the {\it European Photon Imaging Cameras} charged coupled devices (CCDs) onboard XMM-Newton [two MOS and one positive-negative (PN)]~\cite{Turner:2000jy,Struder:2001bh} records a separate exposure, and each camera may have multiple exposures in a single observation if the data taking was interrupted. For each observation we process and reduce the data using the standard tools for extended emission~\cite{SM}. In addition to the photon-count data, we also extract the quiescent particle background (QPB). The QPB is an instrumental background caused by high-energy particles interacting with the detector, rather than true photon counts. The magnitude of the QPB contribution is estimated from parts of the instrument that are shielded from incident X-rays; we refer to this as the QPB data.

We then perform a background-only analysis of each of the exposures to determine properties that are used for further selection. We calculate the QPB contribution and the astrophysical flux over the energy range of 2.85 to 4.2 keV.  The QPB rate is estimated from the QPB data, whereas the astrophysical flux is measured using the likelihood analysis described below.  We rescale the astrophysical flux measured in the restricted energy range to a wider energy range of 2 to 10 keV by assuming a power-law spectrum of $dN/dE \sim E^{-1.5}$ where $N$ is the photon flux and $E$ is energy.  The cosmic X-ray background has a $2$ to $10$ keV intensity of $I_{2-10} \approx  2 \times 10^{-11}$ erg cm$^{-2}$ s$^{-1}$ deg$^{-2}$~\cite{Lumb:2002sw,Moretti:2008hs}.  In our fiducial analysis we remove exposures with $I_{2-10} > 10^{-10}$ erg cm$^{-2}$ s$^{-1}$ deg$^{-2}$ to avoid including exposures with either extended emission or flux from unresolved point sources. Approximately $58$\% of the exposures pass this cut, whereas $\sim$$13$\% of the exposures have $I_{2-10} < 3 \times 10^{-11}$ erg cm$^{-2}$ s$^{-1}$ deg$^{-2}$. Because the individual exposures are in the background-dominated regime and the signal we are searching for is restricted to a narrow energy range, even a clearly detectable DM line would have no effect on this selection criterion. We further remove exposures with anomalously high QPB rates; for our fiducial analysis, we keep the 68\% of exposures with the lowest QPB rates. We apply this criterion separately to the MOS and PN exposures. Lastly, we remove exposures with $<1$ ks of exposure time, because these exposures do not substantially improve our sensitivity and the associated low photon counts reduce the reliability of the background estimates. After these cuts, we are left with $\sim$30.6 Ms of exposure time distributed between 1397 exposures and 752 distinct observations.

We analyze the ensemble of exposures for evidence of the UXL by using a joint likelihood procedure. Individual exposures are not stacked. To evaluate the UXL hypothesis for a given $m_s$, we first construct profile likelihoods for the individual exposures as functions of the DM-induced line flux $F$. The X-ray counts are analyzed with a Poisson likelihood, from the number of counts in each energy channel. The associated model is a combination of the DM-induced flux represented by an X-ray line broadened by the detector response and two independent power laws for the background astrophysical emission and the instrumental QPB, where the normalization and spectral indices of each power law are free parameters. This same QPB power-law contribution is also fitted to the estimated QPB data using a Gaussian likelihood. Both datasets are restricted to the energy range $m_s/2 \pm 0.25\,{\rm keV}$, which was chosen to be wider than the energy resolution of the detector ($\sim$0.1 keV) but small enough that our power-law background models are valid over the whole energy range.

The two likelihoods for the X-ray counts and the QPB estimate are then combined, providing a likelihood that, for a given $m_s$, is a function of five parameters: the DM-induced line flux $F$ and the normalization factors and spectral indices of the astrophysical and QPB power laws. The last four of these are treated as nuisance parameters; that is, we maximize the individual likelihoods over the valid ranges of these parameters. Each dataset was therefore reduced to a profile likelihood as a function of $F$. This flux can be converted to a lifetime and, hence, $\sin^2(2 \theta)$~\cite{PhysRevD.25.766,SM}, once the $D$ factor for this region of the sky is known. In our fiducial analysis we compute the $D$ factors by assuming that the DM density profile of the Milky Way is an NFW profile with a 20 kpc scale radius. We normalize the density profile, assuming a local DM density of $0.4$ GeV$ $cm$^{-3}$~\cite{Catena:2009mf}, and take the distance between the Sun and the Galactic Center to be $8.13$ kpc~\cite{Abuter:2018drb}.

Joining the resulting likelihoods associated with each exposure yields the final joint likelihood that is a function of only $\sin^2(2\theta)$ for a given $m_s$. This likelihood is then used to calculate the one-sided 95\% confidence limit on the mixing angle and to search for evidence for the UXL using the discovery TS, which is defined as twice the log-likelihood difference between the maximum likelihood and the likelihood at the null hypothesis [assuming the likelihood is maximized at a positive value of $\sin^2(2\theta)$]. For statistical consistency, we must include negative values of $\sin^2(2\theta)$ in the profile likelihood, which correspond to under-fluctuations of the data.

To calibrate our expectation for the sensitivity under the null hypothesis, we construct the 68 and 95\% expectations for the limit using the Asimov procedure~\cite{Cowan:2010js}.  The Asimov procedure requires a model for the data under the null hypothesis; we compute this model by performing the likelihood fits described above under the null hypothesis [$\sin^2(2\theta) = 0$].  We use this to set one-sided power-constrained limits~\cite{Cowan:2011an}.  The measured limit is not allowed to go below the 68\% containment region for the expected limit, so as to prevent setting tighter limits than expected because of downward statistical fluctuations.  

\begin{figure}[p]
\includegraphics[width = 0.99\textwidth]{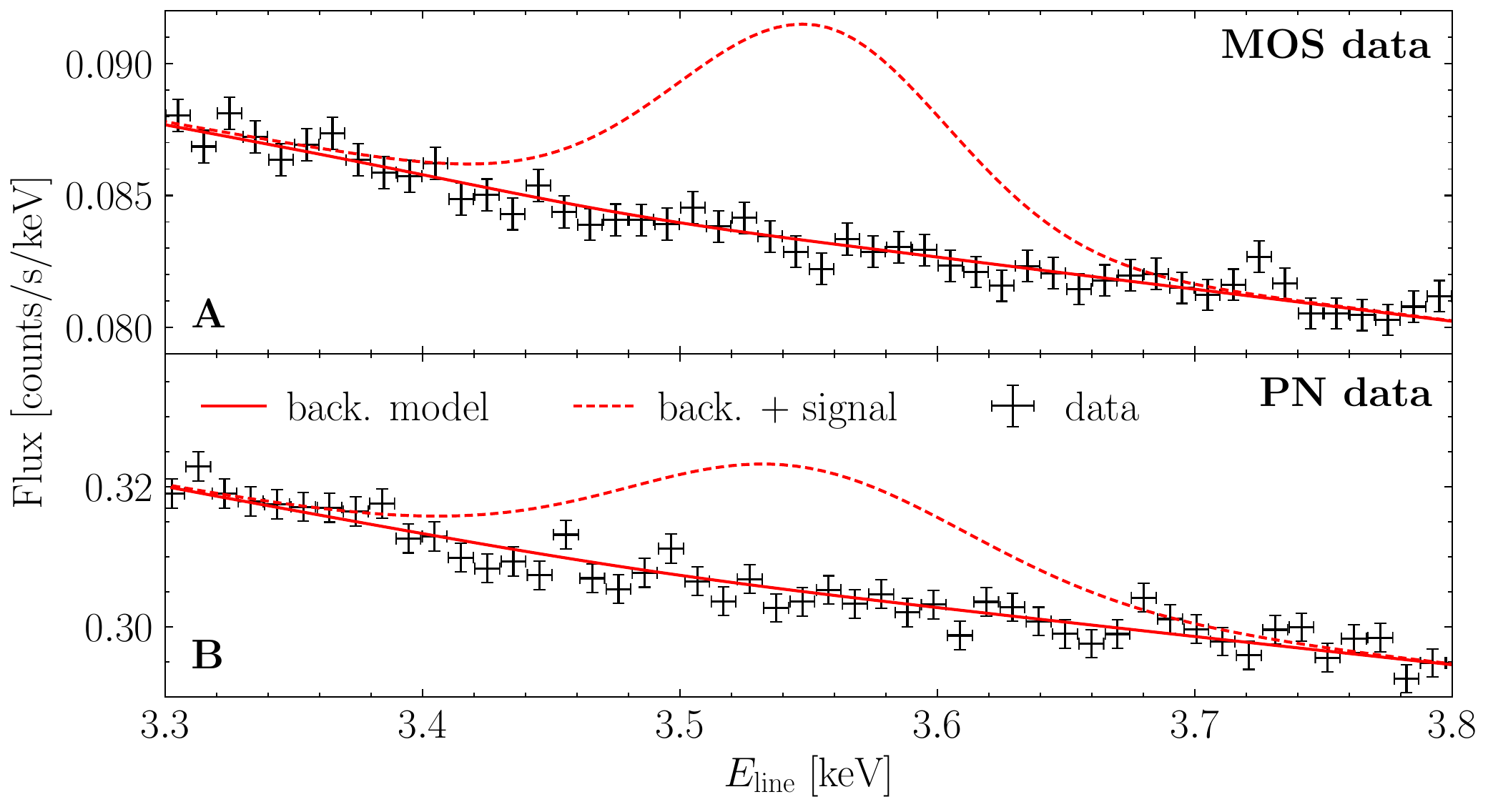}
\caption{\noindent {\bf The summed spectra.} ({\bf A} and {\bf B}) The summed MOS (A) and PN (B) spectra (black data points) for the exposures used in our fiducial analysis.  We also show the summed best-fitting background (back.) models (red solid line) and an example signal contribution with $m_s = 7.105$ keV and  $\sin^2(2\theta) = 10^{-10}$ (red dashed line).}
\label{fig: spectra}
\end{figure}

In Fig.~\ref{fig: spectra}, we show the summed spectra over all exposures included in the analysis for the MOS and PN data separately.  We emphasize that we do not use the summed spectra for our fiducial data analysis; instead we use the joint likelihood procedure described above. However, the summed spectra are shown for illustrative purposes.  We also show the summed best-fitting background models. Because our full model has independent astrophysical and QPB power-law models for each exposure, these curves are not single power laws but sums over 2794 independent power-laws. The summed data closely match the summed background models.  Fig.~\ref{fig: spectra} also shows the expected signal for $m_s = 7.105$ keV and  $\sin^2(2\theta) = 10^{-10}$, which are values we chose to be in the middle of the parameter space for explaining the observed UXL (Fig.~\ref{fig: main-figure}).  Fig.~\ref{fig: spectra} shows that this model is inconsistent with the data.

\begin{figure}[p]
\includegraphics[width = 0.99\textwidth]{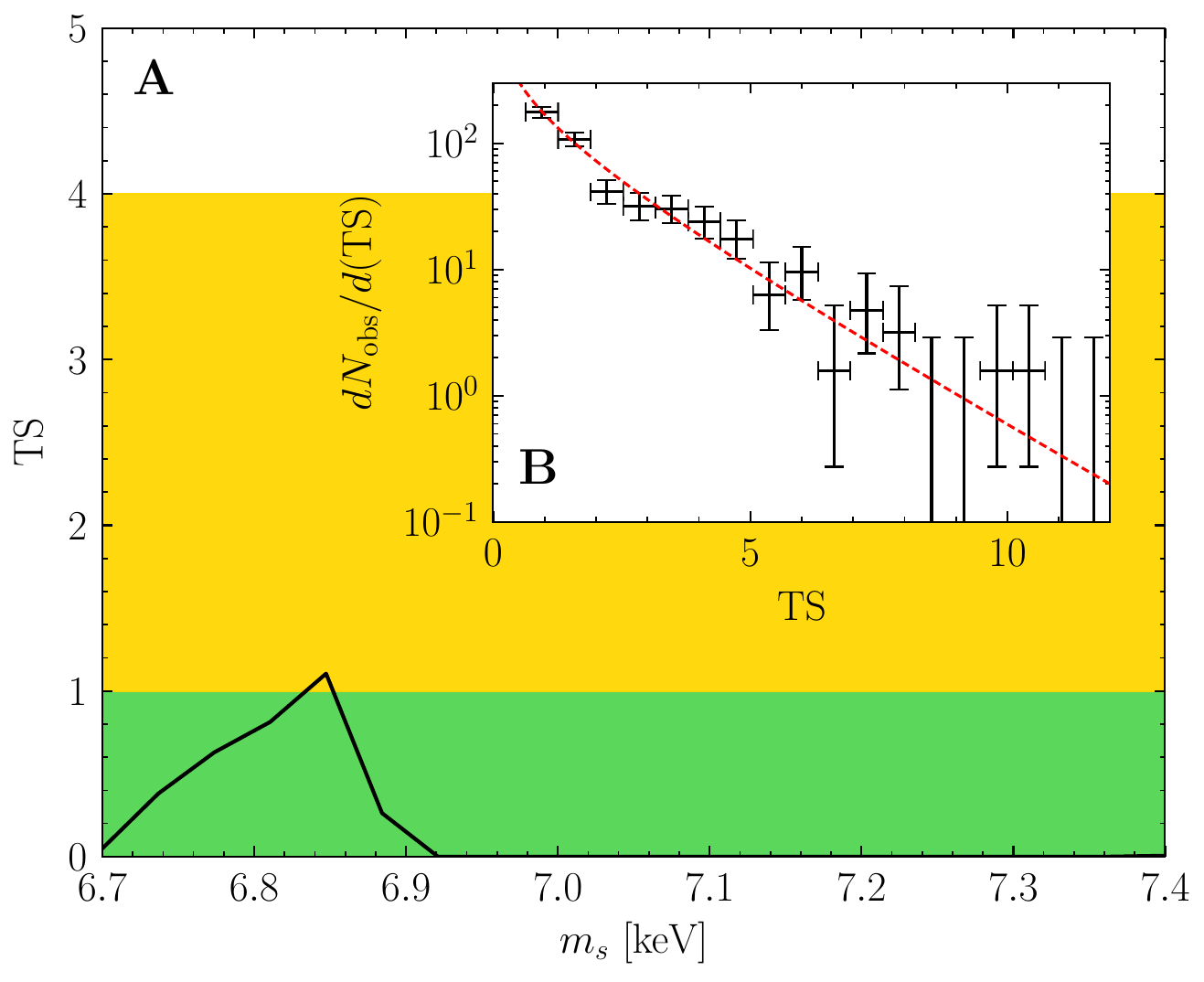}
\caption{\noindent {\bf No evidence for the decaying DM interpretation of the UXL.} ({\bf A}) The TS for the UXL as a function of the DM mass $m_s$ from the joint likelihood analysis.  The black curve shows the result from the data analysis, whereas the green and yellow shaded regions indicate the 1$\sigma$ and 2$\sigma$ expectations, respectively, under the null hypothesis. ({\bf B}) A histogram of the TSs from the individual exposures, with vertical error bars from Poisson counting statistics and horizontal error bars bracketing the histogram bin ranges.}
\label{fig: TS_chi2}
\end{figure}

Our fiducial one-sided power-constrained 95\% upper limit is shown in Fig.~\ref{fig: main-figure} along with mean, 1$\sigma$, and 2$\sigma$ expectations under the null hypothesis.  The upper limit is consistent with the expectation values and strongly disfavors the decaying DM explanation of the UXL. Our results disagree with the parameters required to explain the previous UXL observations as decaying DM by over an order of magnitude in $\sin^2(2\theta)$.  In Fig.~\ref{fig: TS_chi2}, we show the TS for decaying DM as a function of DM mass, with the 1 and 2$\sigma$ expectations under the null hypothesis; we find no evidence for decaying DM. 

Fig.~\ref{fig: TS_chi2} shows the TS for the joint-likelihood analysis over the ensemble of exposures.  However, we can also calculate a TS for decaying DM from each individual exposure. Under the null hypothesis, Wilks' theorem states that the distribution of TSs from the individual exposures should asymptotically follow a $\chi^2$ distribution. In the inset of Fig.~\ref{fig: TS_chi2}, we show the histogram of the number of exposures that are found for a given TS, for our reference mass of $m_s = 7.105$ keV. The distribution matches the expectation under the null hypothesis.  We also performed a Kolmogorov-Smirnov test comparing the observed TSs with the expected one-sided $\chi^2$ distribution, and found a P value of $0.77$, which indicates that the TS data is consistent with the null hypothesis. 

Although Fig.~\ref{fig: TS_chi2} shows that our results appear to be consistent with the expected statistical variability, there remains the possibility that systematic effects such as un-modeled instrumental lines could conspire to hide a real line.  We test for such systematics in Fig.~\ref{fig: mos-pn} by analyzing the data from the individual cameras separately~\cite{SM}, in Fig.~\ref{fig: limits-lines} by explicitly allowing for extra possible instrumental lines in the background model~\cite{SM}, and in Figs.~\ref{fig: var_large} and~\ref{fig: var_small} by looking at the data in sub-regions increasingly far away from the Galactic Center~\cite{SM}.  Accounting for these possible systematics in a data-driven way may weaken our limits to as much as $\sin^2(2\theta)~<~2~\times 10^{-11}$ (Fig.~\ref{fig: var_small}, Reg. 4), which still strongly rules out the decaying DM interpretation of the UXL. We also analyze the summed X-ray count data shown in Fig.~\ref{fig: spectra} directly~\cite{SM}, and found, again, that the decaying DM interpretation of the UXL was excluded (see Fig.~\ref{fig: stacked}). 

We have analyzed $\sim$30 Ms of {\it XMM-Newton} BSOs for evidence of DM decay in the energy range of 3.35 to 3.7 keV.  We found no evidence for DM decay.  Our analysis rules out the decaying DM interpretation of the previously observed 3.5 keV UXL because our results exclude the required decay rate by more than an order of magnitude.


%
%

\section*{Acknowledgments}

 We thank S. Mishra-Sharma for collaboration in the early stages of this work, K. Abazajian, J. Beacom, A. Boyarsky, E. Bulbul, D. Finkbeiner, J. Kopp, K. Perez, S. Profumo, J. Thaler, and C. Weniger for useful discussions and comments on the draft.
We further thank K. Abazajian for
preliminary discussions of this topic, and the members of the \textit{XMM-Newton} Helpdesk for assistance with the data reduction process. 
We downloaded the observations used in this work from the \href{http://nxsa.esac.esa.int/nxsa-web/#home}{XMM archive}.
We provide a full list of exposures used in our fiducial analysis, data-reduction software, our analysis code, and the numerical data plotted in the figures at~\cite{nick_rodd_2020_3669387}.
CD and BRS were supported by the Department of Energy Early Career Grant DE-SC0019225. 
NLR is supported by the Miller Institute for Basic Research in Science at the University of California, Berkeley.
Computational resources and services were provided by Advanced Research Computing at the University of Michigan, Ann Arbor.

\clearpage
\newpage

\begin{center}
{\noindent \LARGE Supplementary Materials for\\
The dark matter interpretation of the 3.5-keV line is inconsistent with blank-sky observations} \\
\centering{\large Christopher Dessert, Nicholas L. Rodd, Benjamin R. Safdi}
\end{center}

\setcounter{equation}{0}
\setcounter{figure}{0}
\setcounter{table}{0}
\setcounter{section}{0}
\setcounter{page}{1}

\renewcommand{\theequation}{S\arabic{equation}}
\renewcommand{\thefigure}{S\arabic{figure}}
\renewcommand{\thetable}{S\arabic{table}}
\newcommand\ptwiddle[1]{\mathord{\mathop{#1}\limits^{\scriptscriptstyle(\sim)}}}
\renewcommand{\thesubfigure}{\Alph{subfigure}}

\section{Materials and Methods}

\subsection{Data Reduction}

The data products were downloaded from the \href{http://nxsa.esac.esa.int}{\textit{XMM-Newton} Science Archive} and processed into the X-ray spectra and QPB flux estimates used in the main text. This process is applied to each exposure individually. We have applied our data reduction pipeline to all 6,350 observations within 90$^{\circ}$ of the Galactic Center, collected by \textit{XMM-Newton} up to 2018 September 5 (the instrument collected a total of 12,044 observations in that time). The data reduction process uses the \textit{XMM-Newton} Extended Source Analysis Software (ESAS) package for modeling extended objects and the diffuse X-ray backgrounds. The ESAS package is part of the \textit{XMM-Newton} Science Analysis System (SAS)~\cite{XMM-SAS}; we used version 17.0.

After selecting an observation, we obtain summary information for this dataset and its associated exposures. The Calibration Index File (CIF) is generated using the task \texttt{cifbuild}, which locates the Current Calibration File (CCF). The CCF provides information about the state of the detector at observation time; for example, it supplies the location of bad pixels on the detector. Next, the task \texttt{odfingest} is used to generate the Observation Data Files (ODF), which contains uncalibrated summary files in addition to general information on the observation including data quality records. The relevant science exposures for each observation ID to use for data reduction are determined from the Pipeline Processing Subsystem summary file. Only PN exposures in submodes \texttt{Full Frame} and \texttt{Extended Full Frame} were chosen to ensure an accurate estimate of the instrumental Quiescent Particle Background (QPB).

From this information, a set of filtered events is then created for both MOS and PN cameras for each available science exposure. The PN pipeline is as follows.  The task \texttt{epchain} is first used to generate an event list. The list of out-of-time events, which are events recorded while the CCD is being read out, is generated with \texttt{epchain withoutoftime=true}. After obtaining the list of events, the task \texttt{pn-filter} is called to record only those events that occurred during a good time interval (GTI). This task calls the SAS routine \texttt{espfilt} to filter the light-curves for periods of soft proton (SP) contamination. An observation affected by SP will typically have a count rate histogram with a peak at the unaffected rate, and a long tail due to the contamination. \texttt{espfilt} establishes thresholds at $\pm 1.5\sigma$ of the count rate distribution, and then creates a GTI file containing the time intervals where the data is contained within those limits. The MOS pipeline is analogous to that for PN, requiring the tasks \texttt{emchain} and \texttt{mos-filter}.

Now that the data has been cleaned, we identify regions of the dataset we wish to mask. The routine \texttt{cheese} is applied to search for any point sources in the field of view for the energy range $3-4$ keV. The resulting mask is then used to exclude these sources from further analysis. Applying this mask also removes the necessity of a pile-up correction, as for extended source analyses this is only a concern near point sources. In addition, MOS CCDs flagged as anomalous are disregarded. For example, a suspected micrometeorite impact caused the loss of MOS1 CCD6 on 2005 March 12, and a similar event caused the loss of MOS1 CCD3 on 2012 December 11. These CCDs are excluded from analysis for observations made after these dates. 

With the cleaned data masked, the final step is the production of the spectra and QPB data. For the PN and MOS cameras this is achieved with the tasks \texttt{pn-spectra} and \texttt{mos-spectra} respectively. These tasks use the filtered event files to create the photon-count data, the QPB data, the Ancillary Response File (ARF), and the source count weighted Redistribution Matrix File (RMF), for the masked region but otherwise the full field of view (FOV). The ARF and RMF account for the detector response, and will be described in more detail in the following subsection.

\subsection{Data Analysis}

For a given exposure, we model the observed number of X-rays as originating from a combination of instrumental effects and conventional astrophysical sources, which we consider backgrounds, and a putative DM decay line as our signal hypothesis.  The DM in the Milky Way is sufficiently non-relativistic ($v \sim 10^{-3}$ in natural units where $v$ is velocity of the DM) that we treat the decay signal as a zero-width line at an energy $m_s/2$. The line-width generated by the finite velocity dispersion of DM within the Milky Way is small compared to the energy resolution of the detector. The flux of this line in counts cm$^{-2}$s$^{-1}$sr$^{-1}$keV$^{-1}$, averaged over the full region of interest (ROI) for this observation, is given by
\begin{equation}
\frac{d\Phi}{d E} = \frac{d\Phi_{\rm pp}}{d E} \times D\,,
\end{equation}
where the particle physics and $D$-factor contributions are given by
\begin{equation}
\frac{d\Phi_{\rm pp}}{d E} = \frac{\Gamma}{4\pi\,m_s}\,\delta(E-m_s/2)\,,\hspace{0.5cm}
D = \int ds\, \rho_{\rm DM}(s,\Omega)\,.
\end{equation}
Above, $\Gamma = \tau^{-1}$ is the DM decay rate (the inverse of the DM lifetime $\tau$), $s$ is an integration variable along the line-of-sight, and $\rho_{\rm DM}$ is the Milky Way DM distribution, which will be discussed further below.  For our searches for DM decay in the ambient MW, we may compute $D$ using any angular position within the ROI, and use this as an estimate for the average $D$ factor. This is because variations in the line-of-sight integral through the Milky Way halo are negligible (at most $\sim$$2$\% for the regions we consider) over the small {\it XMM-Newton} field-of-view. However, if the DM density varies over the scale of the ROI, as is the case when considering extragalactic sources such as galaxy clusters, then the $D$-factor needs to be averaged over the ROI, accounting for the vignetting of the instrument. Note the $D$-factor here is the \textit{D}ecay analogue of the $J$-factor, which is used for DM annihilation. For the specific case of sterile-neutrino DM, the decay rate can be related to the mixing angle between active and sterile neutrinos, $\theta$, as~\cite{PhysRevD.25.766}
\begin{equation}
\Gamma = 1.361 \times 10^{-29}~{\rm s}^{-1} \left( \frac{\sin^2 2 \theta}{10^{-7}} \right) \left( \frac{m_s}{1~{\rm keV}} \right)^5\,.
\label{eq:taumix}
\end{equation}
This expression is valid for a Majorana neutrino, for a Dirac neutrino it is a factor of 2 smaller.

For each observation, we only included the contribution from the MW halo DM column density. Each region will also include a large column density from extragalactic DM. We can neglect the extragalactic contribution because an extragalactic line emitted over cosmological distances will be smeared out by redshifting, and the resulting smooth emission will be more than an order of magnitude smaller than the line from the MW halo.

We have concentrated on sterile neutrino DM, but our results apply to any model of decaying DM which produces an X-ray line. Alternate models for the 3.5 keV UXL have been proposed, however, that involve the decay of DM into an ultralight axion-like particle, which converts to photons within the galactic and/or cluster magnetic fields~\cite{Cicoli:2014bfa}.  Our results do not directly apply to these models because the spatial morphology of the signal is a convolution of the DM density distribution and the magnetic field distribution. Estimating the size of the effect~\cite{Conlon:2014xsa} indicates that our results also constrain this DM.

By restricting our attention to relatively blank regions of the sky and a narrow energy range, we reduce the number of backgrounds that need to be considered. As discussed in the main text, we model the contributions to the X-ray counts using a power-law instrumental QPB rate and a power-law astrophysical spectrum, which may also describe the soft proton background if present~\cite{Kuntz:2008}. In principle, the soft proton background an unfolded power-law that has not been passed through the instrument response, however we find that including such an additional model has minimal impact on our results. Physical astrophysical emission may be present within the ROIs from the cosmic X-ray background, extended emission regions, or unresolved populations of Galactic sources.  We model the QPB spectrum using a power-law in counts, while the astrophysical emission is modeled by a power-law in flux.  A flux power-law is, in principle, not directly equivalent to a counts power-law because of the energy-dependent detector response.  However, over the narrow energy ranges we consider the distinction is small.  Still, for consistency we model the spectra in these different ways.

Given the signal and astrophysical background models, we calculate the predicted number of model counts in each of the camera channels.  Let us define $S(E,\mathbf{\theta}^e_{\rm \bf phys})$ in units of counts cm$^{-2}$s$^{-1}$sr$^{-1}$keV$^{-1}$, as the signal and astrophysical background spectrum, as a function of energy $E$. Within this expression, the index $e$ is used to enumerate the different exposures. The parameters $\mathbf{\theta}^e_{\rm \bf phys}$ denote the astrophysical background parameters and the signal parameters for the given exposure $e$. The signal parameters can be separated out by writing $\mathbf{\theta}^e_{\rm \bf phys} = \{ m_s,\,\Gamma,\,\mathbf{\theta}_B^e \}$, where $\mathbf{\theta}_B^e$ are the background astrophysical power-law parameters. The QPB is not included here but will be incorporated separately, as described below. By using $\Gamma$ we keep our discussion appropriate for a general decaying DM scenario, but the analysis can immediately be specialized to the sterile neutrino scenario using equation~\eqref{eq:taumix}. For the decaying DM hypothesis the DM parameters do not vary throughout the Milky Way or over time, and thus must be identical across exposures, so they do not carry an index $e$. The background parameters do vary between exposures and thus must be treated independently. Explicitly, the expected flux can be written as follows:
\begin{equation}
S(E,\mathbf{\theta}^e_{\rm \bf phys}) = \frac{\Gamma\,D}{4\pi\,m_s} \delta(E-m_s/2) + A^e_{\rm astro} \left( \frac{E}{1~{\rm keV}} \right)^{n_{\rm astro}^e}\,.
\label{eq:modelFlux}
\end{equation}

To compare this predicted spectrum to the observed number of X-rays in counts, we use forward modeling to incorporate the instrument response. The predicted number of counts in a given energy bin indexed by $i$ is given by
\begin{equation}
\mu_{i, {\rm phys}}^e(\mathbf{\theta}^e_{\rm \bf phys}) = t^e \Delta \Omega^e \int dE'\,{\rm RMF}^e_i(E')\,{\rm ARF}^e(E')\,S(E',\mathbf{\theta}^e_{\rm \bf phys})\,,
\label{eq:modelcounts}
\end{equation}
where $t^e$ is the observation time for the given exposure in s, $\Delta \Omega^e$ is the angular area of the ROI, the ARF provides the effective area of the detector as a function of energy in cm$^2$, and the dimensionless RMF accounts for the energy resolution and detector gain effects. All of these detector quantities vary between exposures and so carry an explicit $e$ index.  We now add to equation~\eqref{eq:modelcounts} the contribution from the QPB rate as a power law in reconstructed (rather than true) energy
\begin{equation}
\mu^e_{i,{\rm QPB}}(\mathbf{ \theta}^e_{\rm \bf QPB}) =  A^e_{\rm QPB} \times E_i^{n^e_{\rm QPB}} \,, 
\label{eq:QPB_counts}
\end{equation}
where $\mathbf{ \theta}^e_{\rm \bf QPB} = \{ A^e_{\rm QPB},\, n^e_{\rm QPB} \}$ are the model parameters defining the power-law. The separate treatment for the QPB arises as its flux is not folded with the detector response.

For the given exposure we now have the total predicted model counts $\mu_i^e$ in each energy bin as a function of the model parameters $\mathbf{\theta}^e = \{\mathbf{\theta}^e_{\rm \bf phys},\,\mathbf{ \theta}^e_{\rm \bf QPB}\}$:
\begin{equation}
\mu_i^e(\mathbf{ \theta}^e) =
 \mu_{i, {\rm phys}}^e(\mathbf{\theta}^e_{\rm \bf phys})  + \mu_{i,{\rm QPB}}^e(\mathbf{ \theta}^e_{\rm \bf QPB}) \,. 
\end{equation}
The data collected in this exposure can be identically binned, such that we can represent the X-ray dataset for each exposure by a set of integers $d_{\rm X\mhyphen ray}^e = \{ k_i^e \}$, where explicitly $k_i^e$ is the number of X-rays in energy bin $i$ for this exposure. With the data and model in identical forms, we can now compare the two by constructing a joint likelihood over all energy bins as follows
\begin{equation}
\mathcal{L}^e_{\rm X\mhyphen ray} (d_{\rm X\mhyphen ray}^e \vert \mathbf{\theta}^e) = \prod_i \frac{\mu_i^e(\mathbf{\theta}^e)^{k_i^e} e^{-\mu_i^e(\mathbf{\theta}^e)}}{k_i^e!}\,.
\end{equation}

The above likelihood accounts for the X-ray data collected during a given exposure, but there is additional information collected by the cameras that we incorporate into our model. This arises in the form of an estimate for the QPB background during the given exposure, as determined from pixels on the CCD that were shielded and therefore unexposed to direct X-rays. The ESAS tools provide this information as the mean and standard deviation on the (non-integer) QPB counts in each energy bin, which we denote by $\lambda^e_{i, {\rm QPB}}$ and $\sigma^e_{i,{\rm QPB}}$ respectively. We then construct a Gaussian likelihood for the QPB dataset $d_{\rm QPB}^e = \{ \lambda^e_{i, {\rm QPB}},\, \sigma^e_{i,{\rm QPB}} \}$ as
\begin{equation}
\mathcal{L}_{\rm QPB}^e(d_{\rm QPB}^e \vert \mathbf{\theta}^e_{{\rm \bf QPB}}) = \prod_i \frac{1}{\sigma^e_{i,{\rm QPB}}\,\sqrt{2\pi}} \exp \left[ -\frac{(\mu^e_{i,{\rm QPB}}(\mathbf{\theta}^e_{ {\rm \bf QPB}})- \lambda^e_{i, {\rm QPB}} )^2}{2(\sigma^e_{i,{\rm QPB}})^2} \right]\,.
\end{equation}

To account for both the X-ray and QPB data simultaneously, we form the joint likelihood as
\begin{equation}
\mathcal{L}^e(d^e \vert m_s, \Gamma, \mathbf{\theta}_B^e) = \mathcal{L}^e_{\rm X\mhyphen ray} (d^e_{\rm X\mhyphen ray} \vert \mathbf{\theta}^e_{\rm \bf phys}) \times \mathcal{L}^e_{\rm QPB}(d^e_{\rm QPB} \vert \mathbf{\theta}^e_{\rm \bf QPB})\,,
\end{equation}
where $d^e = \{d^e_{\rm X\mhyphen ray},\,d^e_{\rm QPB}\}$ and where $ \mathbf{\theta}_B^e$ denotes the four model parameters that describe the background astrophysical power-law and the power-law QPB model. 

In a similar manner we can construct likelihoods for each exposure, recalling that the signal parameters will not vary between them. We therefore remove these background parameters at the level of individual exposures, using the standard frequentist technique of profiling. At fixed $m_s$ we construct the profile likelihood as a function of $\Gamma$~\cite{Rolke:2004mj}. The profile likelihood is given by
\begin{equation}
\mathcal{L}^e(d^e \vert m_s, \Gamma) = \mathcal{L}^e(d^e \vert m_s, \Gamma, \hat{\mathbf{\theta}} {\vphantom{\theta}}_B^e)\,,
\end{equation}
with $\hat{\mathbf{\theta}} {\vphantom{\theta}}_B^e$ denoting the value of each of the background parameters that maximizes the likelihood for the specific values of $m_s$ and $\Gamma$ under consideration. This technique does not involve fixing the background to its value under the null hypothesis or the signal hypothesis. Instead, we determine a new value for $\mathbf{\theta}_B^e$ for each value of $\Gamma$ considered, at fixed $m_s$ using \texttt{minuit}~\cite{James:1975dr}.

We construct a profile likelihood for each exposure, leaving a likelihood depending only on the DM parameters. The information from each of these exposures can then be combined into the joint likelihood, which depends on the entire dataset $d = \{d^e\}$:
\begin{equation}
\mathcal{L}(d \vert m_s, \Gamma) = \prod_e \mathcal{L}^e(d^e \vert m_s, \Gamma)\,.
\label{eq:jplike}
\end{equation}
We reiterate that the signal parameters do not vary between exposures.

Using the likelihood in equation~\eqref{eq:jplike} we perform hypothesis testing between a signal model containing a DM decay line at fixed $m_s$ and the null hypothesis without the DM line. Following frequentist standards, we will quantify the significance of any excess using a test statistic (TS) for discovery
\begin{equation}
{\rm TS}(m_s) = \left\{ 
\begin{array}{ll}
2 \left[ \ln \mathcal{L}(d \vert m_s, \hat \Gamma) - \ln \mathcal{L}(d \vert m_s, \Gamma=0) \right] & \hat \Gamma \geq 0\,,  \vspace{0.05cm}\\
0 &  \hat \Gamma < 0 \,.
\end{array} \right.
\label{eq:TS}
\end{equation}
Here $\hat \Gamma$ is the value of $\Gamma$ that maximizes the likelihood at fixed $m_s$, and asymptotically ${\rm TS}(m_s) = \sigma^2$, where $\sigma$ is the significance of the excess. We may also construct a test statistic appropriate for establishing one-sided limits on $\Gamma$ for a fixed $m_s$.  Note that $\Gamma$ is physically constrained ($\Gamma \geq 0$), though for consistency we must consider negative values of $\Gamma$ as well, so we define~\cite{Cowan:2010js}
\begin{equation}
q(m_s,\Gamma) =
 \left\{ 
\begin{array}{ll}
2 \left[ \ln \mathcal{L}(d \vert m_s, \hat \Gamma) - \ln \mathcal{L}(d \vert m_s, {\Gamma}) \right] &  \hat \Gamma \leq \Gamma\,, \vspace{0.05cm}\\
 0 & \hat \Gamma > \Gamma \,.
 \end{array} \right.
\end{equation}
This statistic then allows us to determine the one-sided 95\% limit on the decay rate $\Gamma_{95\%}$ by solving $q(m_s, \Gamma_{95\%}) = 2.71$.  We also power-constrain the limits, to avoid setting stronger limits than expected due to statistical fluctuations~\cite{Cowan:2011an}, as discussed in the main text. To obtain the expected value for $q(m_s,\Gamma)$, we apply the Asimov procedure~\cite{Cowan:2010js} to the null hypothesis. 
The interpretation of the square root of the discovery TS as being the significance of the line and the precise TS threshold used to calculate the one-sided 95\% limit rely on the TS in equation~\eqref{eq:TS} following a one-sided $\chi^2$ distribution under the null hypothesis.
This is justified by the fact that our photon statistics are sufficient to invoke Wilks' theorem.

\section{Supplementary Text}

In this section, we provide extended results for the fiducial analysis presented in the main text and test variations to the procedure.  This section is organized as follows.  First, we subject our fiducial analysis to a key statistical test by injecting a synthetic signal into the data.  Then, we present results from individual exposures and determine which observations contribute most to our limits. In the following section, we consider how our limits depend on assumptions for the DM profile of the Milky Way. In the final sub-section we explore how our sensitivity varies for different selection criteria on the exposures.

\subsection{Synthetic Signal}

The limit on decaying DM, shown in Fig.~\ref{fig: main-figure}, is tighter than in previous studies.  We therefore subject our analysis to test that it is statistically meaningful.  For example, it is possible that systematic effects cause the limit to appear stronger than it should be and that a real signal, if present, would be excluded by our analysis.  To test this possibility, we add a synthetic signal to the real data and verify that our limit does not exclude the signal that we inject.

\begin{figure}[p]
\centering
\begin{subfigure}{0.49\textwidth}
\includegraphics[width = \textwidth]{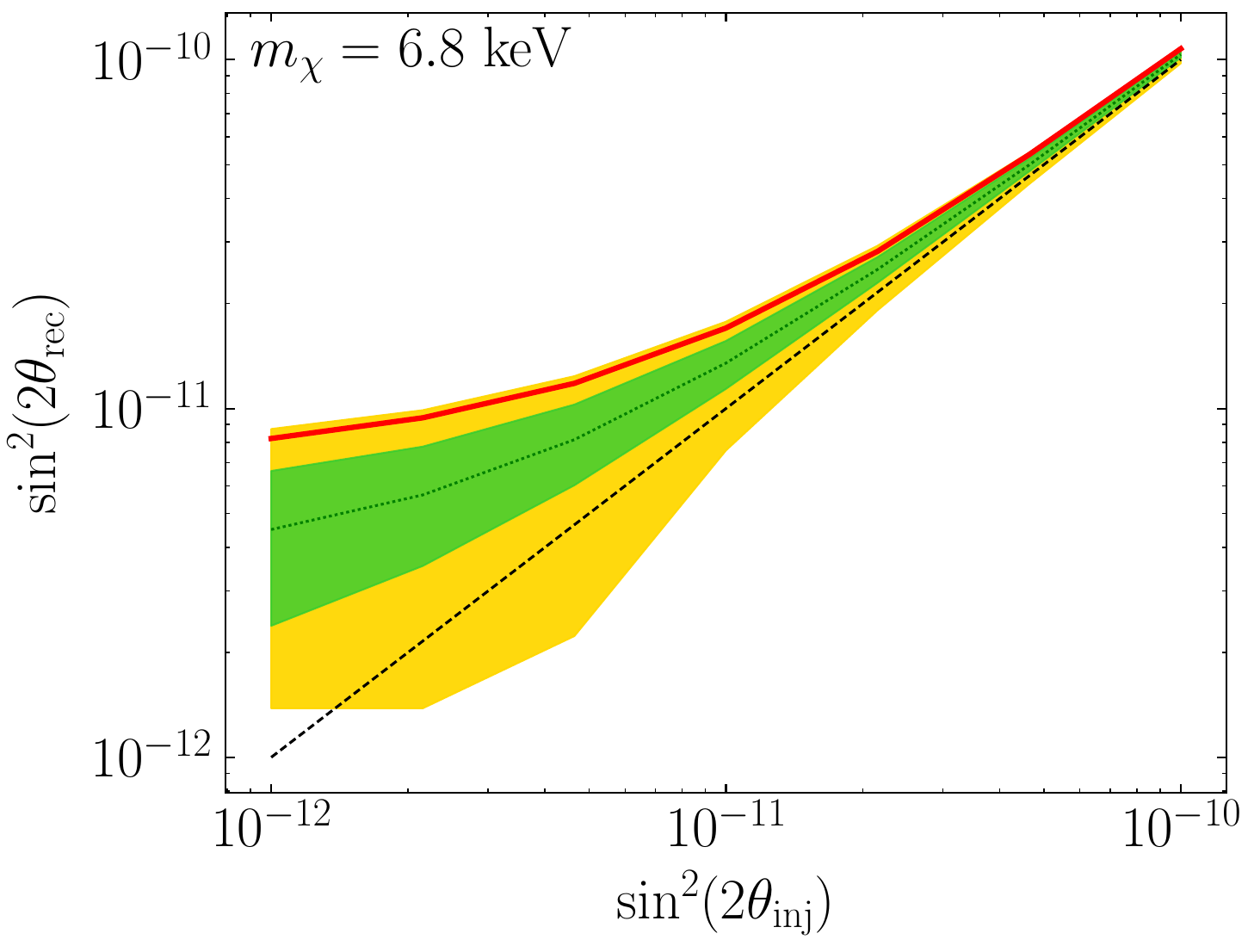} 
    \subcaption{}
\end{subfigure}
\begin{subfigure}{0.49\textwidth}
\includegraphics[width = \textwidth]{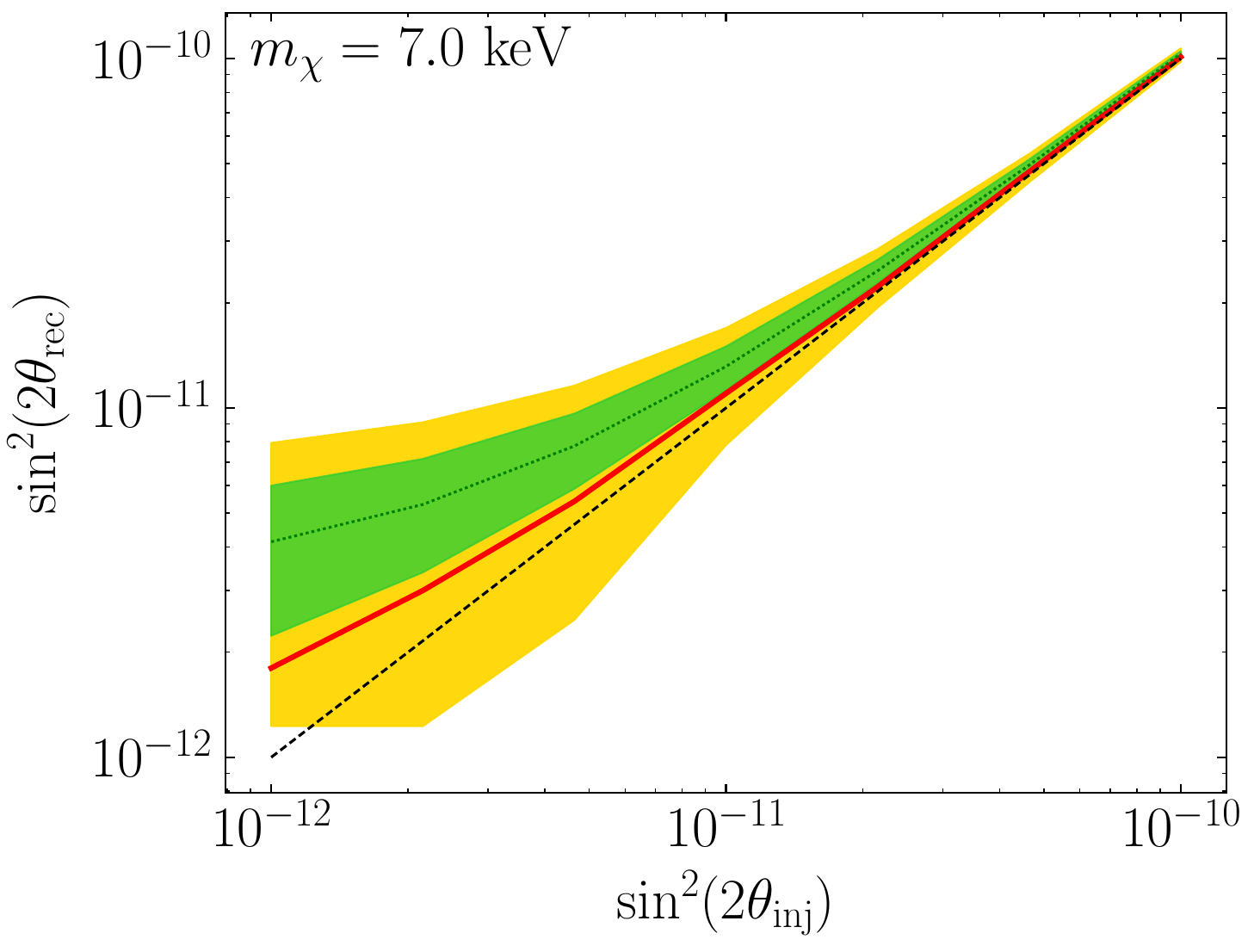}
    \subcaption{}
\end{subfigure}
\begin{subfigure}{0.49\textwidth}
\includegraphics[width = \textwidth]{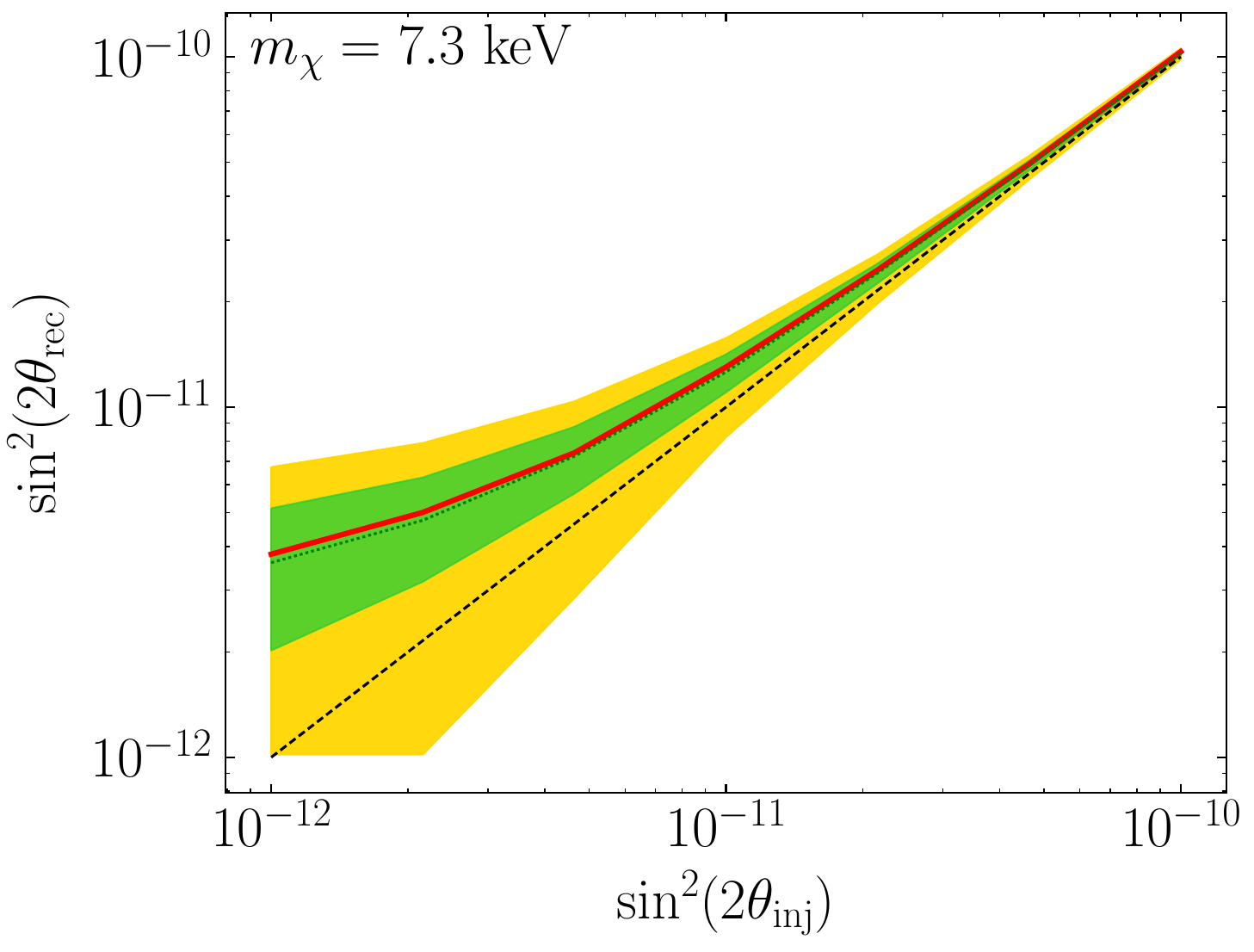}
    \subcaption{}
\end{subfigure}
\vspace{-0.2cm}
\caption{\textbf{Results of the synthetic signal test.} We inject an artificial DM signal to the data, with mixing angle $\sin^2(2 \theta_\text{inj})$ as indicated on the x-axes, and recover values $\sin^2(2 \theta_\text{rec})$, shown on the y-axes. In (A), we show the results for 6.8 keV; (B), for 7.0 keV; (C), for 7.3 keV. The red curves show the power-constrained 95\% one-sided upper limits that we find on the analysis of the hybrid datasets, consisting of the real data plus the synthetic signal. The bands show the mean (black), 1$\sigma$ (green) and 2$\sigma$ (yellow) expectations for the 95\% one-sided upper limit. The injected signal strength is never excluded, as indicated by the red line never dropping below the dashed black diagonal line.
}
\label{fig: synthetic}
\end{figure}

We perform this analysis for our fiducial selection criterion described in the main text.  The results of the test, for three different assumed DM masses, are shown in Fig.~\ref{fig: synthetic}.  That figure shows the value of the mixing angle for the synthetic signal injected into the data, $\theta_{\rm inj}$, and the mixing angle recovered by our analysis, $\theta_{\rm rec}$.  The 95\% one-sided limits are computed as the injected signal strength is varied. The limits never fall below the true value of the injected signal for any of the mass points shown. The mean, 1 and 2$\sigma$ expectations for the 95\% one-sided limit under the signal hypothesis were computed from the Asimov procedure~\cite{Cowan:2010js}.  Our limits are consistent with the real data being a realization of the null hypothesis and the only signal contribution comes from that which we inject.  

There is no inconsistency in that the lower 2$\sigma$ band for the 95\% one-sided limit falls below the injected value.  This is expected because the one-sided 95\% limit and the 2$\sigma$ bands for the 95\% limit have different statistical interpretations, because the 2$\sigma$ band is a 2-sided interval while the one-sided 95\% limit is a statement about a one-sided interval.  Figure~\ref{fig: synthetic} also shows that the lower 2$\sigma$ bands flatten at low injected mixing angles.  This is because we are showing power-constrained limits~\cite{Cowan:2011an}, and those values reach the constraints.

We also test the effect of assuming the wrong DM density profile. We consider how our limits change for different assumed DM profiles  below.  Here, we address the question of whether the evidence for a real DM-induced line may be obscured if an incorrect DM profile is used in the profile likelihood analysis.  We follow the same procedure described above to construct a hybrid dataset consisting of the real data and a synthetic signal at $m_s = 7.0$ keV.  That synthetic signal is constructed assuming our canonical NFW DM profile.  We then analyze the synthetic data assuming the DM profile follows the alternative Burkert DM profile~\cite{Burkert:1995yz,Salucci:2000ps}.  That profile is an extreme departure from the NFW DM profile, in that it has a roughly 9 kpc core.  The difference between the spatial morphologies of the NFW profile and the Burkert profile encapsulate the largest mismatch between the  DM profiles we test in this work and the real profile of the Milky Way.  

\begin{figure}[p]
\centering
\includegraphics[width = 0.5\textwidth]{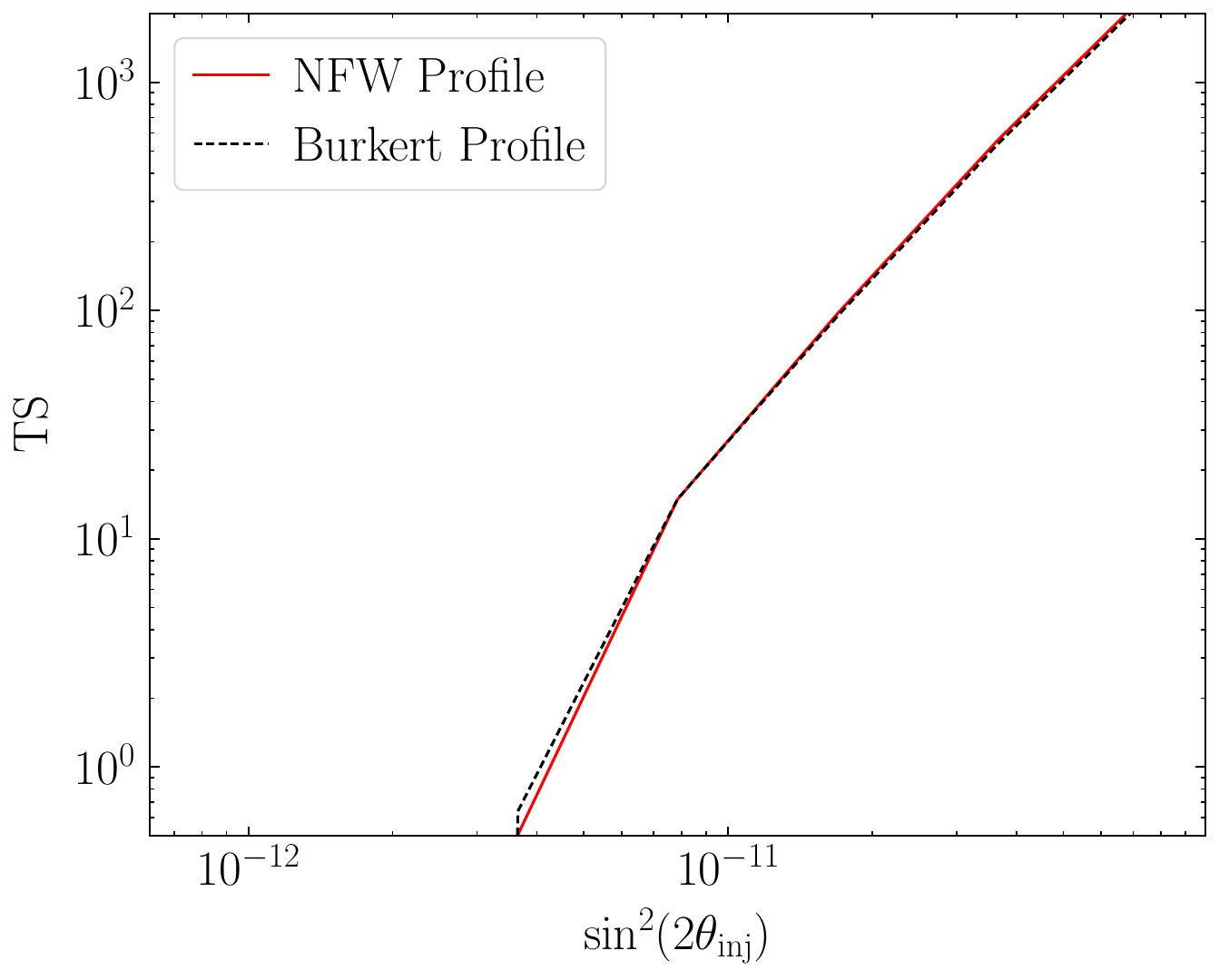} 
\vspace{-0.2cm}
\caption{\textbf{The effects of a different DM profile.} As in Fig.~\ref{fig: synthetic}, we add a fake DM signal to the real data, with mixing angle $\sin^2(2 \theta_\text{inj})$ as indicated on the x-axis.  Here we have fixed $m_s = 7.0$ keV.  We show the TS assuming the NFW DM profile (red), which was used in the production of the synthetic signal, and the Burkert profile (black dashed) with a 9 kpc core.  The TS is almost insensitive to the DM profile assumed.}
\label{fig: synthetic TS}
\end{figure}

In Fig.~\ref{fig: synthetic TS} we show the resulting TS in favor of DM as a function of the synthetic injected mixing angle, for an analysis that assumes the NFW DM profile and one that assumes the Burkert DM profile, with the NFW DM profile injected.  The two TS curves are extremely similar, so we conclude that a real signal is not going undetected because we do not have the correct DM density profile. In both cases the $D$-factor does not change appreciably between different exposures in our region of interest.  In both cases the TS at $\sin^2(2 \theta_{\rm inj}) \approx 10^{-10}$ is $\sim$$10^3$, meaning that at this signal strength a DM-induced line would have been detected at approximately 30$\sigma$. 

If the DM profile used in the analysis is not correct then the limit will be systematically biased.  However, Fig.~\ref{fig: synthetic TS} shows the true limit, constructed with the correct DM profile, may be obtained by rescaling the limit obtained with an incorrect DM profile by the appropriate ratio of mean $D$-factors, where the means are constructed from the ensemble of exposures used in the joint-likelihood analysis. 

As an additional cross-check, below we compute the limit on the DM-induced line in regions consisting of narrow annuli centered around the Galactic Center.  For all the DM profiles considered, these annuli are small enough that the DM density does not change appreciably between exposures in these subregions. 

\subsection{Individual Exposures}

Our fiducial result relied on the construction of the joint likelihood over 1,397 independent exposures.  We now consider the sensitivity of the most constraining individual exposures, and their individual properties. We begin by describing how the individual exposures in our fiducial analysis are distributed.

\subsubsection{Spatial Distribution of exposures}

\begin{figure}[p]
\centering
\includegraphics[width = 0.7\textwidth]{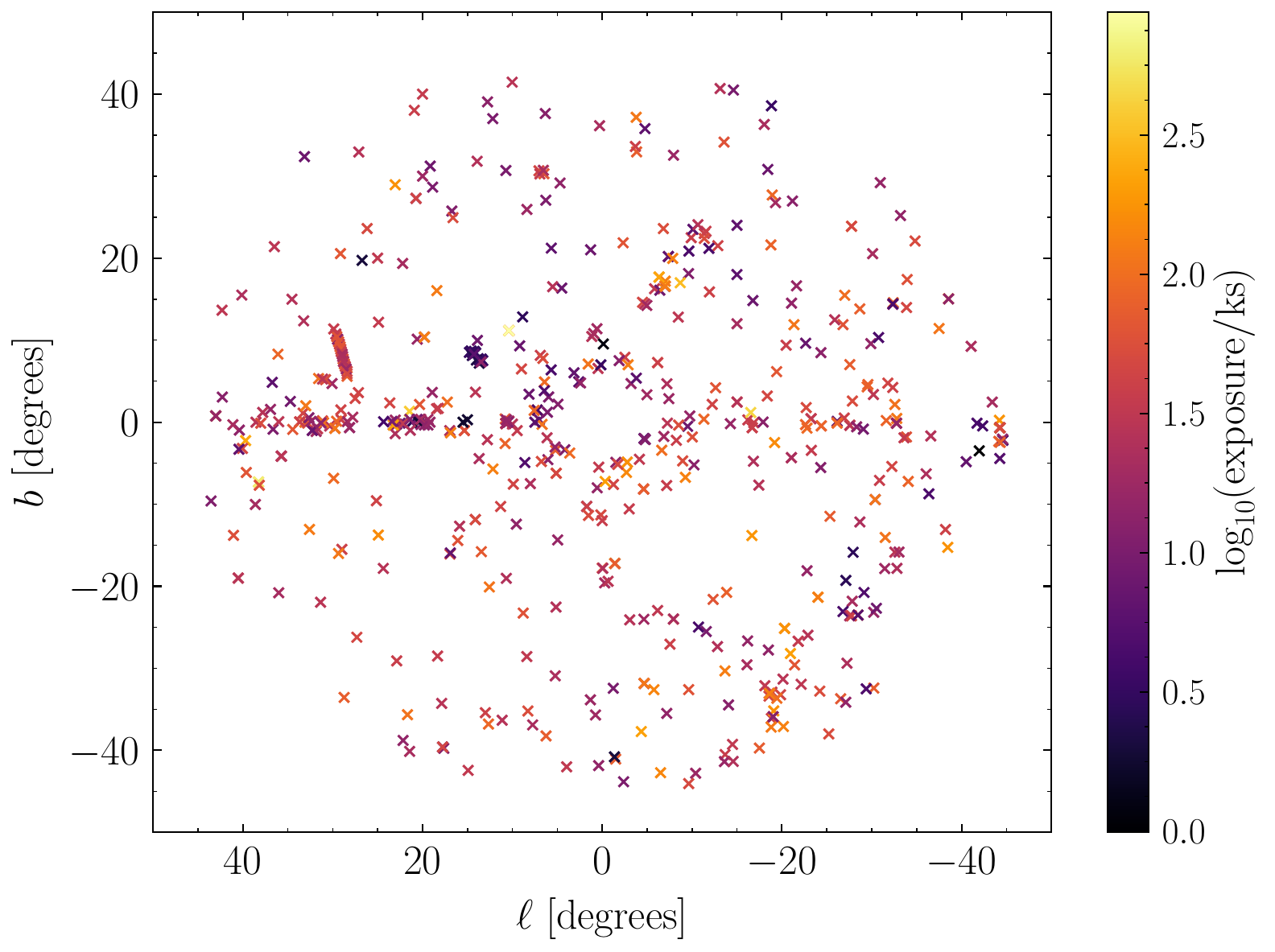} 
\vspace{-0.2cm}
\caption{\textbf{A map of the exposure times.} Exposure times for the exposures included in the fiducial analysis on a map of galactic coordinates, where $l$ is longitude and $b$ is latitude.  In cases where multiple exposures occur at the same position, we only show the longest exposure time.}
\label{fig: exp_map}
\end{figure}

\begin{figure}[p]
\centering
\includegraphics[width = 0.49\textwidth]{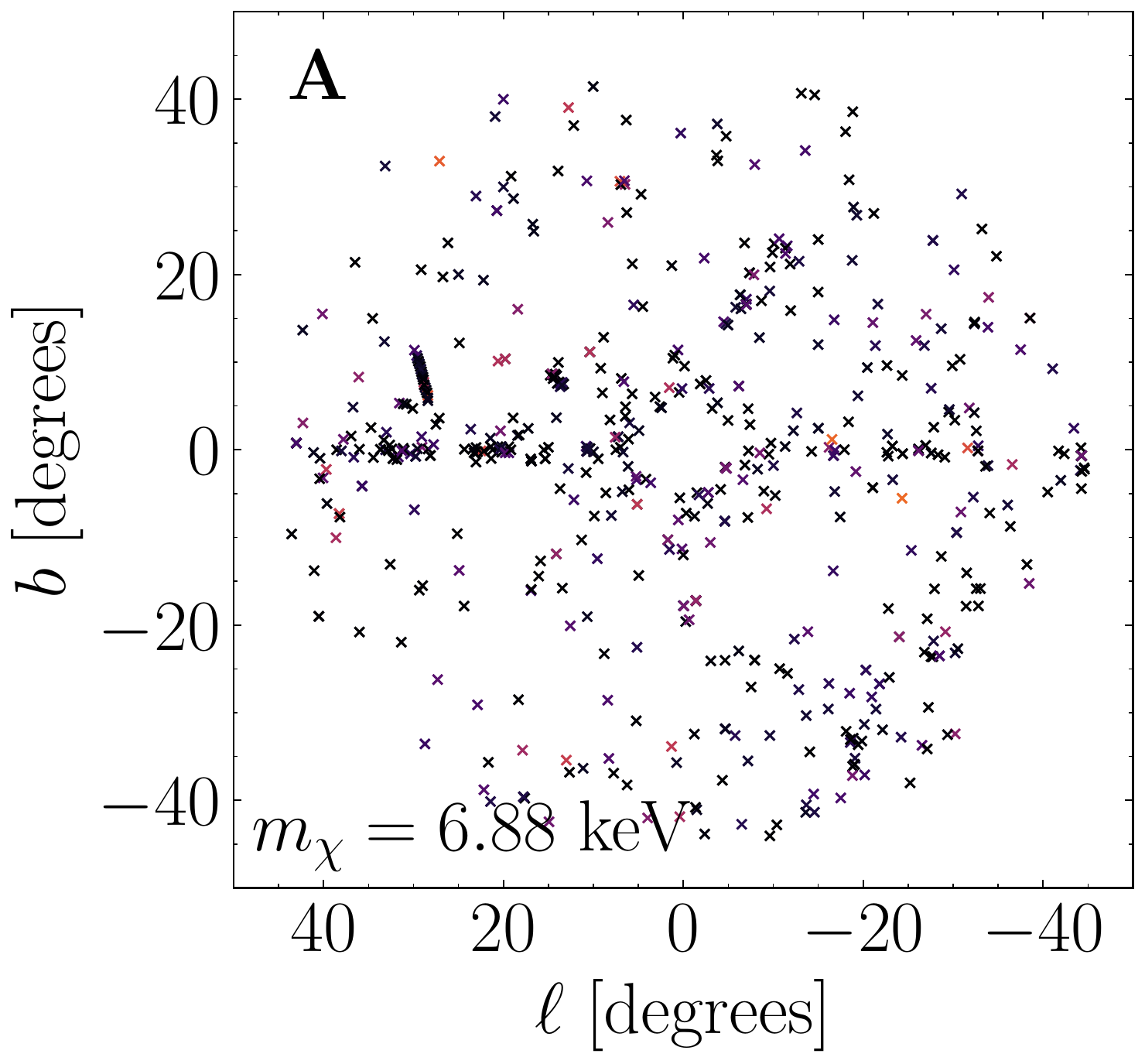}
\includegraphics[width = 0.49\textwidth]{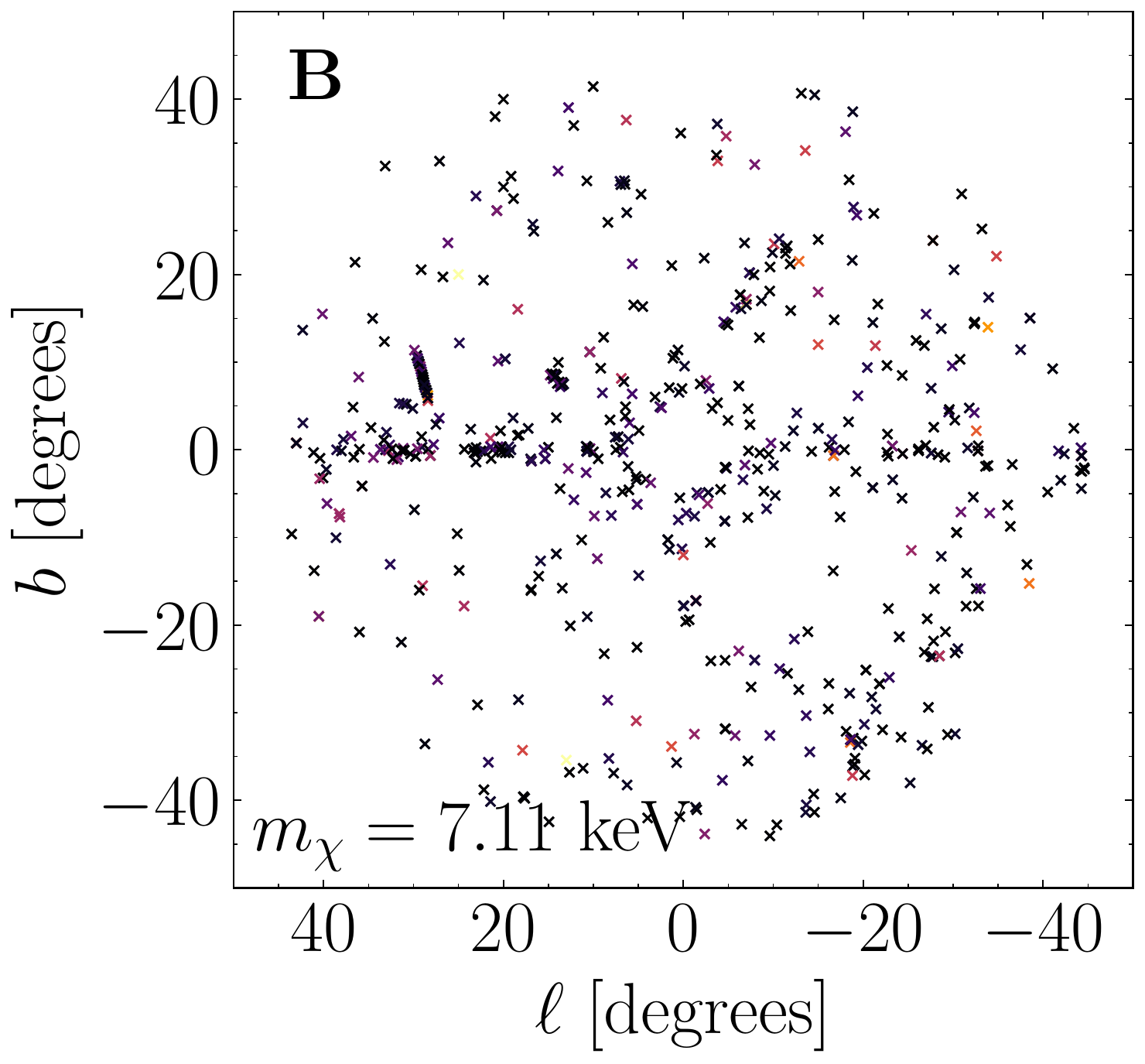}
\includegraphics[width = 0.49\textwidth]{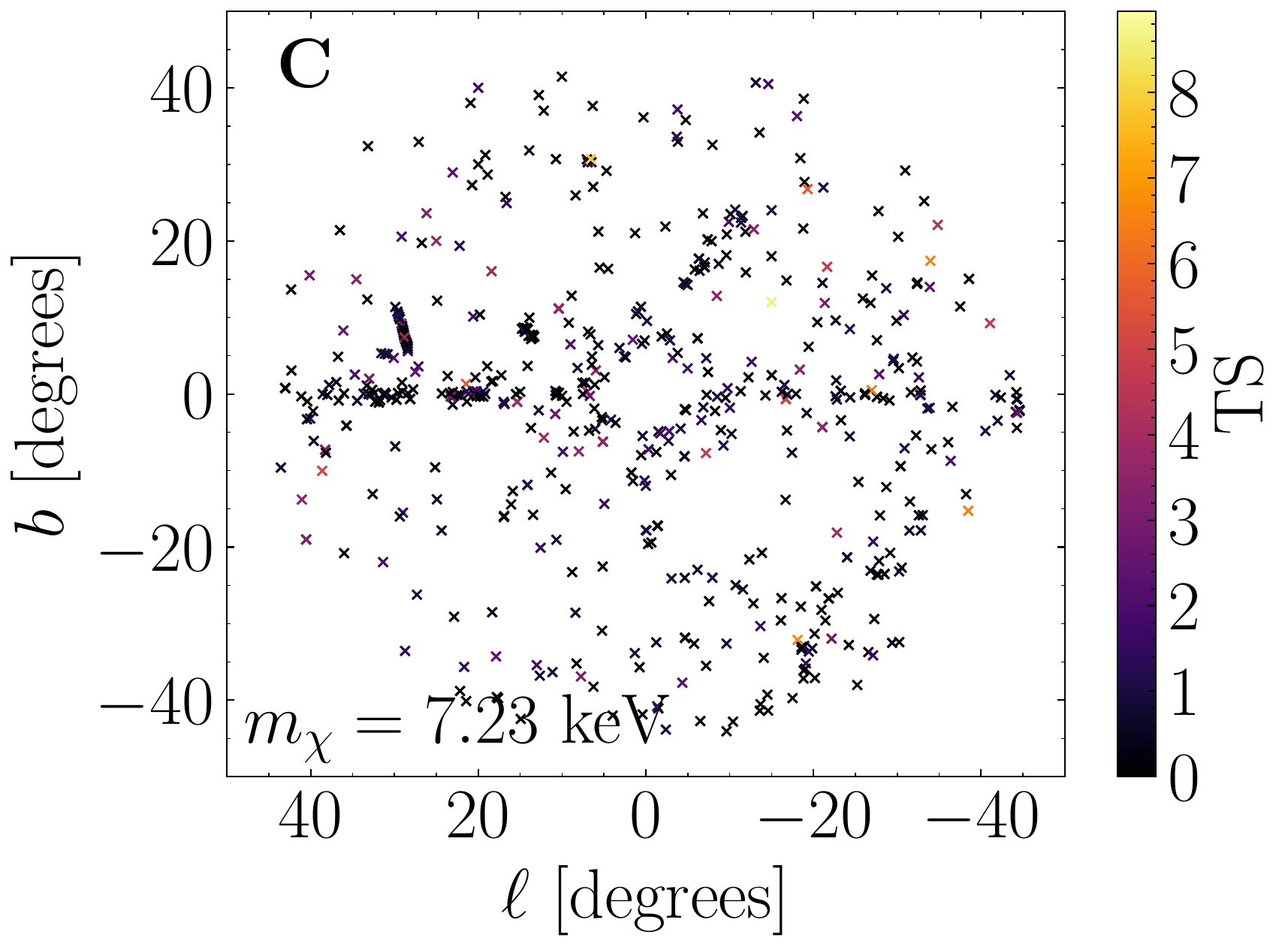} 
\vspace{-0.2cm}
\caption{\textbf{Maps of the maximum TSs.} The maximum TSs for the individual exposures illustrated in Fig.~\ref{fig: exp_map} at three different mass points 6.88 keV (A), 7.11 keV (B), and 7.28 keV (C).  The high-TS exposures appear randomly distributed about the region.}
\label{fig: TS_map}
\end{figure}

Fig.~\ref{fig: exp_map} shows the spatial distribution of the exposures included in the fiducial analysis about the Galactic Center.  In cases where there are multiple exposures at the same location, we have only shown the highest exposure case.  This can be compared with Fig.~\ref{fig: TS_map}, which shows the TS at three different mass points for the exposures illustrated in Fig.~\ref{fig: exp_map}.  The high-TS exposures do not correlate with distance from the Galactic Center and appear randomly distributed about the region.

\clearpage

\subsubsection{Goodness of fit for individual exposures}

 We quantify the goodness of fit for an individual exposure through the $\delta \chi^2$ per degree of freedom.  In calculating  $\delta \chi^2$ we only include the X-ray count data, and we also take the degrees of freedom to be the number of data points minus two to account for the two degrees of freedom in the astrophysical power-law.  We assume that the QPB model parameters are already fixed by the QPB data, for the purpose of counting model parameters.  There are typically $\sim$100 energy bins in the 0.5 keV energy window around the putative line energy considered in the analysis.  The exact number of energy bins varies slightly as a function of the line energy.  We present results for $m_s = 7.1$ keV, though the results at other masses are similar.  In this case, there are 100 energy bins included in the MOS analyses and 97 in the PN analyses. Thus, we take 98 (95) degrees of freedom for the MOS (PN) exposures.  

\begin{figure}[p]
\centering
\begin{subfigure}{0.49\textwidth}
\includegraphics[width = \textwidth]{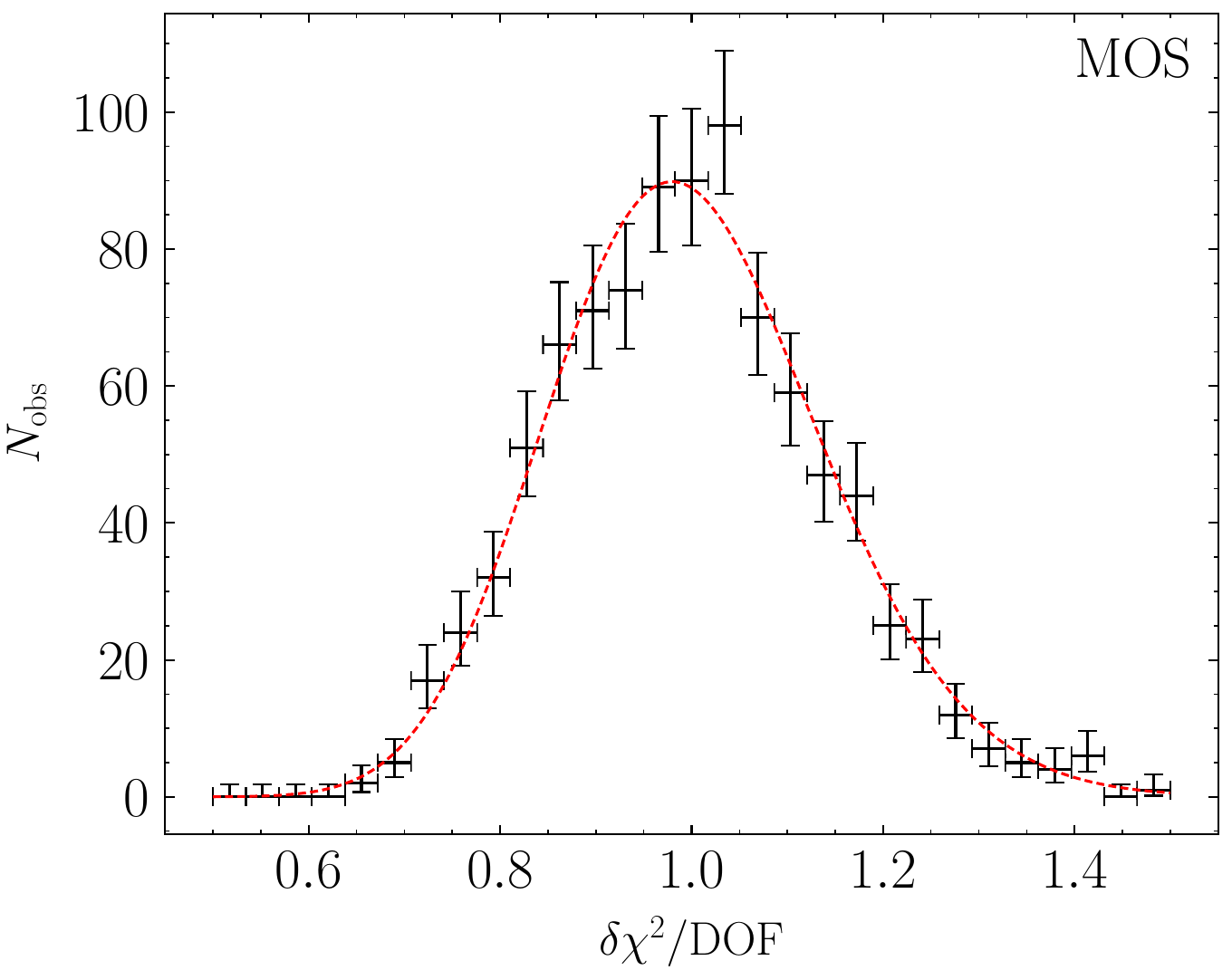} 
    \subcaption{}
\end{subfigure}
\begin{subfigure}{0.49\textwidth}
\includegraphics[width = \textwidth]{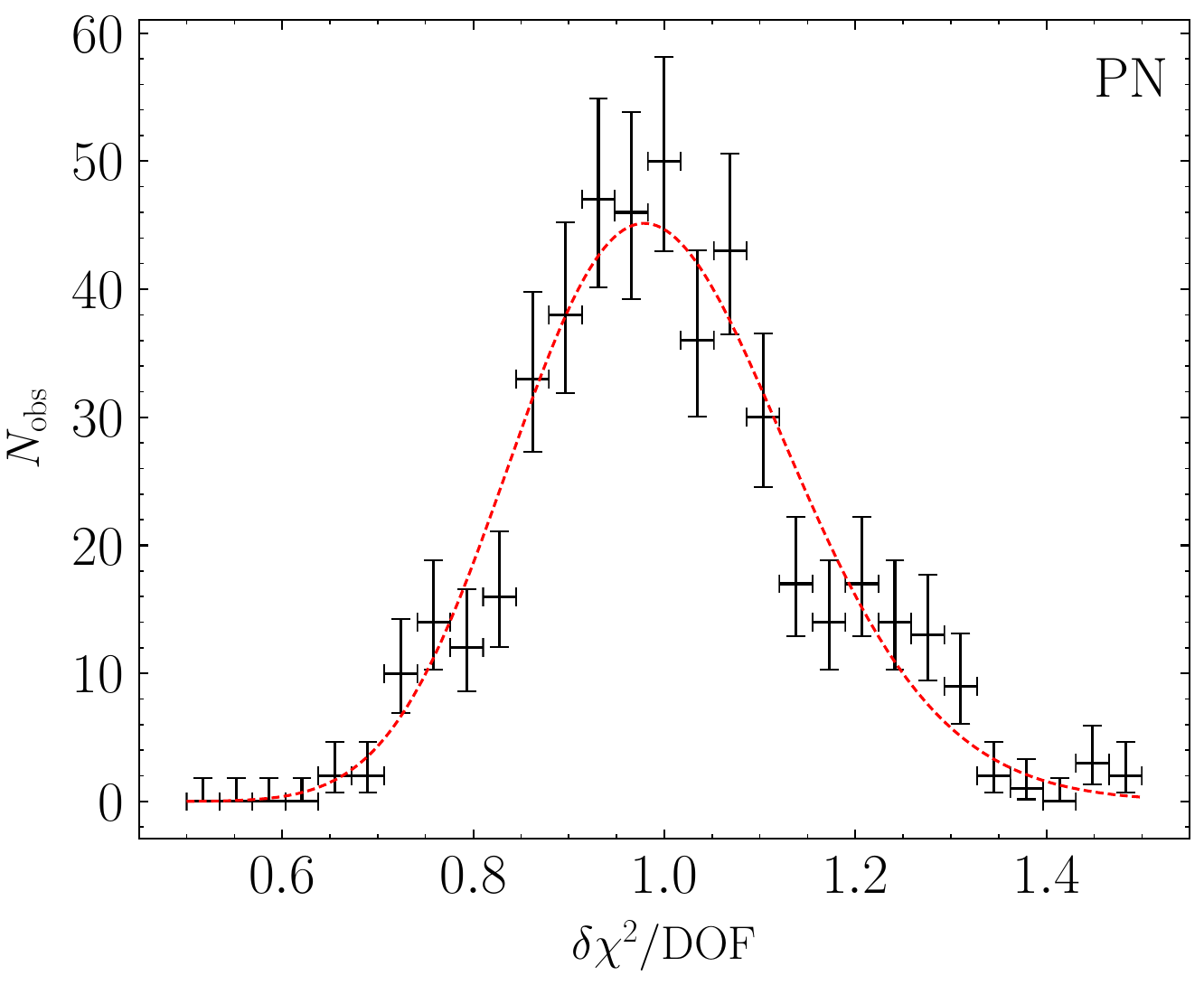}
    \subcaption{}
\end{subfigure}
\vspace{-0.2cm}
\caption{\textbf{The distribution of $\delta \chi^2 / \text{DOF}$ for the MOS (A) and PN (B) exposures considered in our fiducial analysis for $m_s = 7.1$ keV.}  The number of degrees of freedom is 98 (95) for the MOS (PN) exposures.  Under the null hypothesis, these distributions should follow the appropriate $\chi^2$-distributions, which are shown in dashed red. The vertical error bars on the black data points are the 1$\sigma$ Poisson counting uncertainties, while the horizontal error bars show the bin ranges.  The observed data are consistent with the null hypothesis model. }
\label{fig: deltachi2}
\end{figure}

In Fig.~\ref{fig: deltachi2} we show the distribution of $\delta \chi^2$ per degree of freedom (DOF) over all of the MOS and PN exposures in the fiducial analysis.  The vertical error bars show the 1$\sigma$ Poisson counting uncertainties.  The data histograms are consistent with expectations under the null hypothesis.  Under the null hypothesis these distributions should follow the $\chi^2$ distribution with the appropriate number of degrees of freedom. 

\subsubsection{Top 10 Exposures}

\begin{figure}[p]
\centering
\begin{subfigure}{0.49\textwidth}
\includegraphics[width = \textwidth]{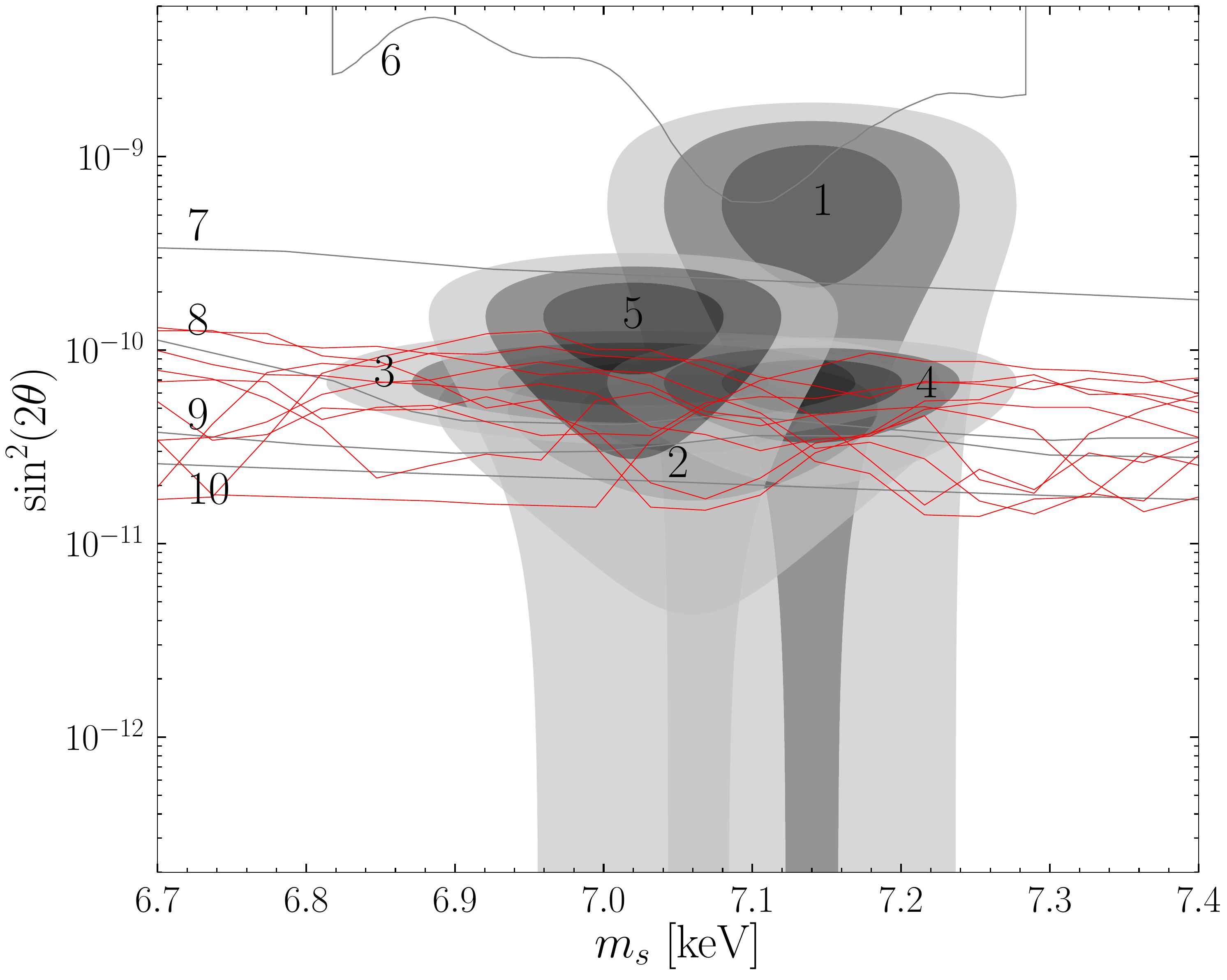} 
    \subcaption{}
\end{subfigure}
\begin{subfigure}{0.49\textwidth}
\includegraphics[width = \textwidth]{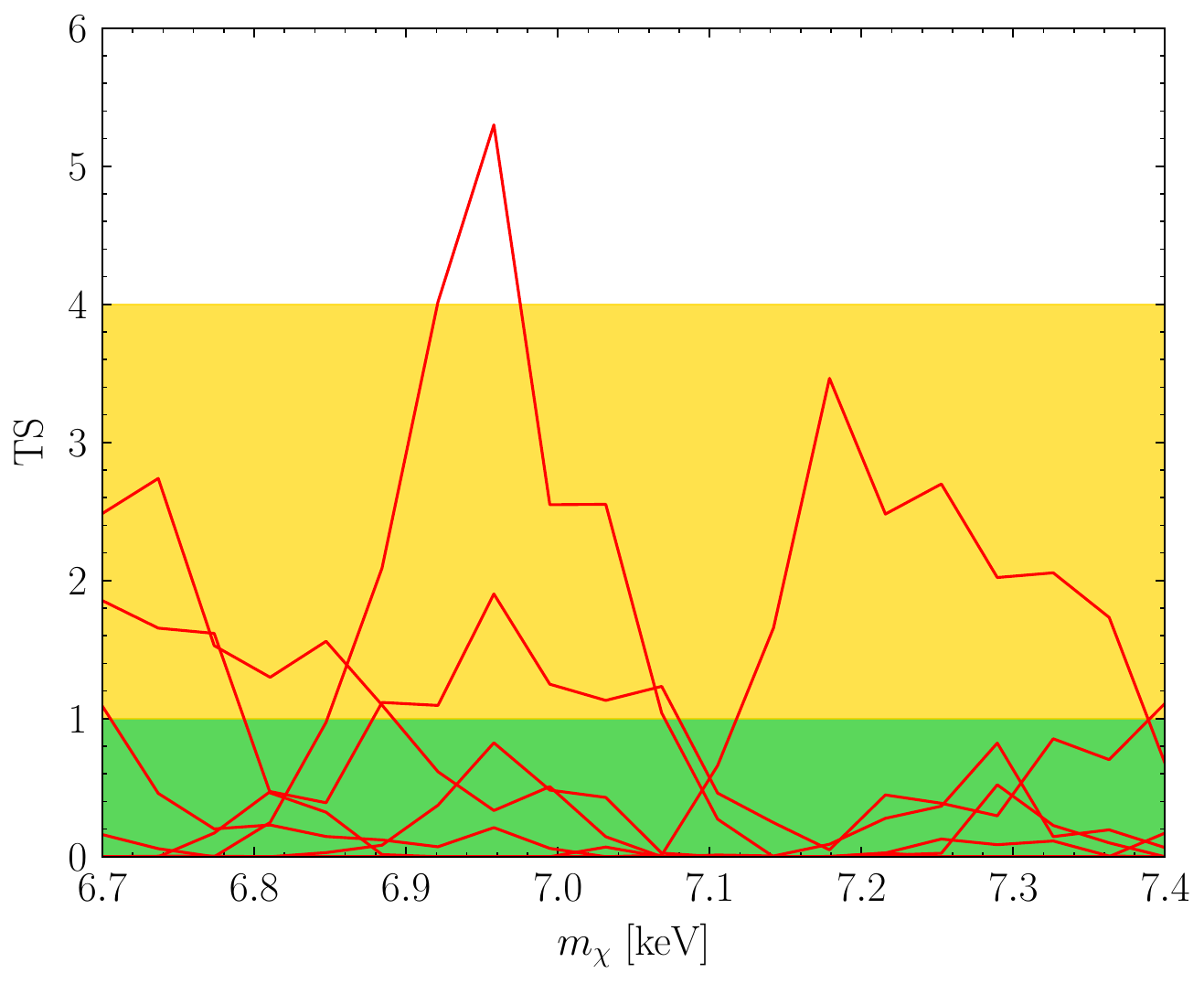}
    \subcaption{}
\end{subfigure}
\caption{\textbf{Limits from individual exposures}. (A) The one-sided power-constrained 95\% limits (red) from the 10 most constraining exposures, which are listed in Table~\ref{tab: IC}.  The shaded regions and pre-existing constraints are as labeled in Fig.~\ref{fig: main-figure}.  (B) The maximum TSs (red) for the 10 exposures.  The distribution of TSs observed is consistent with the null hypothesis. The green and yellow regions indicate 1$\sigma$ and 2$\sigma$ detections, respectively.}
\label{fig: top10}
\end{figure}

We show the limits obtained from the top 10 exposures individually in Fig.~\ref{fig: top10}. These exposures are listed in Table~\ref{tab: IC}, ranked in order of the strongest predicted limit under the null hypothesis, from the Asimov analysis at $m_s = 7.0$ keV.  None of the top 10 exposures were proposed to search for extended emission.  These ten observations were all looking at specific astrophysical sources, which we mask in our analysis.  

\begin{table}[p]
\caption{\textbf{The 10 most constraining \textit{XMM-Newton} exposures in the fiducial analysis.}  The exposures are ranked by their predicted limits under the null hypothesis at $m_s = 7.0$ keV from the Asimov analysis. The ``Identifier'' column denotes the specific exposure within an observation. LMXRB stands for ``low-mass X-ray binary.''}
\centering
\begin{tabular}{C{2.2cm}C{1.4cm}C{1.7cm}C{1.8cm}C{1.2cm}C{1.2cm}C{3.0cm}C{0.0cm}}
\toprule
\Xhline{3\arrayrulewidth}
Observation ID & Camera & Identifier & Exposure [ks] & $l$ [deg] & $b$ [deg] & Target Type & \vspace{0.2cm} \\
\midrule \hline
0653550301 & PN & S003 & 63.2 & 5.1 & -6.2 & Quiescent Novae & \vspace{0.1cm} \\
0203750101 & PN & S003 & 33.7 & -2.8 & -4.9 & LMXRB Black Hole & \vspace{0.1cm} \\
0152750101 & PN & S001 & 30.1 & 1.6 & 7.1 & Dark Cloud & \vspace{0.1cm} \\
0203750101 & MOS2 & S002 & 43.4 & -2.8 & -4.9 & LMXRB Black Hole & \vspace{0.1cm} \\
0781760101 & PN & S003 & 46.0 & -2.7 & -6.1 & LMXRB Burster & \vspace{0.1cm} \\
0761090301 & PN & S003 & 95.2 & -8.7 & 17.0 & B2III Star & \vspace{0.1cm} \\
0206610101 & PN & S003 & 35.4 & -2.9 & 7.0 & Dark Cloud & \vspace{0.1cm} \\
0412601501 & MOS2 & S002 & 90.2 & -1.4 & -17.2 & Neutron Star & \vspace{0.1cm} \\
0727760301 & MOS2 & S003 & 67.9 & -1.4 & -17.2 & Neutron Star & \vspace{0.1cm} \\
0761090301 & MOS2 & S002 & 107.4 & -8.7 & 17.0 & B2III Star & \vspace{0.1cm} \\
\bottomrule
\Xhline{3\arrayrulewidth}
\end{tabular}

\label{tab: IC}
\end{table}

Fig.~\ref{fig: top10}A shows the one-sided 95\% power-constrained limits obtained from these exposures.  Many of these top 10 exposures are themselves strong enough to independently disfavor the decaying DM interpretation of the UXL.  None of these exposures show evidence for an UXL.  This is illustrated in Fig.~\ref{fig: top10}B, which shows the TSs as a function of mass for the top 10 exposures.  There is only one exposure whose TS exceeds the 2$\sigma$ expectation.  This, however, is not surprising, considering that there are 10 independent exposures and each exposure has roughly three independent mass points across the mass range considered.

\subsubsection{Profile Likelihood for the Top Exposure}

We describe in detail our most constraining exposure, observation ID 0653550301 in Table~\ref{tab: IC}.  Our goal is to illustrate the profile likelihood procedure at the level of the individual exposures.

\begin{figure}[p]
\centering
\begin{subfigure}{0.49\textwidth}
\includegraphics[width = \textwidth]{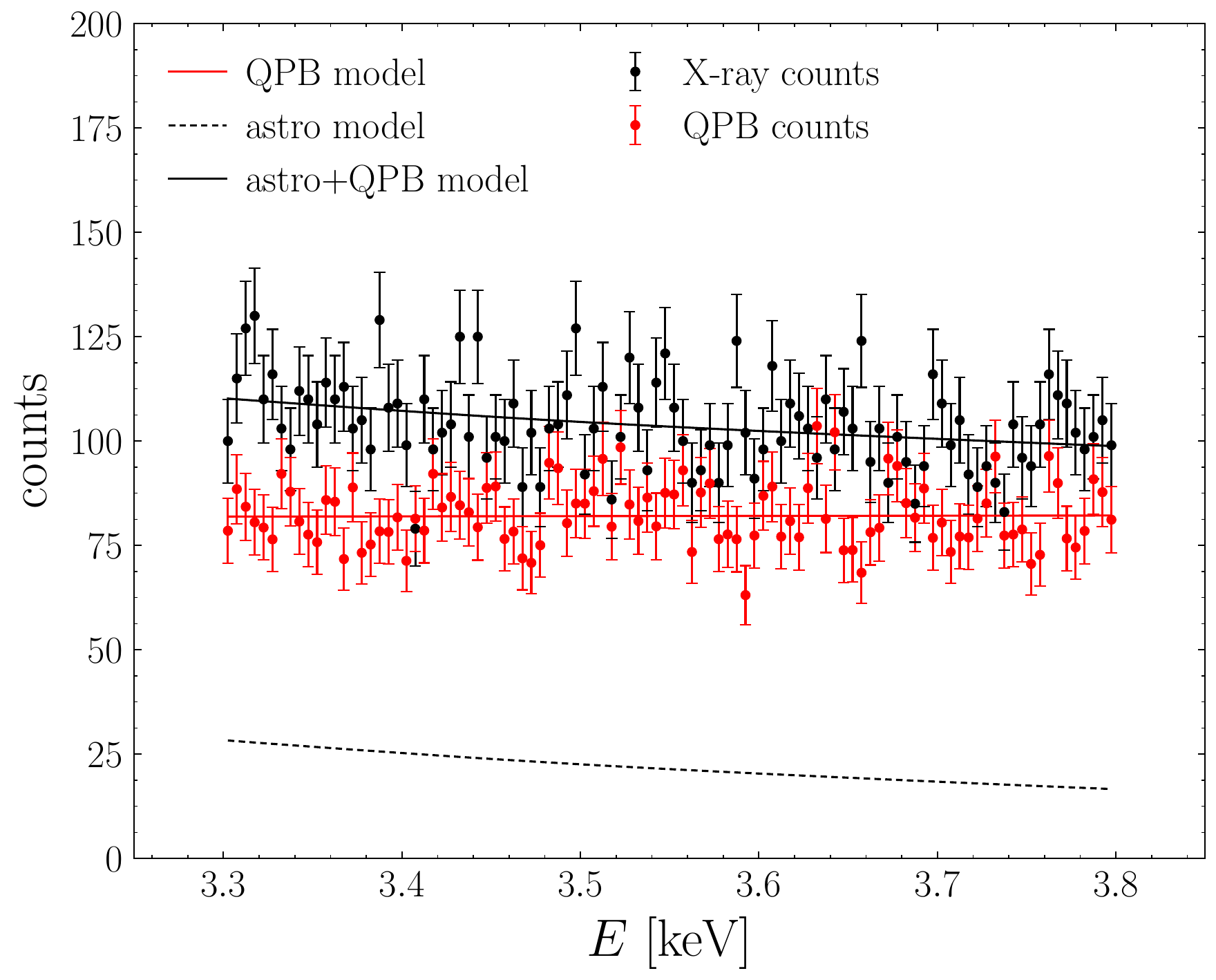} 
    \subcaption{}
\end{subfigure}
\begin{subfigure}{0.49\textwidth}
\includegraphics[width = \textwidth]{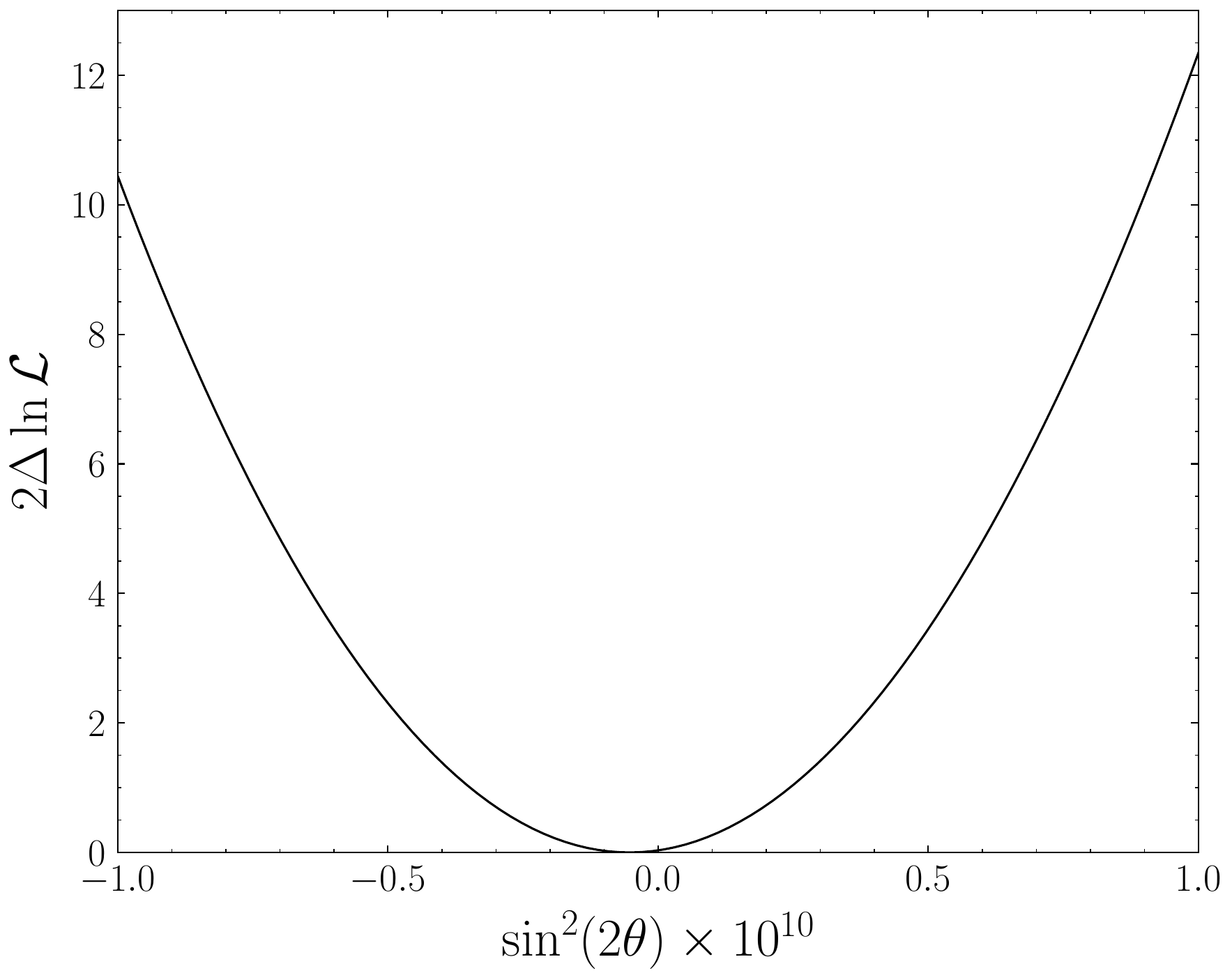}
    \subcaption{}
\end{subfigure}
\caption{\textbf{Results for our most constraining exposure.} (A) An example spectrum obtained from the PN camera of observation ID 0653550301, our most constraining exposure, as listed in Table~\ref{tab: IC}.  In addition to the data (black circles) we show the best-fitting QPB and astrophysical models (red line and dashed black line respectively), under the assumption of no UXL.  The energy range shown corresponds to that in our fiducial analysis, and the individual energy bins are $0.015$ keV wide. (B) The profile likelihood for the strength of the 3.55 keV signal for the dataset shown on the left, in terms of the mixing angle.}
\label{fig: example}
\end{figure}

The X-ray count and QPB data for this exposure are shown Fig.~\ref{fig: example}A. The data are shown over a $0.5$ keV energy range centered around the example line energy of $3.55$ keV. The figure also shows the best-fitting QPB and astrophysical models under the assumption of no signal.  The models match the data within the statistical noise, which can be quantified by calculating the $\chi^2$ per degree of freedom: $\chi^2/\text{DOF} \approx 1.016$.  We then construct the profile likelihood for the putative line signal at 3.55 keV. In constructing the profile likelihood we re-fit for the best-fitting nuisance parameters for each value of the line signal strength, as is mandated by the profile likelihood procedure.  The resulting profile likelihood is shown in Fig.~\ref{fig: example}B.  We show the profile likelihood as twice the difference in log likelihood with the convention $2 \Delta \ln \mathcal{L} = 2 [ \ln {\mathcal L}(\sin^2(2\hat{\theta})) -   \ln {\mathcal L}(\sin^2(2 \theta)) ]$, where $\hat{\theta}$ is the value of $\theta$ that maximizes the likelihood. Note that the best-fitting mixing angle is slightly negative in this case. To convert from counts to flux to $\sin^2(2 \theta)$ within this exposure and for $m_s =3.55$ keV, we use the following properties. First, the average $D$-factor within this region for the fiducial NFW profile is $9.15\times 10^{28}$ keV/cm$^2$. Second, the channel bin widths are $0.015$ keV wide because this is a PN exposure.  And third, a spectral value of 1 count/cm$^2$/s/sr/keV at 3.55 keV produces, on average, $\sim$22 counts, distributed across a $\sim$$0.2$ keV-wide window in channels about 3.55 keV, due to the energy resolution of the camera.

We compare the $2-10$ keV intensity and the QPB rate for this exposure to sets of cuts on our fiducial analysis.  Under the null hypothesis, we infer $F_{2-10} \approx 3.47 \times 10^{-11}$ erg$/$cm$^2/$s$/$deg$^2$ and a QPB rate that is in the lower 57\% percentile, with a rate of $\sim$0.127 QPB counts$/$s.

\subsection{Dependence on the Dark Matter Profile}

Here we consider how our results depend on the assumed DM profile for the Milky Way.  For our fiducial analysis we used the NFW DM profile~\cite{Navarro:1995iw,Navarro:1996gj} for the DM density $\rho_\text{DM}(r)$
\begin{equation}
\rho_\text{NFW}(r) = \frac{\rho_0}{r/r_s\,\left( 1  + r/r_s\right)^2 } \,,
\end{equation}
with $r$ the distance from the Galactic Center and $r_s$ the scale radius. The density normalization parameter $\rho_0$ is fixed to give the measured local DM density $\rho_\text{local}$ at the solar radius $r_\odot$. In our fiducial analysis, we took $\rho_\text{local} = 0.4$ GeV$/$cm$^3$, $r_\odot = 8.13$ kpc, and $r_s = 20$ kpc.

\begin{figure}[p]
\centering
\includegraphics[width = 0.49\textwidth]{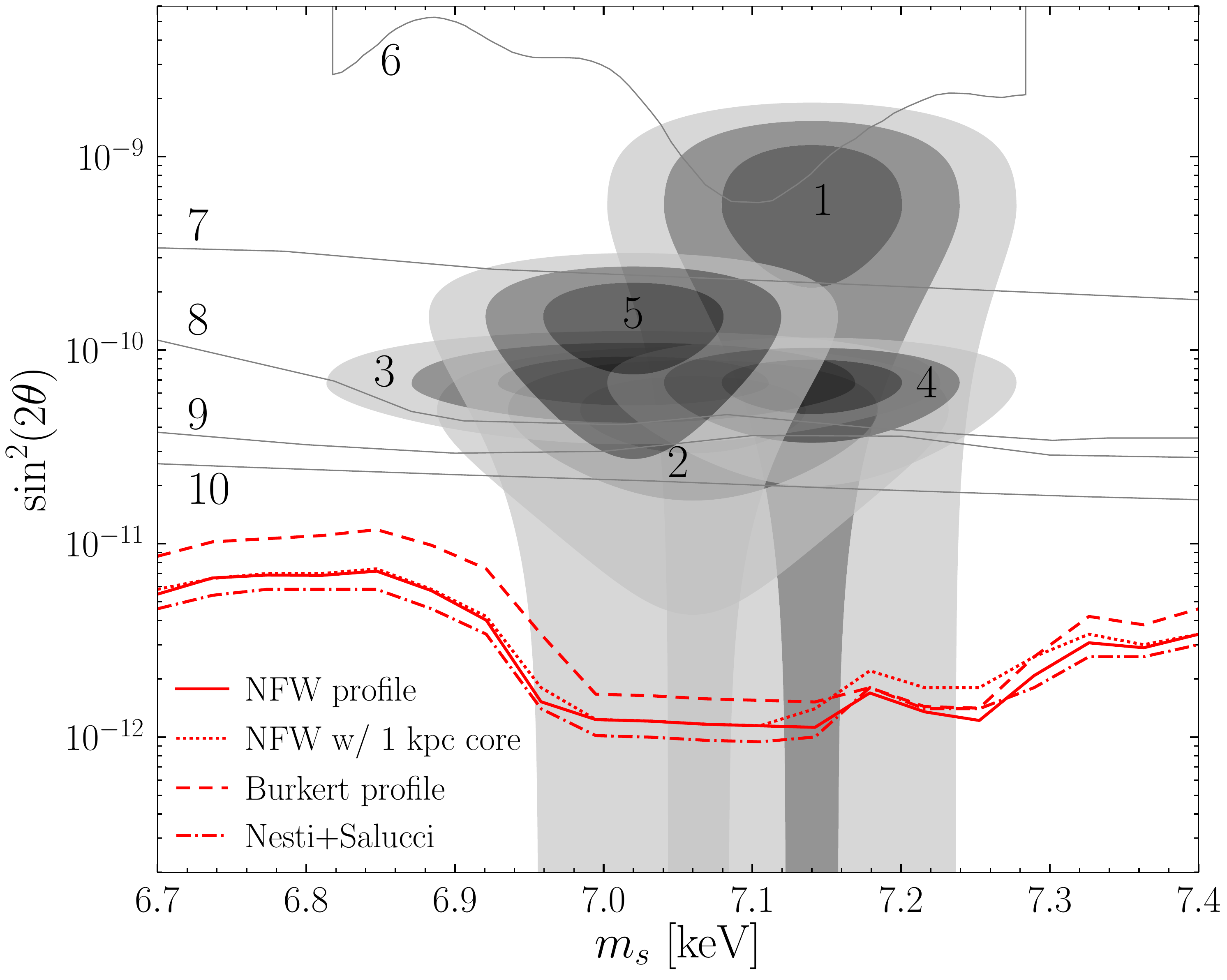}
\caption{\textbf{Limits with different DM density profiles.} The parameter space from Fig.~\ref{fig: spectra}, compared to our limits for different assumptions about the DM density profile.  In additon to the fiducial NFW profile (solid red), we consider the NFW profile with a 1 kpc core (dot-short dashed red), an NFW with $r_s = 16$ kpc and $\rho_0 = 0.47~{\rm GeV/cm}^3$~\cite{Nesti:2013uwa} (dot-long dashed red), and the Burkert profile with a 9 kpc core (dashed red). See text for details.}
\label{fig: DM syst}
\end{figure}

We consider the effects of departing from the assumed NFW profile.  One possibility is that baryonic feedback in the inner regions of the Milky Way leads to the formation of a dark matter core in the inner part of the DM halo. Hydrodynamic simulations of Milky Way size galaxies suggest that feedback could increase DM density~\cite{Hopkins:2017ycn}, which would increase our sensitivity.  Taking the more pessimistic scenario, we consider the possibility that within the inner 1 kpc of the Milky Way, the DM density profile is flat:
\begin{equation}
\rho_\text{core}(r) = \left\{ 
\begin{array}{ll}
\rho_\text{NFW}(r) & r > r_c\,, \\
\rho_\text{NFW}(r_c) & r \leq r_c \,,
\end{array} \right.
\end{equation}
where $r_c = 1$ kpc is the core radius. Figure~\ref{fig: DM syst} compares the limit we obtain with this DM profile to our fiducial limit; the difference between the two limits is small.  This is because we mask the inner $5^\circ$ of the Milky Way, which covers almost the entire region that would be affected by the 1 kpc core. Varying other fiducial NFW parameters has little impact. For example, using $r_s = 16$ kpc and $\rho_0 = 0.47~{\rm GeV/cm}^3$~\cite{Nesti:2013uwa}, we find only minor impact on our results.

We also consider the Burkert DM profile~\cite{Burkert:1995yz,Salucci:2000ps}:
\begin{equation}
\rho_\text{Burk}(r) = \frac{\rho_0}{( 1 + r/r_c ) (1 + (r/r_c)^2 ) } \,,
\end{equation}
where $r_c$ is the core radius and again $\rho_0$ is fixed by matching $\rho_\text{local}$. To be conservative, we take the core radius to be $r_c = 9$ kpc~\cite{Nesti:2013uwa}, which effectively corresponds to coring the DM density profile within the solar radius.  While there is no evidence that this density profile describes our own Milky Way, we consider it the largest plausible deviation from the NFW profile.  The limit is also shown in Fig.~\ref{fig: DM syst}.  Even with such an extreme DM profile we still find that the best-fitting parameters for DM to explain the UXL remain inconsistent with our results.  Changing between the NFW and Burkert profiles has a small effect on the limit, as within our fiducial region the difference between the $D$-factors computed between the two profiles is relatively small. 

\subsection{MOS and PN Independent Analyses}

\begin{figure}[p]
\centering
\begin{subfigure}{0.49\textwidth}
\includegraphics[width = \textwidth]{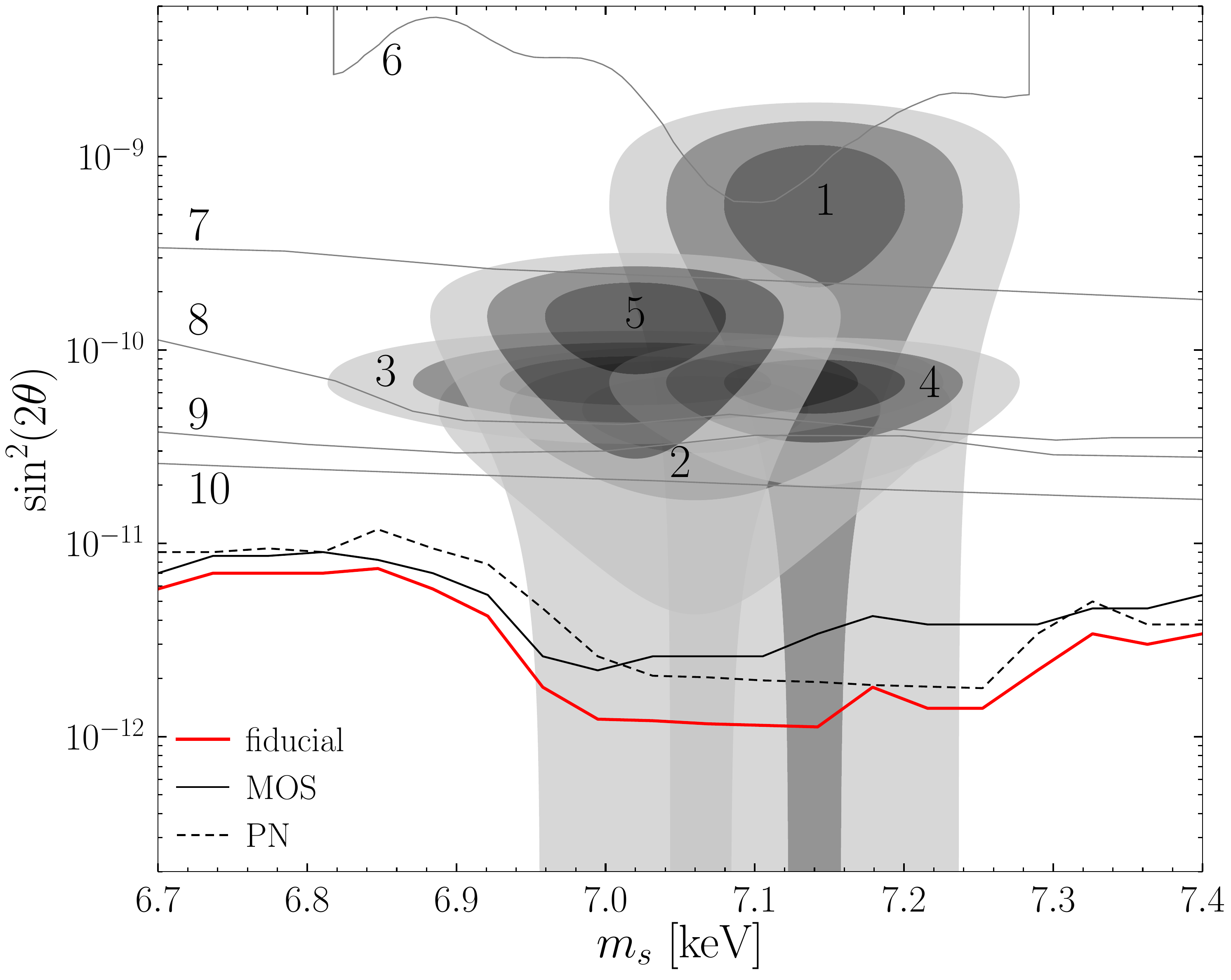} 
    \subcaption{}
\end{subfigure}
\begin{subfigure}{0.49\textwidth}
\includegraphics[width = \textwidth]{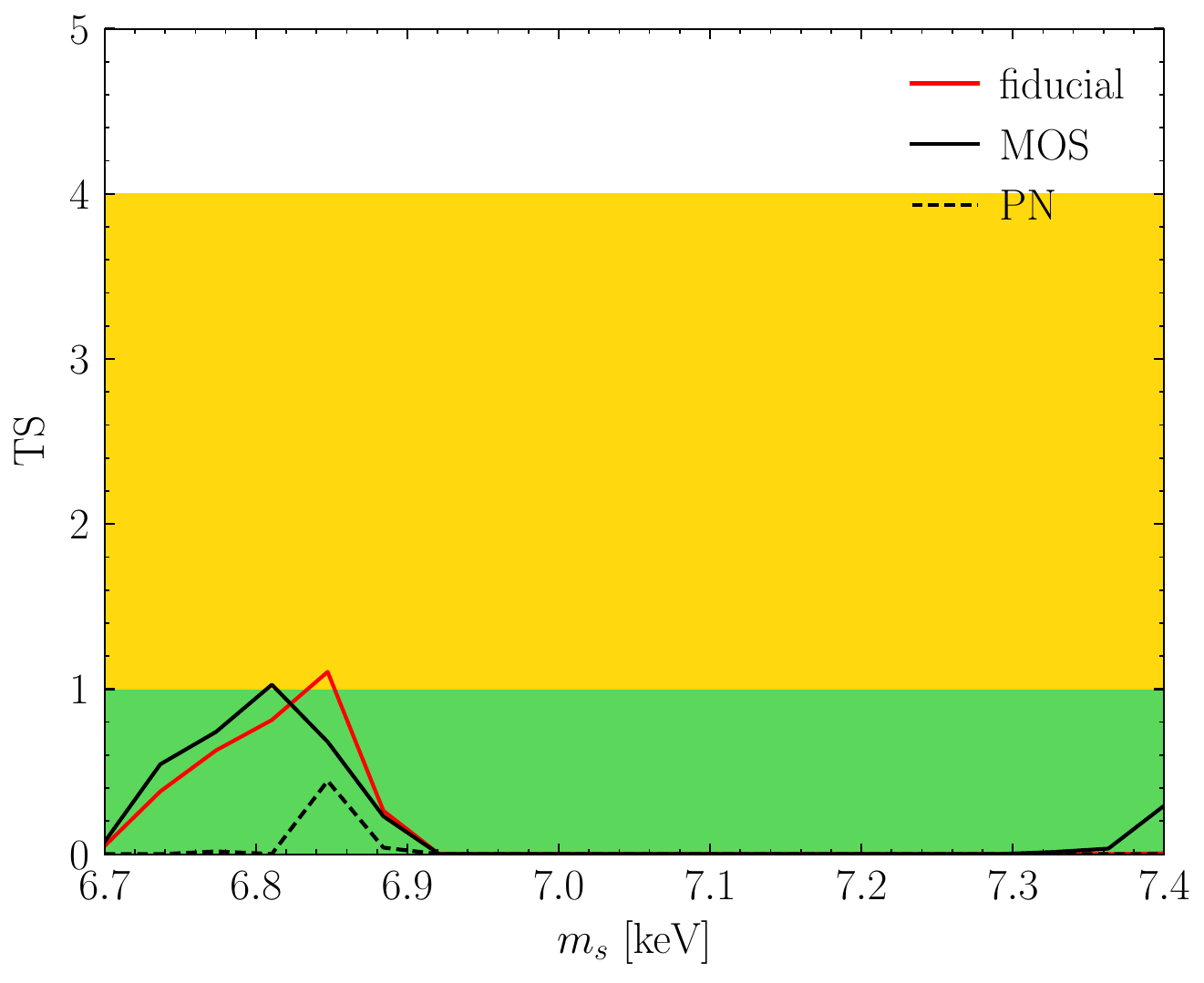}
    \subcaption{}
\end{subfigure}
\caption{\textbf{Exploring the results from individual cameras.} Variations to the limit (A) and TS (B) arising from performing independent analyses on the MOS (solid black) and PN (dashed black) datasets, to test for possible systematic effects present in one camera but not in the other. These can be compared to the fiducial results (red). We find that both limits are independently inconsistent with the decaying DM interpretation of the UXL.}
\label{fig: mos-pn}
\end{figure}

We analyze the data from the MOS and PN cameras separately to test for systematic issues in the cameras.  Because these are different detectors, we also expect their instrumental systematic uncertainties (such as effective area uncertainties and possible instrumental lines) to also be largely independent.  In Fig.~\ref{fig: mos-pn} we show the limits and TS distributions from independent analyses of the MOS and PN datasets, as compared to the joint (fiducial) analysis.  Neither dataset shows evidence for DM decay, and both constrain the decaying DM interpretation of the UXL. A low-significance feature (TS $\sim$1) is seen in both datasets at $m_s \sim 6.8$ keV.  While this feature could be a statistical fluctuation, it is also possible that it is due to a common systematic, such as a feature in the background emission, that affects both analyses.  Nevertheless, this feature is not distinguishable from statistical fluctuations.

\subsection{Variations to Selection Criteria}

We tested how our limits change when we vary the selection criteria for the exposures used to produce the joint likelihood.  We summarize the various combinations of the criterion that we consider in Table~\ref{tab: syst}. The region of interest extends from $r_\text{min}$ to $r_\text{max}$ from the Galactic Center, with the Galactic plane masked at $|b|_\text{min}$.  We include all exposures with 2-10 keV intensity less than $I_{2-10}^\text{max}$.  Similarly, we include exposures with QPB rates in the lower $F_\text{QPB}^\text{max}$ percentile, separately determined for MOS and PN exposures. For two of the analyses, we mask either the northern or southern hemispheres as well.

\begin{table}[p]
\caption{\textbf{The different selection criteria that we test in Fig.~\ref{fig: syst}.} These are variations to the cuts we have chosen in our fiducial analysis.}
\centering
\begin{tabular}{C{1.7cm}C{0.8cm}C{0.8cm}C{0.8cm}C{3.0cm}C{0.8cm}C{2.0cm}C{2.2cm}C{0.0cm}}
\toprule
\Xhline{3\arrayrulewidth}
& $r_\text{min}$ [deg] & $r_\text{max}$ [deg] & $|b|_\text{min}$ [deg] & $I_{2-10}^\text{max}$ [erg$/$cm$^2/$s$/$deg$^2$] & $F_\text{QPB}^\text{max}$ [\%] & Exposure [Ms] & other & \vspace{0.2cm} \\
\midrule \hline 
fiducial & 5 & 45 & 0 & $10^{-10}$ & 68 & 30.6 & - & \vspace{0.1cm} \\
$r \geq 10^\circ$ & 10 & 45 & 0 & $10^{-10}$ & 68 & 27.9 & - & \vspace{0.1cm} \\
$r \leq 60^\circ$ & 5 & 60 & 0 & $10^{-10}$ & 68 & 56.9 & - & \vspace{0.1cm} \\
$b \geq 1.5^\circ$ & 5 & 45 & 1.5 & $10^{-10}$ & 68 & 24.8 & - & \vspace{0.1cm} \\
north & 5 & 45 & 0 & $10^{-10}$ & 68 & 12.5 & mask $b<0^{\circ}$ & \vspace{0.1cm} \\
south & 5 & 45 & 0 & $10^{-10}$ & 68 & 18.1 & mask $b>0^{\circ}$ & \vspace{0.1cm} \\
$F_{2-10}^{\rm low}$ & 5 & 45 & 0 & $5 \times 10^{-11}$ & 68 & 18.8 & - & \vspace{0.1cm} \\
$F_{2-10}^{\rm high}$ & 5 & 45 & 0 & $5 \times 10^{-10}$ & 68 & 35.7 & - & \vspace{0.1cm} \\
low QPB & 5 & 45 & 0 & $10^{-10}$ & 16 & 6.3 & - & \vspace{0.1cm} \\
high QPB & 5 & 45 & 0 & $10^{-10}$ & 95 & 45.6 & - & \vspace{0.1cm} \\
$t > 10$ ks & 5 & 45 & 0 & $10^{-10}$ & 68 & 28.2 & require $t^e > 10$ ks & \vspace{0.1cm} \\
\bottomrule
\Xhline{3\arrayrulewidth}
\end{tabular}

\label{tab: syst}
\end{table}

\begin{figure}[p]
\centering
\begin{subfigure}{0.49\textwidth}
\includegraphics[width = \textwidth]{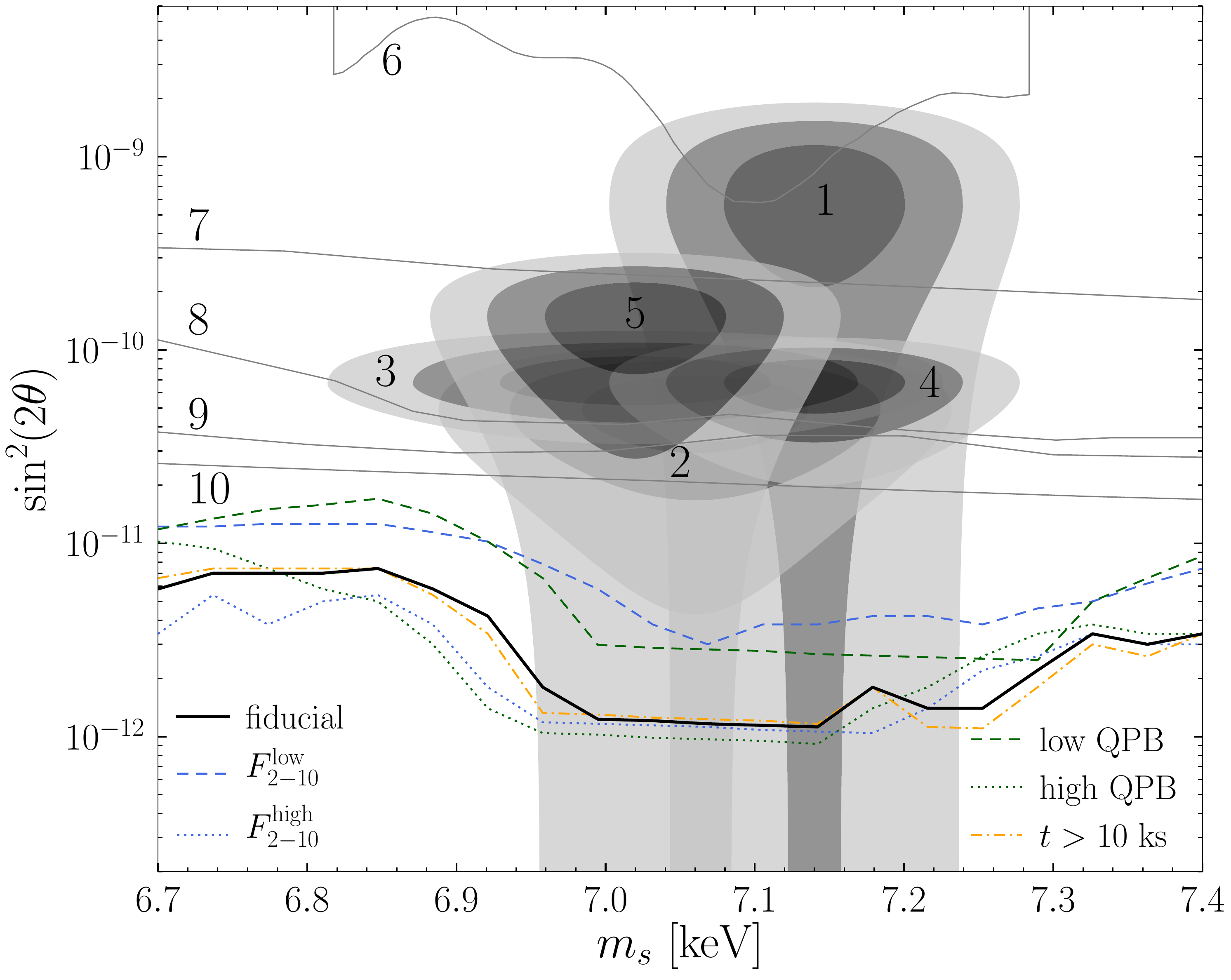} 
    \subcaption{}
\end{subfigure}
\begin{subfigure}{0.49\textwidth}
\includegraphics[width = \textwidth]{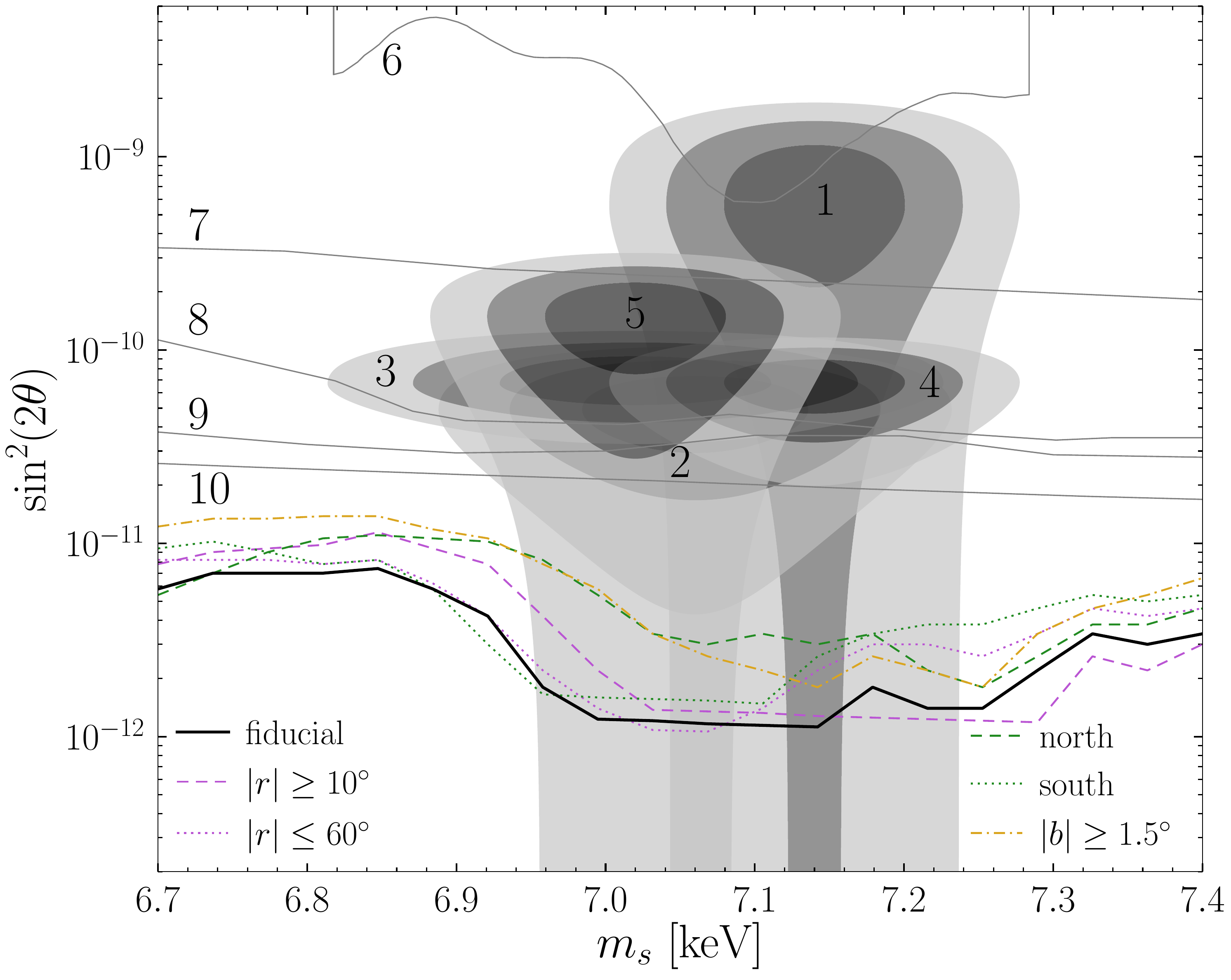}
    \subcaption{}
\end{subfigure}
\caption{\textbf{Variations to the limits arising from different selection criteria that determine which exposures are included in the joint likelihood.} In (A) we vary the cuts on the exposures while in (B) we vary the regions considered. The various criteria are summarized in Table~\ref{tab: syst}. In all cases the decaying DM origin of the UXL is inconsistent with the resulting limits.}
\label{fig: syst}
\end{figure}

In Fig.~\ref{fig: syst} we show the limits obtained with the criteria given in Table~\ref{tab: syst}.  Our main conclusion -- that the decaying DM interpretation of the UXL is inconsistent with our results -- is insensitive to these variations in the selection criterion.

\subsection{Energy binning}

In our fiducial analysis we used the default energy binning recommended by the {\it XMM-Newton} analysis software.  The PN forward modeling matrix uses 15 eV input energy bins, for physical flux, in our energy range of interest and outputs predicted detector counts in 5 eV output energy channels.  The MOS forward modeling matrix maps 5 eV input energy bins to 5 eV output energy channels.  Because the output energy channel widths are much smaller than the detector energy resolution, our results should not depend on the energy binning.  We test this by down-binning the detector responses to lower energy resolutions.  
  
\begin{figure}[p]
\centering
\begin{subfigure}{0.49\textwidth}
\includegraphics[width = \textwidth]{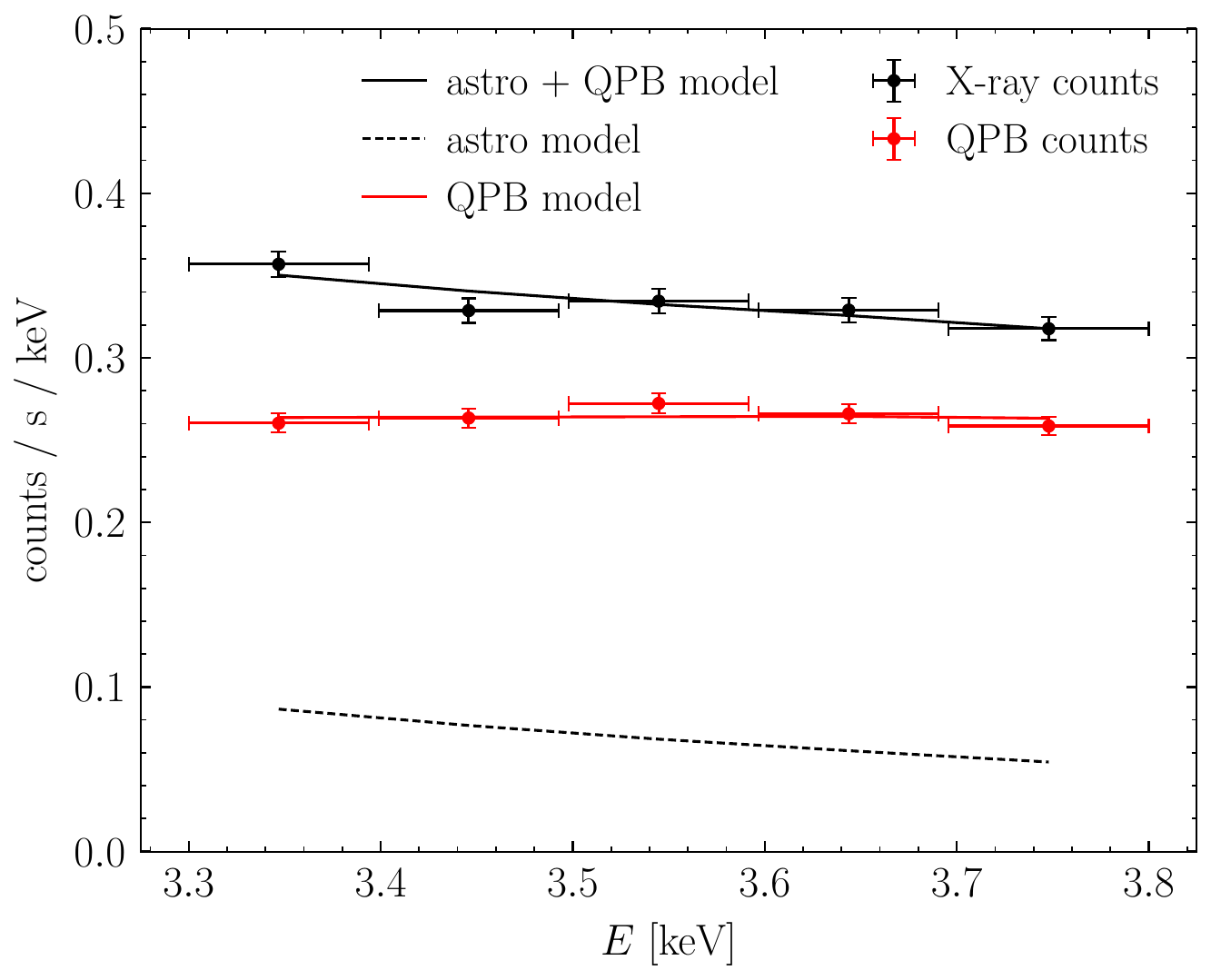} 
    \subcaption{}
\end{subfigure}
\begin{subfigure}{0.49\textwidth}
\includegraphics[width = \textwidth]{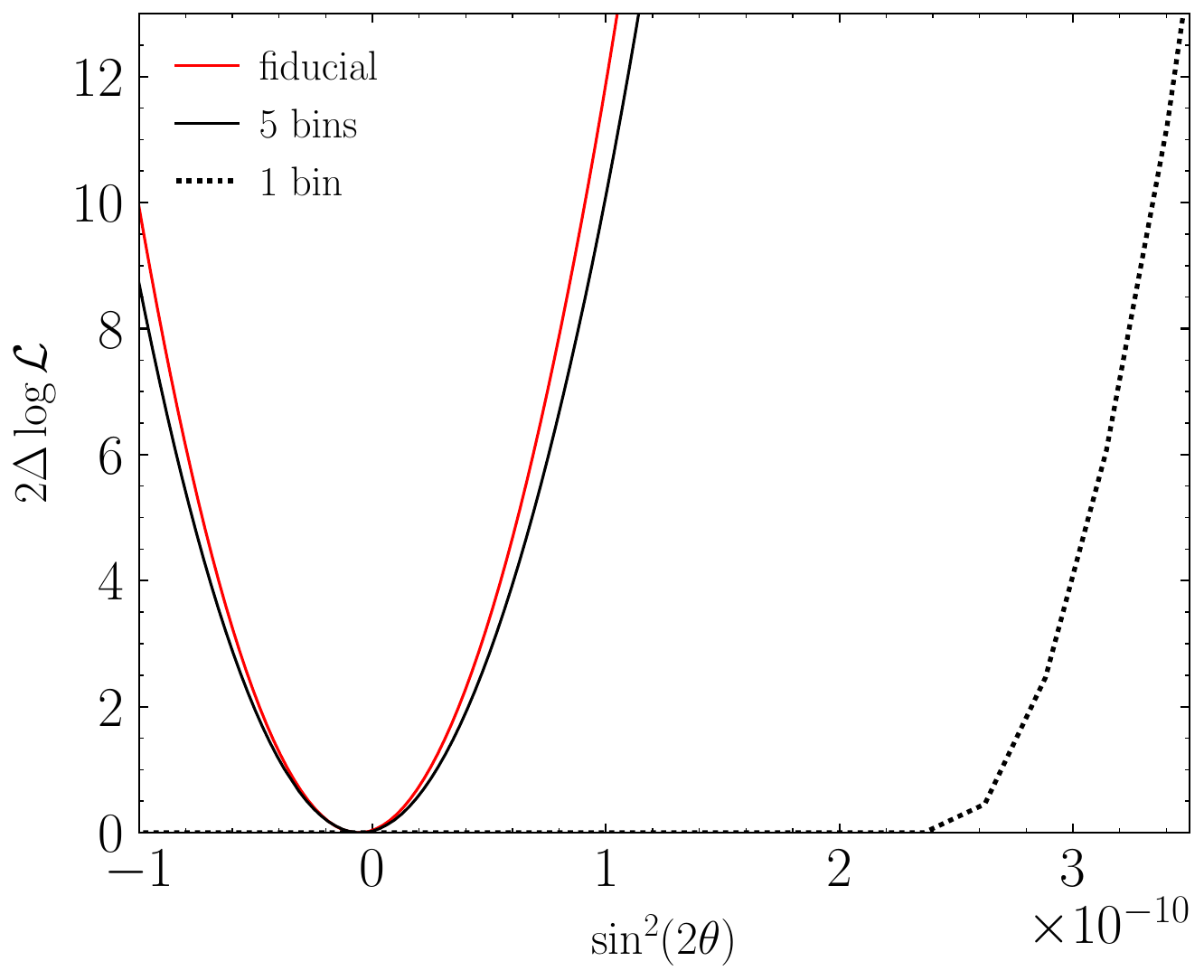}
    \subcaption{}
\end{subfigure}
\caption{\textbf{The effects of down-binning the data.}
As in Fig.~\ref{fig: example}, except here we have down-binned the output energy channels of the PN detector forward modeling matrix. (A) we down-bin to to 0.1 keV output energy channels across the 0.5 keV energy window and show the fitted model.  In red we show the QPB counts as data points and the model as a solid line, respectively. The X-ray data we show as black data points, and we show the model for the astronomical counts in dotted black. The sum of the two models, which is fitted to the X-ray counts, is shown in solid black. (B) we compare the profile likelihood for a UXL at 3.55 keV obtained from this analysis (solid black line), compared to the fiducial profile likelihood in Fig.~\ref{fig: example} (red).  The two profile likelihoods are very similar, as expected given that the PN energy resolution is $\sim$0.1 keV.  We also show the result of using a single 0.15 keV wide output energy bin (dotted black line) centered at 3.55 keV.  
}
\label{fig: example-5bin}
\end{figure}

As a demonstration, we focus on the individual PN observation with observation ID 0653550301 used in Fig.~\ref{fig: example}.  In our fiducial analysis there are 97 output energy channels across the 0.5 keV energy window of the analysis.  However, the energy resolution of the PN camera, which is captured by the forward modeling procedure that we use, is around 0.1 keV.  Thus we expect that down-binning the forward modeling matrix to $\sim$5 energy channels, each of 0.1 keV, should have little effect on the resulting profile likelihood for the putative UXL.  This is demonstrated in Fig.~\ref{fig: example-5bin}, which shows the X-ray and QPB data down-binned to 5 $\sim$0.1 keV energy channels, with the best-fitting astrophysical power-law and QPB power-law models from the joint likelihood fit using the down-binned data.  The profile likelihood for the UXL at 3.55 keV is shown compared to the fiducial profile likelihood.  The differences between the two profile likelihoods are small, with the fiducial profile likelihood being slightly more sensitive.  This is because the 0.1 keV energy bins are at the limit of the detector energy resolution.  

Down-binning further leads to a large decrease in sensitivity, since the UXL then becomes degenerate with the other (nuisance) model parameters.  As an extreme example, we show the single-bin profile likelihood in Fig.~\ref{fig: example-5bin}B, where we have adopted a single output energy bin with a width of 0.15 keV centered around 3.55 keV.  Because we are using a single energy bin, the UXL model parameter is completely degenerate with the model parameters describing the QPB and X-ray power-laws.  As such, it would not be possible to find evidence for a UXL with this analysis, but the limit obtained is maximally conservative.  The limit is $\sin^2(2\theta) \lesssim 2.8 \times 10^{-10}$ at 95\% confidence, so any larger value of $\sin^2(2\theta)$ would overproduce the entire observed X-ray data in this energy bin, without any modeling.  The profile likelihoods in Fig.~\ref{fig: example-5bin} show that the fiducial and 5-bin likelihoods are symmetric and quadratic about their minimum, the 1-bin profile likelihood is zero at low $\sin^2(2\theta)$ and then rises steeply at high $\sin^2(2\theta)$.  Any model flux from the UXL that is below the observed X-ray counts can be compensated by the nuisance models (astrophysical and QPB models), which implies that all such models will fit the data equally well. As  $\sin^2(2\theta)$ becomes arbitrarily negative the other models are allowed to become arbitrarily large to compensate.  However, as $\sin^2(2\theta)$ increases and the UXL model flux begins to surpass the observed data counts, the flux from the nuisance parameters is driven to zero (the nuisance astrophysical and QPB models are restricted to positive flux).  Comparing the 1-bin profile likelihood to the fiducial profile likelihood, we see that modeling the astrophysical and QPB emission improves the limit on the UXL by almost an order of magnitude in this case.

\subsection{Additional Degrees of Freedom in the Background Model}

We test how our limits are impacted if there are additional degrees of freedom in the background model.  We consider the possibility of two lines in the energy range of interest, at $3.31$ keV and $3.69$ keV.  The existence of a weak instrumental line at $3.31$ keV was suggested in PN data by~\cite{Struder:2003} and of both lines in the MOS data~\cite{Ruchayskiy:2015onc}. The existence of these lines has not been conclusively established, nevertheless we can test the impact their inclusion would have on our analysis.

Analysis of {\it XMM-Newton} data in a region of the inner Galaxy that partially overlaps with ours~\cite{Boyarsky:2018ktr} found evidence for sterile neutrino DM at a mixing angle $\sin^2(2\theta) \sim 2 \times 10^{-11}$ and mass $m_s \approx 7.0$ keV.  That work used a broad energy range in their analysis of the stacked data and included additional lines in their background model in the vicinity of 3.5 keV, at 3.12 keV, 3.31 keV, 3.69 keV, and 3.90 keV. We analyze how our results change with the addition of these lines.  We only include the lines at 3.31 keV and 3.69 keV because these are the relevant lines for DM masses in the vicinity of 7.0 keV when restricting to a narrow energy range 0.5 keV wide about the line center.

Generically, as we include more degrees of freedom in the background model we expect the limits to weaken, as there is more potential for degeneracy between the signal model and background model, which we profile over.  As an extreme illustration, consider the possibility that the background model has a spectral template that is identical to the signal spectral template and that the normalization of the background spectral template is allowed to float negative in the profile likelihood process.  Then, for a fixed signal value $S$ the signal template can always be completely canceled by the background model with a background signal template normalization of $-S$.  This means that the likelihood profile as a function of $S$ will be completely flat for all $S$. We can never find evidence for DM with this background model, but we will also set limits that are arbitrarily weak, meaning that we will not rule out a real signal if one were present.  The background model we use with the addition of the two extra lines, whose normalizations are allowed to go negative, does not have a complete degeneracy with the signal model, but there is a partial degeneracy which leads to weaker limits as compared to the power-law model we use in our fiducial analysis.  

\begin{figure}[p]
\centering
\begin{subfigure}{0.49\textwidth}
\includegraphics[width = \textwidth]{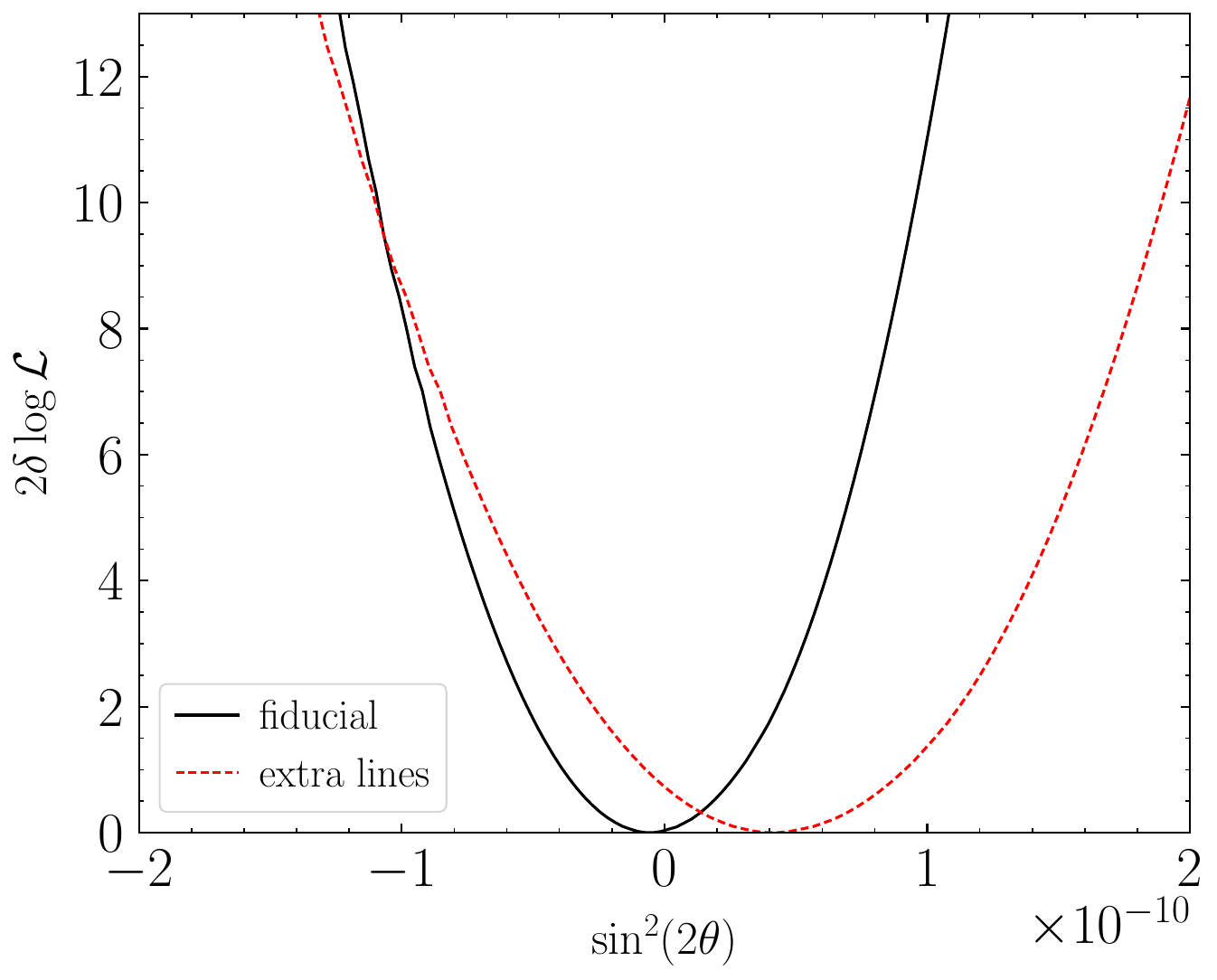} 
    \subcaption{}
\end{subfigure}
\begin{subfigure}{0.49\textwidth}
\includegraphics[width = \textwidth]{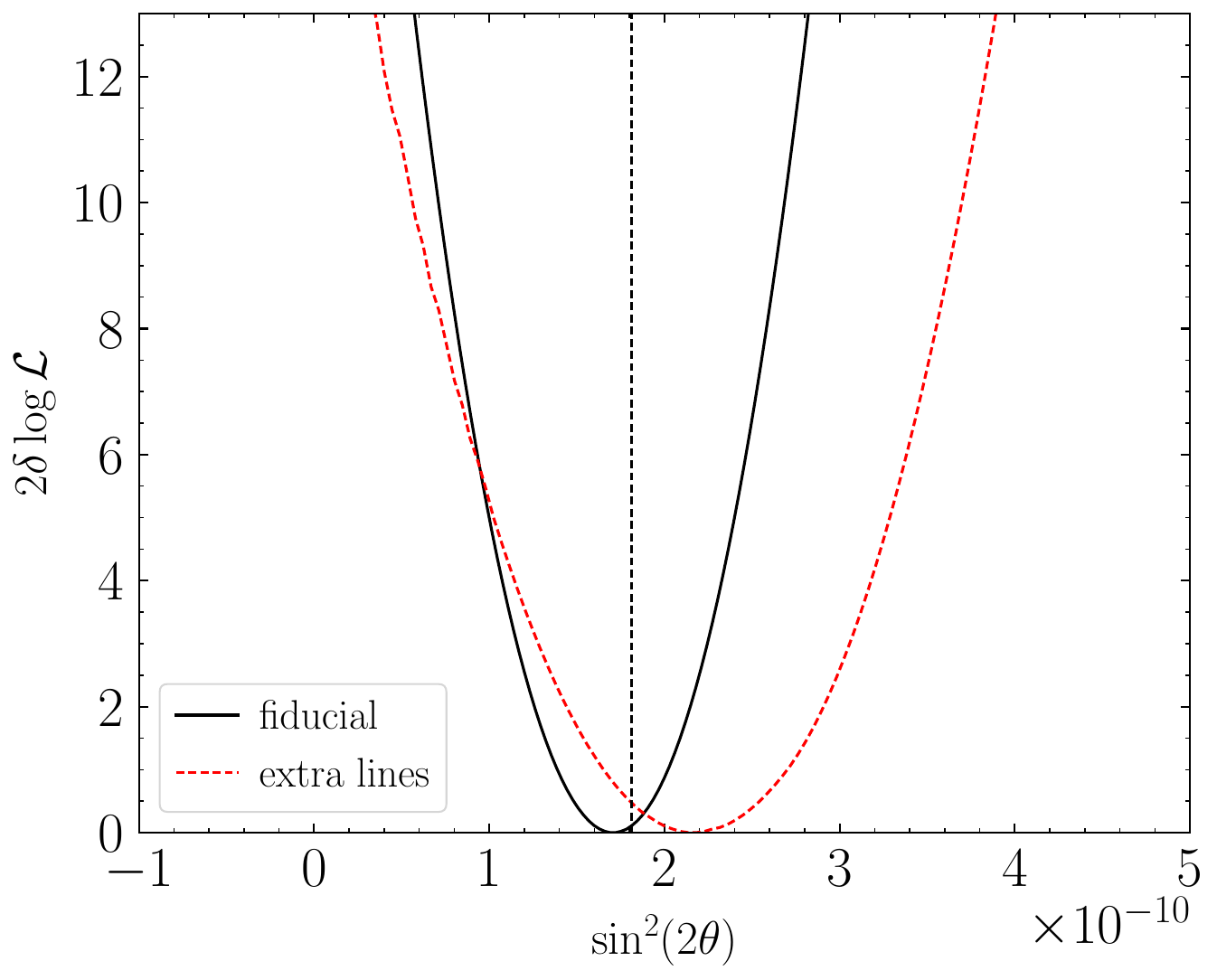}
    \subcaption{}
\end{subfigure}
\caption{\textbf{The effects of adding extra lines on the profile likelihood.} (A) The profile likelihood for a Monte Carlo-generated dataset with no DM decay signal. The solid black line indicates the results using our fiducial analysis, while the dotted red line indicates the results when including the lines at 3.31 keV and 3.69 keV. The analysis with extra lines sets weaker limits on the sterile-active mixing angle, due to the partial degeneracy between the background and signal models. (B) The profile likelihood for a Monte Carlo-generated dataset with an injected signal of $\sin^2(2\theta) = 1.8 \times 10^{-10}$, indicated by the vertical dotted black line. Again, the analysis with the extra lines sets weaker limits.}
\label{fig: example-2lines}
\end{figure}

We illustrate the difference between the two background models with a simple example.  We generate Monte Carlo data using the best-fitting background model shown in Fig.~\ref{fig: example} for  the PN camera of observation ID 0653550301, our most constraining exposure, as given in Table~\ref{tab: IC}.  The Monte Carlo is generated under the null hypothesis, so there is no signal in the data.  We then analyze the simulated data for a signal at $m_s = 7.1$ keV with our fiducial background model and with the background model that has two extra lines at 3.31 keV and 3.69 keV.  The profile likelihoods for these different analyses are shown in Fig.~\ref{fig: example-2lines}A. The profile likelihood with extra lines is broader than the fiducial profile likelihood.  As a result, the limit obtained with the background model containing extra lines is weaker by approximately a factor of two in this case.

\begin{figure}[p]
\centering
\begin{subfigure}{0.49\textwidth}
\includegraphics[width = \textwidth]{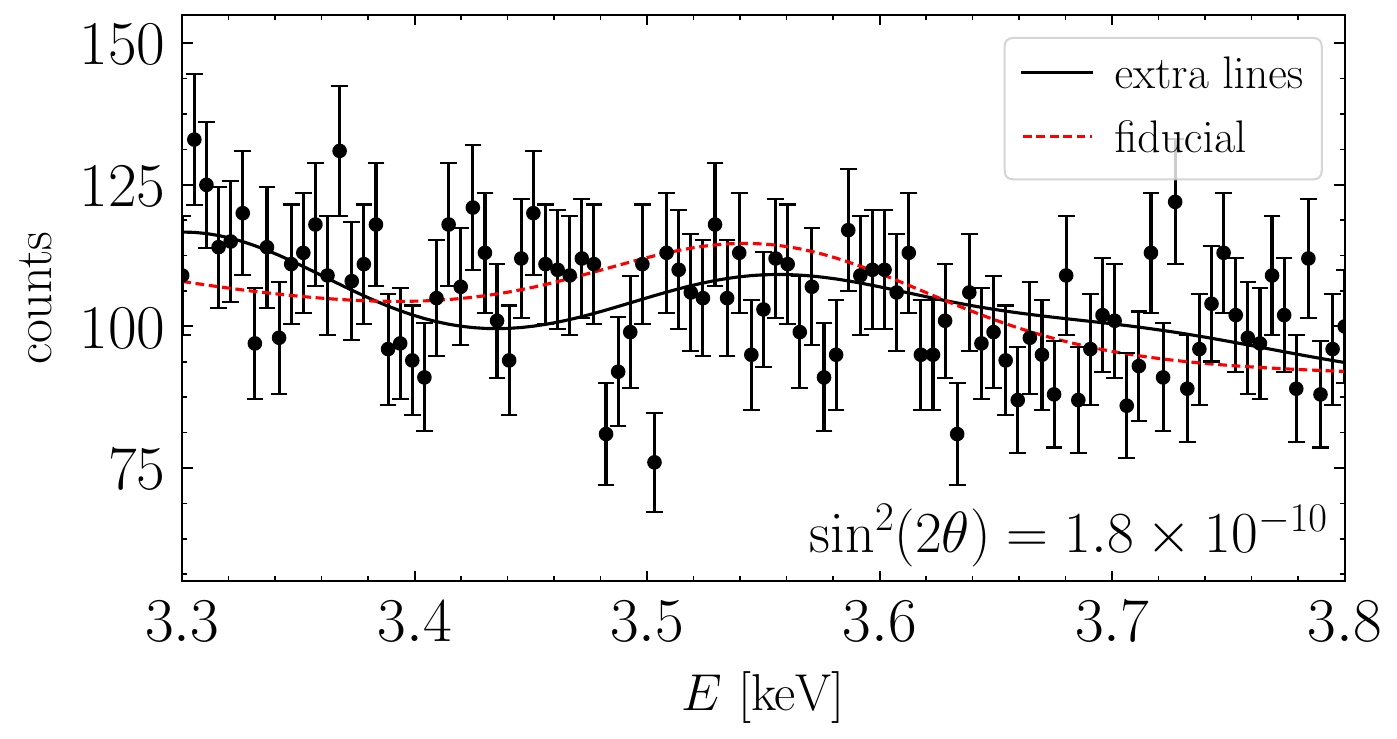} 
    \subcaption{}
\end{subfigure}
\begin{subfigure}{0.49\textwidth}
\includegraphics[width = \textwidth]{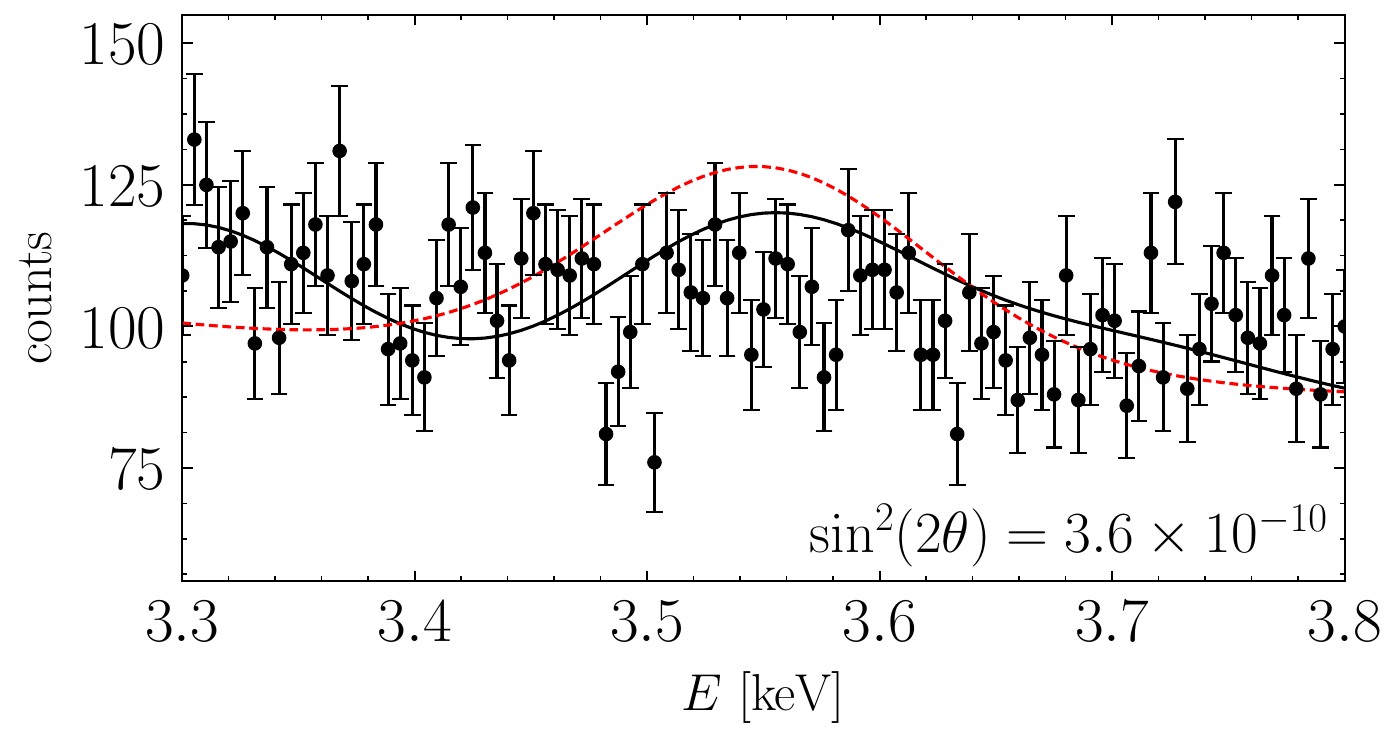}
    \subcaption{}
\end{subfigure}
\begin{subfigure}{0.49\textwidth}
\includegraphics[width = \textwidth]{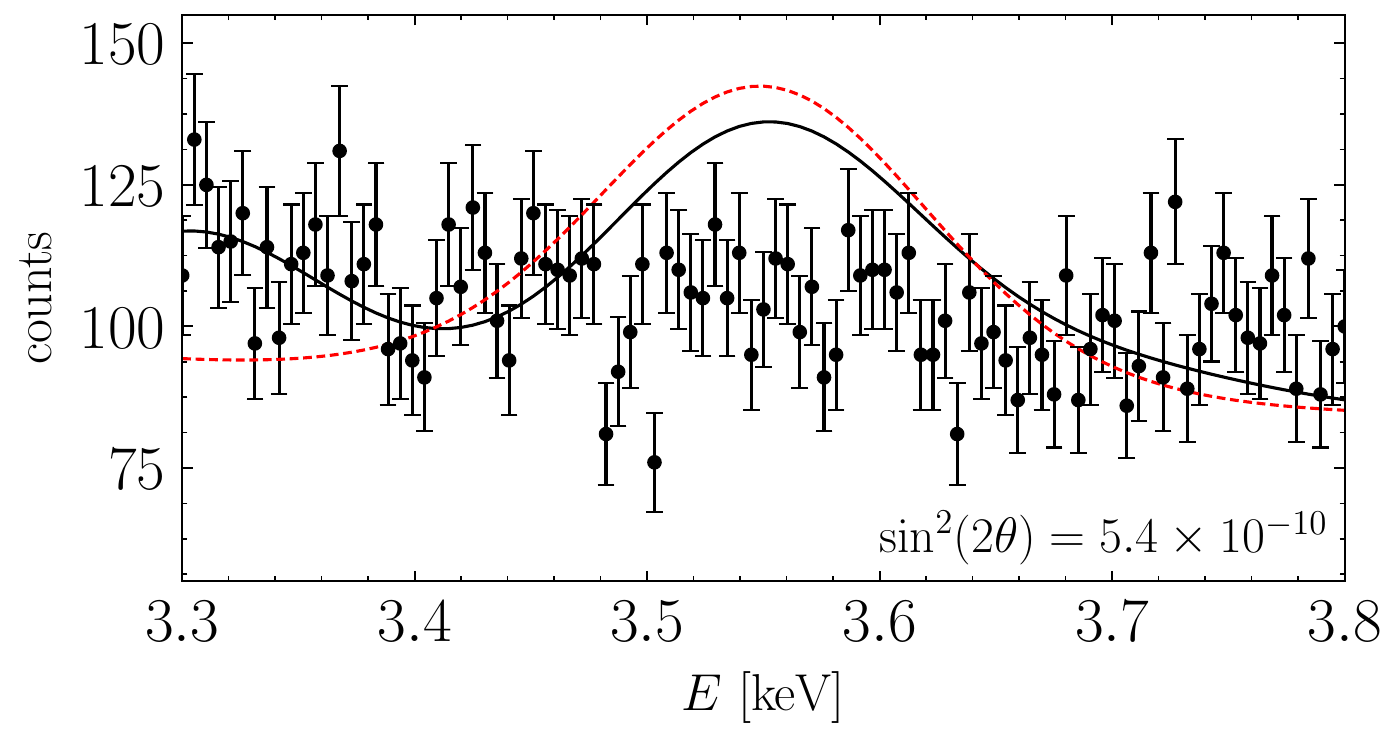}
    \subcaption{}
\end{subfigure}
\caption{\textbf{The effects of adding extra lines on the fitted model.} The best-fitting models assuming a DM mass of 7.1 keV for fixed signal strengths (A, $\sin^2(2\theta) = 1.8 \times 10^{-10}$; B, $\sin^2(2\theta) = 3.6 \times 10^{-10}$; C, $\sin^2(2\theta) = 5.4 \times 10^{-10}$) for the fiducial background model (red line) and the background model with extra lines (black line). The data shown  (black points) is the Monte Carlo data analyzed in the left side of Fig.~\ref{fig: example-2lines}.
}
\label{fig: example-2lines-profile}
\end{figure}

The degeneracy between the signal and background model in the profile likelihood is illustrated in Fig.~\ref{fig: example-2lines-profile}, where we show the best-fitting models for fixed signal fluxes with both the fiducial background model and the background model including the extra lines.
In the presence of a signal, the analysis including the extra lines still recovers the injected signal strength, but the significance of the detection is reduced because the background model is semi-degenerate with the signal. Fig.~\ref{fig: example-2lines}B shows the profile likelihoods for analyses of the same simulated datasets used in the example but now with the addition of a synthetic signal with $\sin^2(2\theta) = 1.8 \times 10^{-10}$.  Both the fiducial analysis and the analysis including the extra lines find the injected signal strength within 68\% confidence, but the significance of the detection is reduced for the analysis with extra lines, as may be seen from the likelihood profile being broader in this case.   

\begin{figure}[p]
\centering
\begin{subfigure}{0.49\textwidth}
\includegraphics[width = \textwidth]{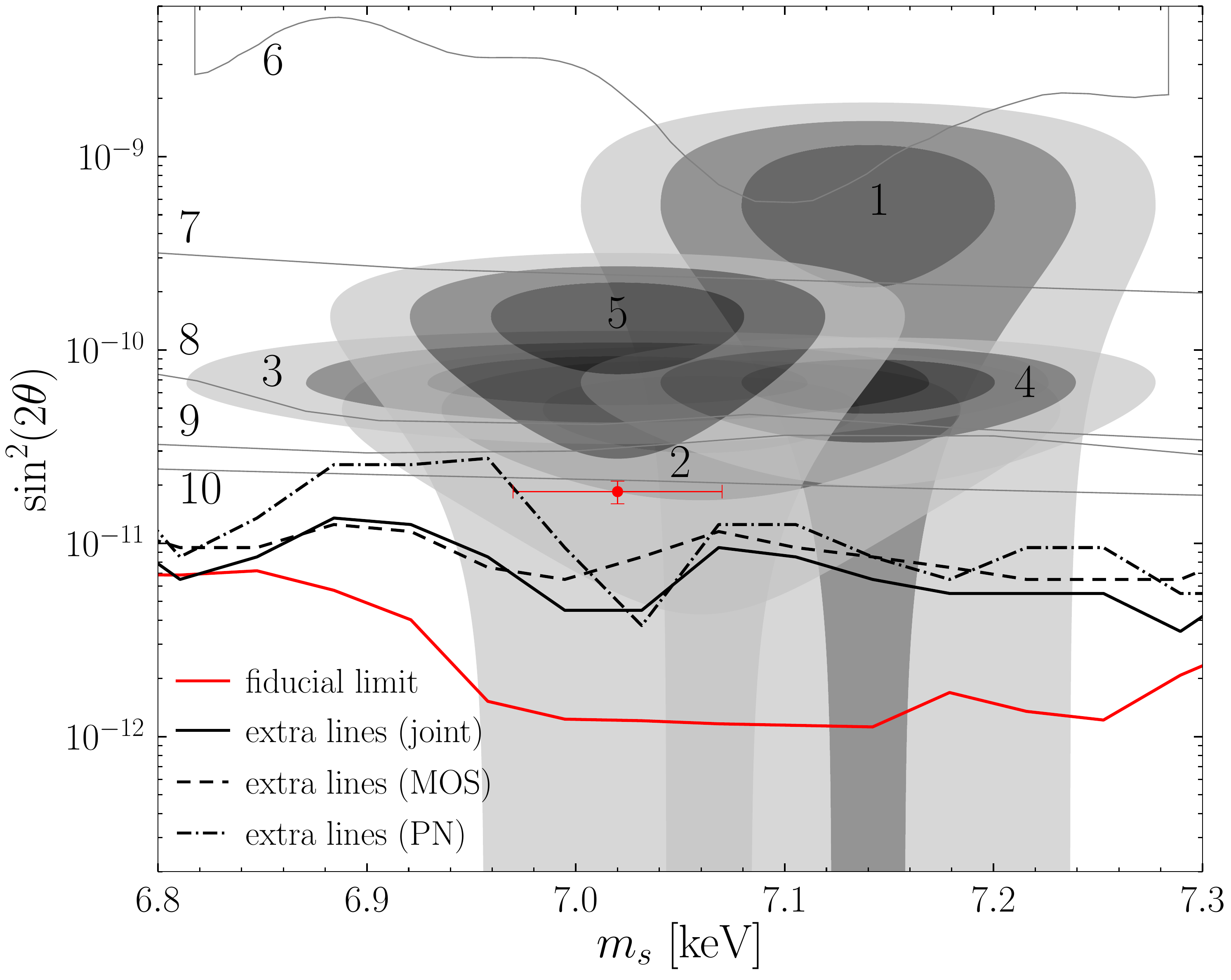} 
    \subcaption{}
\end{subfigure}
\begin{subfigure}{0.49\textwidth}
\includegraphics[width = \textwidth]{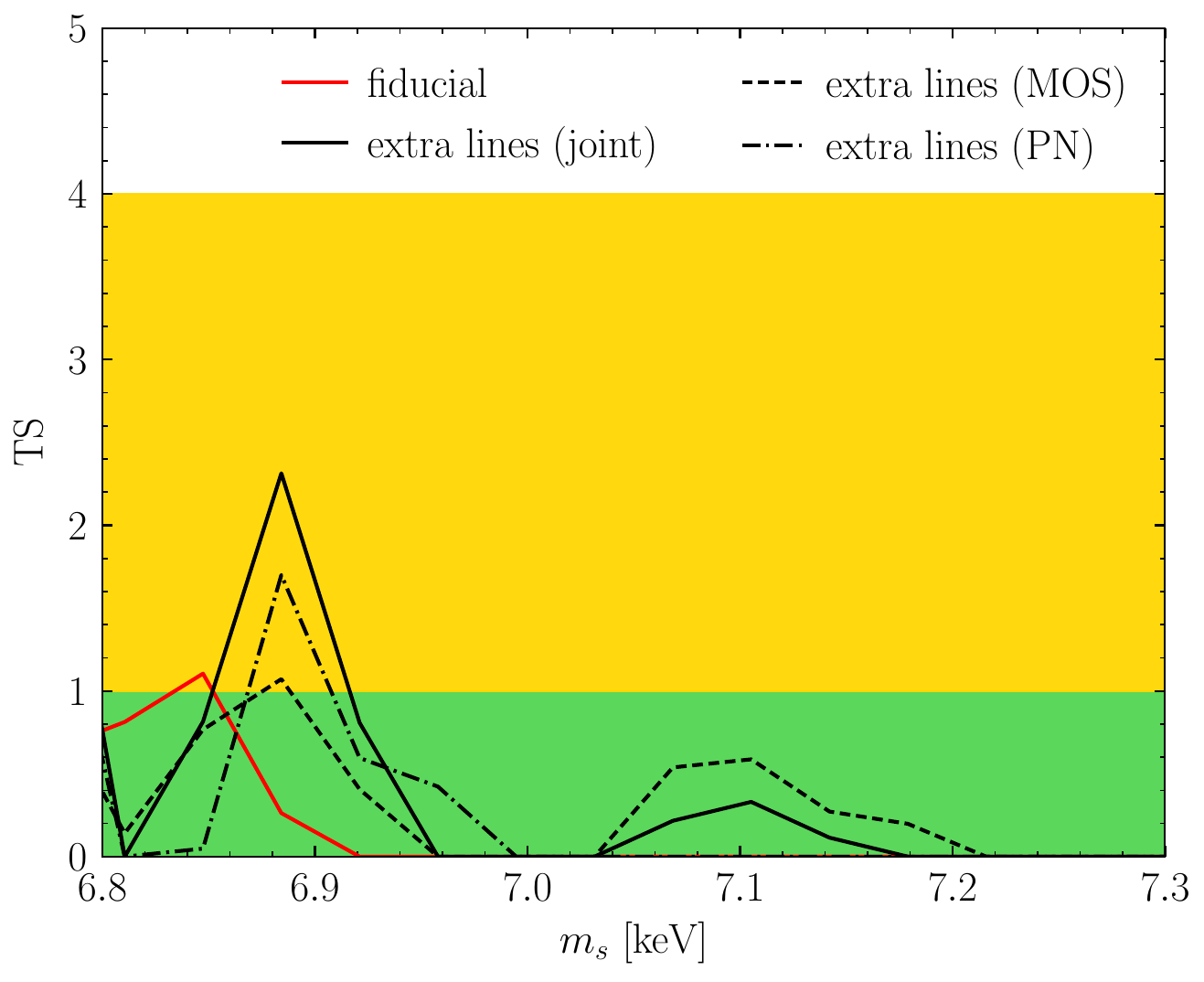}
    \subcaption{}
\end{subfigure}
\caption{\textbf{The effects of adding extra lines on the limits and TS.} (A) The one-sided power-constrained 95\% limits on $\sin^2(2\theta)$ as a function of the DM mass $m_s$ with the analysis with extra lines, compared to the fiducial limit and the parameter space from Fig.~\ref{fig: spectra}.  We show the extra-line limits for the MOS and PN datasets independently and combined. The fiducial limit is stronger than the limit when including the extra lines. Nevertheless, the latter limit remains inconsistent with the detections shown in Fig.~\ref{fig: spectra}, and the best-fitting parameters in~\cite{Boyarsky:2018ktr} (red point). Our conclusions are not dependent on our choice of background model. (B) The test statistic (TS) in favor of a decaying DM interpretation of the UXL as a function of the DM mass $m_s$ with the analysis with extra lines (black), compared to the fiducial TS (red). The green and yellow regions indicate 1$\sigma$ and 2$\sigma$ detections, respectively. No evidence for decaying DM in either analysis is found.}
\label{fig: limits-lines}
\end{figure}

We perform this analysis across our ensemble of exposures and present the limits in Fig.~\ref{fig: limits-lines}A.  We show the limits for the MOS datasets, the PN datasets, and the combination of both datasets (joint).  We show the MOS and PN limits separately to account for the possibility that the extra lines affect one detector but not the other.  As expected, the limits are weaker when including the extra lines in the background model.  However, the limits (from both MOS and PN datasets independently) remain inconsistent with the DM interpretation of the UXL and the best-fitting parameters from the blank-sky analysis in~\cite{Boyarsky:2018ktr}.  In the vicinity of 7.0 keV we find no evidence for DM, as in shown in Fig.~\ref{fig: limits-lines}B. For this analysis we restrict the exposure times of the individual exposures to be greater than $10$ ks to ensure the model fitting of the individual exposures is well converged.  This reduces the total exposure time to $\sim$27.2 Ms, which does not substantially affect the projected sensitivity.

\begin{figure}[p]
\centering
\includegraphics[width = 0.49\textwidth]{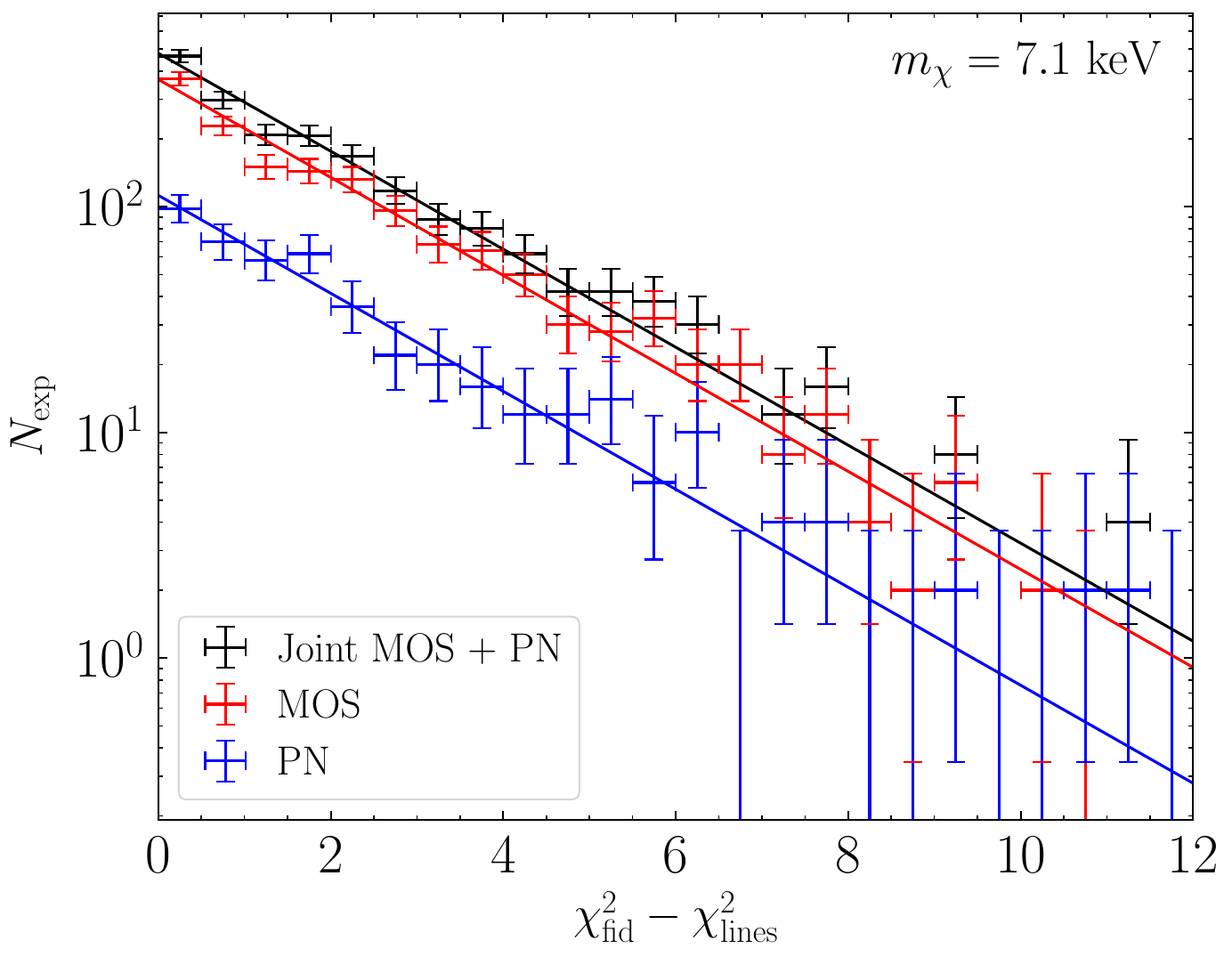} 
\caption{\textbf{The effects of adding extra lines on the $\chi^2$ .} For the 7.1 keV mass point, we show a histogram of the individual $\chi^2_{\textrm{fid}}-\chi^2_{\textrm{lines}}$ values from each individual exposure in our fiducial analysis, with error bars calculated from Poisson statistics.  We show the MOS (red) and PN (blue) data separately and our fiducial data (black) from Fig.~\ref{fig: TS_chi2}B, with the appropriate $\chi^2$ distributions with 2 degrees of freedom illustrated by the solid curves.}
\label{fig: delta-chi2-lines}
\end{figure}

We investigate whether the model fitting favors the background model with the two extra lines over the fiducial model.  Adding two extra degrees of freedom to the background model should lead to an improved fit to the data.  If the data is well described by the fiducial background model then we might expect that using the model with extra lines, which has two extra degrees of freedom, would cause a decrease in the $\chi^2$ for the best-fitting background model that follows a $\chi^2$-distribution with two degrees of freedom.  This is what we observe, as is shown in Fig.~\ref{fig: delta-chi2-lines} for the particular case of $m_\chi = 7.1$ keV. We show the average $\chi^2$ difference, over all of the exposures included in the analysis, between the fiducial background model and the model with two extra lines, for the MOS and PN datasets independently and combined.
The distributions follow the appropriate $\chi^2$-distributions with two degrees of freedom.  

\subsection{Analysis of Disjoint Regions}

We performed our analysis separately in multiple disjoint regions. We determine the TS in favor of DM within each region, and investigate whether it is consistent with the null hypothesis. Fig.~\ref{fig: TS_chi2}B shows the distribution of TSs for the individual exposures combined into the joint likelihood used in our fiducial analysis.  However, in the presence of a small signal of strength near our limiting value, this distribution may resemble the null hypothesis $\chi^2$ distribution, because the effect on each individual exposure is small. Here we extend this check, by determining the TS distribution in different regions.

\begin{figure}[p]
\centering
\begin{subfigure}{0.49\textwidth}
\includegraphics[width = \textwidth]{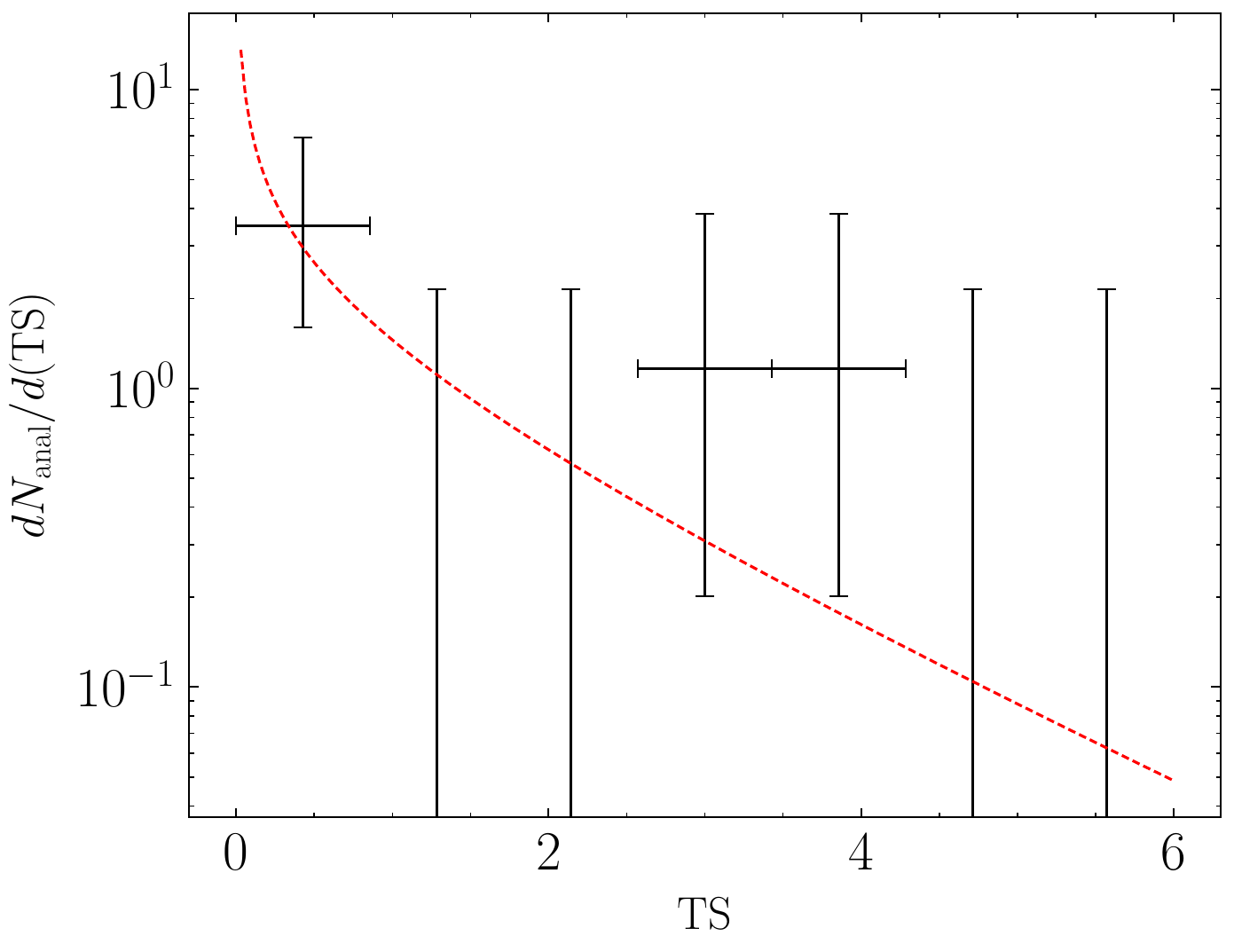} 
    \subcaption{}
\end{subfigure}
\begin{subfigure}{0.49\textwidth}
\includegraphics[width = \textwidth]{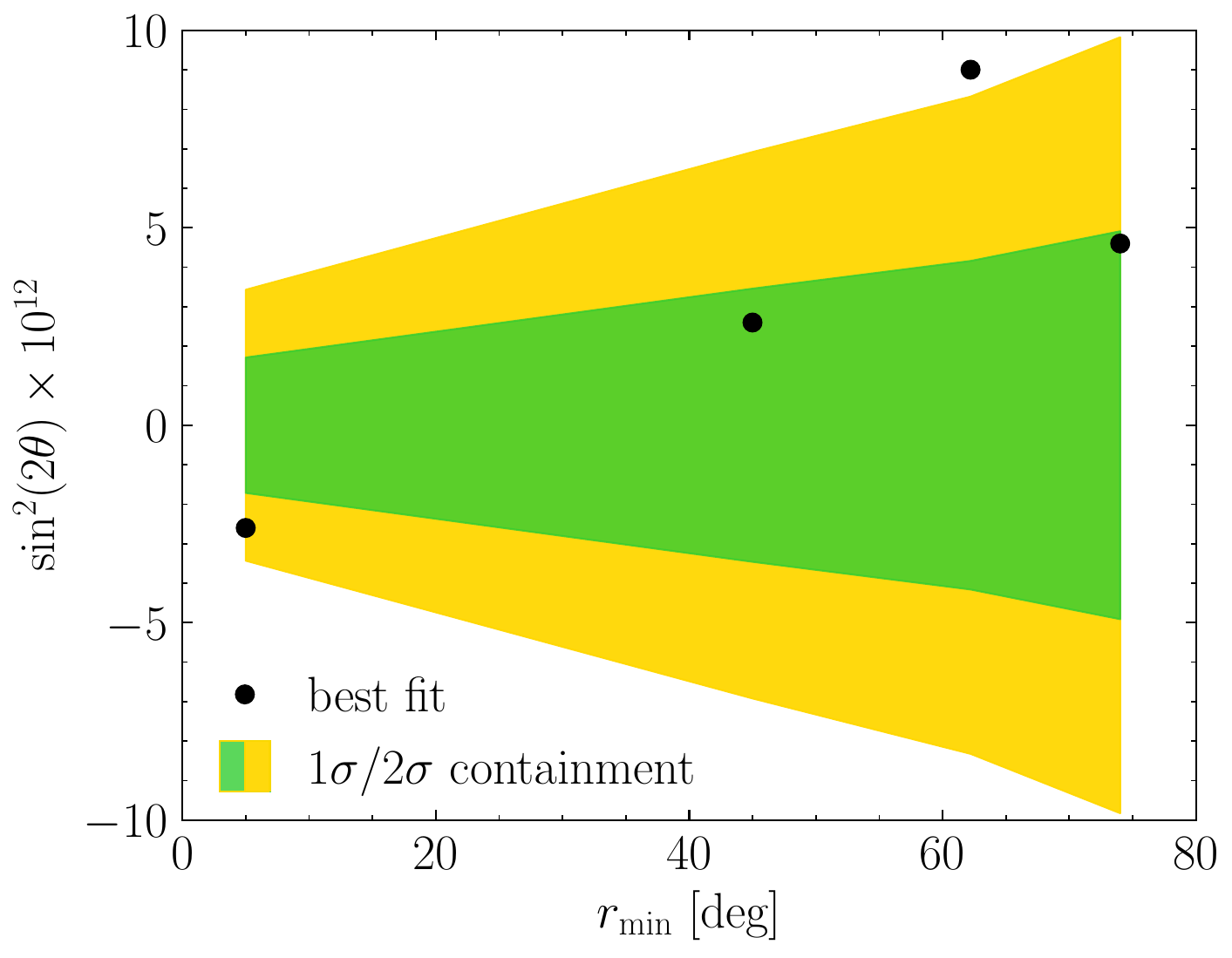}
    \subcaption{}
\end{subfigure}
\caption{\textbf{Dividing the analysis into 4 concentric regions around the Galactic Center.} (A) Distribution of TS values obtained from four independent regions, with three mass points considered for each, are shown in black, with error bars from Poisson statistics. The different regions are our fiducial region used in the main text, as well as observations with $45^\circ < r < 62.2^\circ$, $62.2^\circ < r < 74.0^\circ$, and  $74.0^\circ <  r < 83.4^\circ$, where $r$ is the angle from the Galactic Center.  These regions have approximately 30.6 Ms of exposure each, with our fiducial set of flux and QPB cuts. We also show in red the expectation from statistical fluctuations of the null hypothesis, as determined by the $\chi^2$ distribution. Note the bin with ${\rm TS}=0$ is excluded from the figure, but the values are stated in the text. (B) The best fitting values (black circles) of $\sin^2(2\theta)$ for each of our four regions, identified by their minimum angle from the Galactic Center, for $m_s = 7.1$ keV. The green and yellow regions indicate 1$\sigma$ and 2$\sigma$ containment for these values, respectively. 
}
\label{fig: var_large}
\end{figure}

Our fiducial analysis includes exposures within $5^\circ$ and $45^\circ$ of the Galactic Center. We now consider dividing the region between $5^\circ$ and $\sim$$90^\circ$ into four approximately equal exposure regions, with the first region being our fiducial region, the second region extending from $45^\circ$ to $62.2^\circ$, the third region from $62.2^\circ$ to $74.0^\circ$, and the fourth from $74.0^\circ$ to $83.4^\circ$.  All of these regions have approximately 30.6 Ms of exposure, by construction.  These regions would also be approximately equal area, though in practice we see that the area of the concentric circular regions decreases with distance from the Galactic Center.  This is due to our flux cut on individual exposures, which is less frequently satisfied for observations closer to the Galactic Center.   We compute the TS for analyses in each of these regions for three different mass points: $m_s = 6.9$, 7.1, and 7.3 keV.  We use the three independent mass points to improve the statistics when constructing the TS distribution. We then combine the TSs from the different mass points; the resulting distribution of TSs is shown in Fig.~\ref{fig: var_large}. The distribution TSs follows the one-sided $\chi^2$ distribution, as expected under the null hypothesis, though the number of independent analyses $N_{\rm anal}$ is limited.  This disfavors the presence of large systematic uncertainties, indicating that our uncertainty is is dominated by statistical variations. The bin with ${\rm TS} = 0$ is not shown in Fig.~\ref{fig: var_large}; instead, we state it here.  We expect that half of our 12 analyses should produce ${\rm TS} = 0$ and find $7_{-2.6}^{+3.8}$, consistent with the predicted number of $6$.

In Fig.~\ref{fig: var_large}B, we show how, at $m_s = 7.1$ keV, our best-fitting mixing angle $\sin^2(2\theta)$ changes between the analyses in the four independent regions. We show the expectations under the null hypothesis; for statistical consistency, we must include the possibility of negative $\sin^2(2\theta)$. If our background model systematically under predicts the counts at energies $\sim$$m_s /2$, but also a real signal is present, the evidence for that real signal could be obscured.  However, in that case regions further from the Galactic Center would have a smaller signal contribution, so we would expect the best-fit mixing angle to become more and more negative.  This is not what we see (there is a slight trend in the other direction).  This indicates that our analysis is statistically limited and that there is not a systematic effect at $m_s/2$ obscuring the presence of a signal.

\begin{figure}[p]
\centering
\begin{subfigure}{0.49\textwidth}
\includegraphics[width = \textwidth]{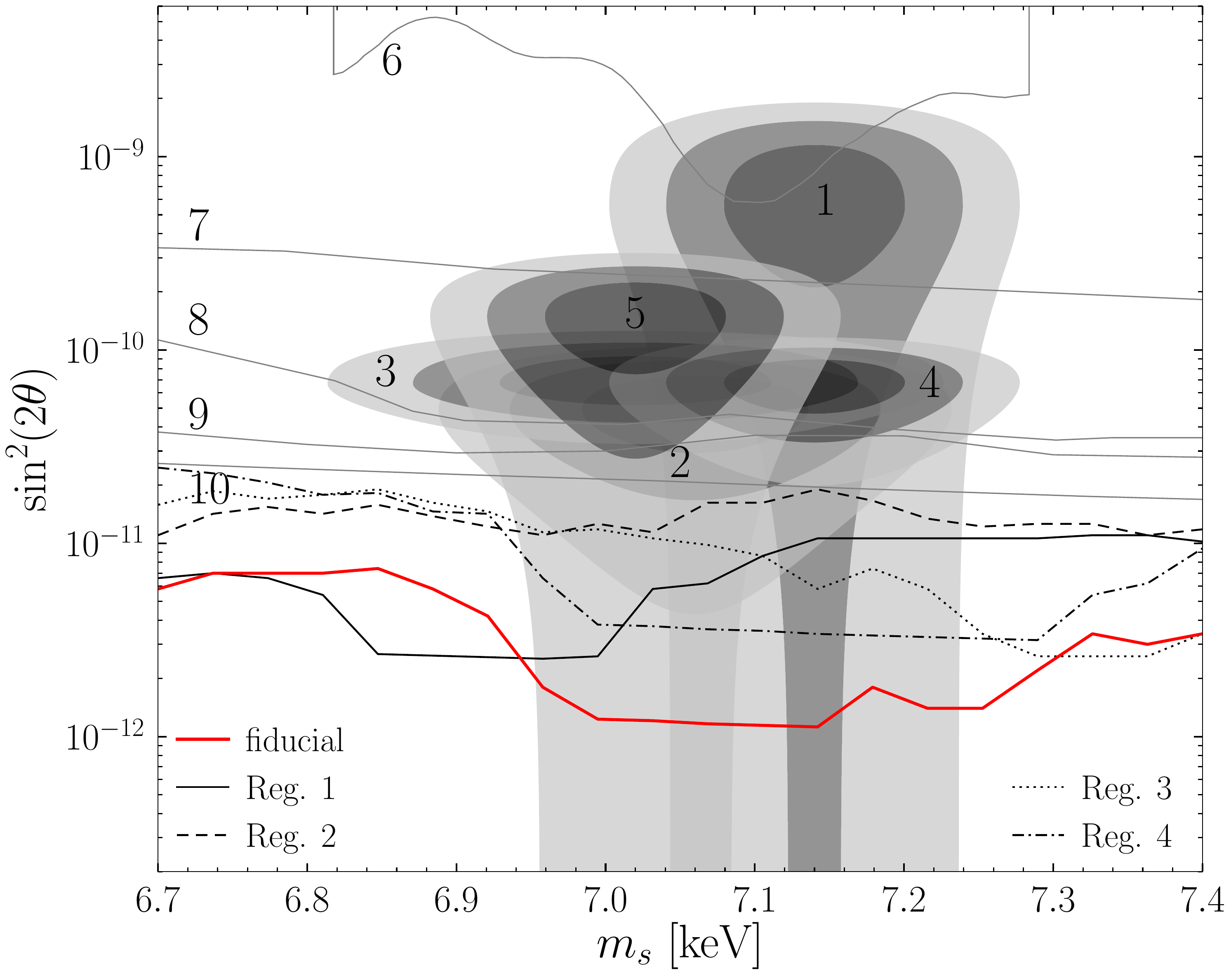} 
    \subcaption{}
\end{subfigure}
\begin{subfigure}{0.49\textwidth}
\includegraphics[width = \textwidth]{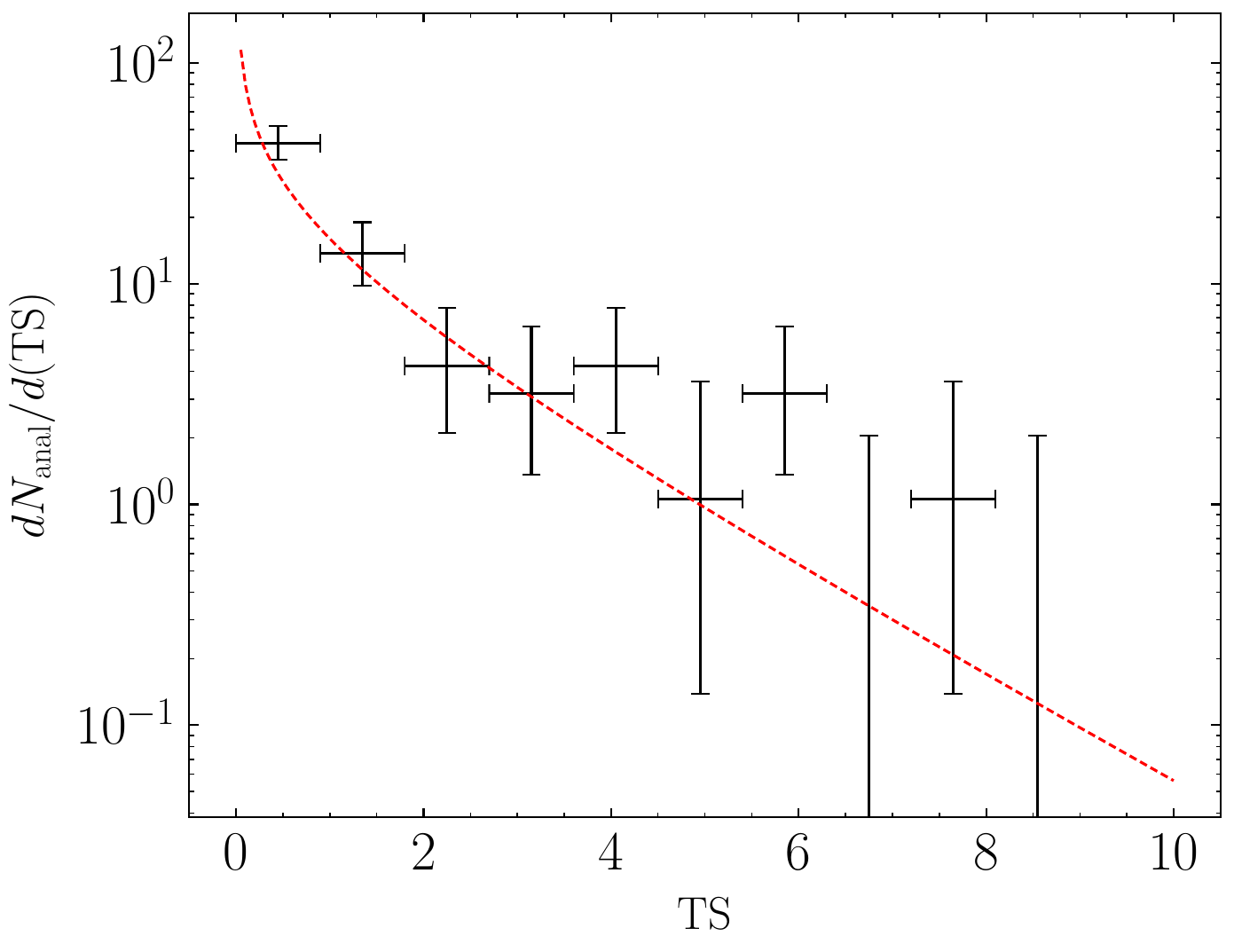}
    \subcaption{}
\end{subfigure}
\caption{\textbf{Dividing the analysis into 45 concentric regions around the Galactic Center.} (A) The limits from the inner four rings, obtained from analyses in sub-regions consisting of concentric rings starting at $5^\circ$ from the Galactic Center that have approximately 3 Ms of exposure per ring, compared with the parameter space from Fig.~\ref{fig: main-figure}. The limits from the first ring is presented in solid black, the second in dashed black, the third in dotted black, and the fourth in dashed-dotted. These can be compared to our fiducial limit, plotted in red.
(B)  As in Fig.~\ref{fig: var_large}, but for analyses in the concentric circle sub-regions used in the left panel.  }
\label{fig: var_small}
\end{figure}

By choosing regions with similar exposure times as our fiducial analysis, the number of regions considered above $N_{\rm anal}$ is limited. We can seek to find an intermediary regime between this result and Fig.~\ref{fig: TS_chi2}B by decreasing the size of our concentric regions. This, however, comes at the cost of reduced sensitivity, because each individual analysis does not have as much exposure. In the extreme limit of individual exposures, the limits obtained are shown in Fig.~\ref{fig: top10}. As an illustration, we consider taking 45 approximately equal-exposure regions, with the first region extending from $5^\circ$ to $10^\circ$, the second from $10^\circ$ to $15.4^\circ$, and the last from $88.6^\circ$ to $88.9^\circ$.  All of the sub-regions have approximately 3 Ms of exposure.  The distribution of TSs at $m_s = 6.9$ keV, 7.1 keV, and 7.3 keV over these analyses is shown in the right panel of Fig.~\ref{fig: var_small}B.  The TS distribution is consistent with the expectation under the null hypothesis.  This again indicates that the dominant source of uncertainty is statistical and not systematic.  The limits obtained from the inner four rings are shown in Fig.~\ref{fig: var_small}A.  While these limits are slightly weaker than in our fiducial analysis, they are each inconsistent with the decaying DM interpretation of the 3.5 keV line.

\subsection{Analysis of Stacked Data}

Our fiducial data analysis is based on the joint likelihood that combines the profile likelihoods from the disjoint analyses of the individual exposures.  An alternative analysis, as described in the main text, is to instead stack the data from the individual exposures and then analyze the stacked data.  The limits obtained from this procedure are consistent with those obtained from our fiducial data analysis and, in particular, are in tension with the DM interpretation of the UXL.  In this section we provide additional details behind this analysis.  

\begin{figure}[p]
\centering
\includegraphics[width = 0.59\textwidth]{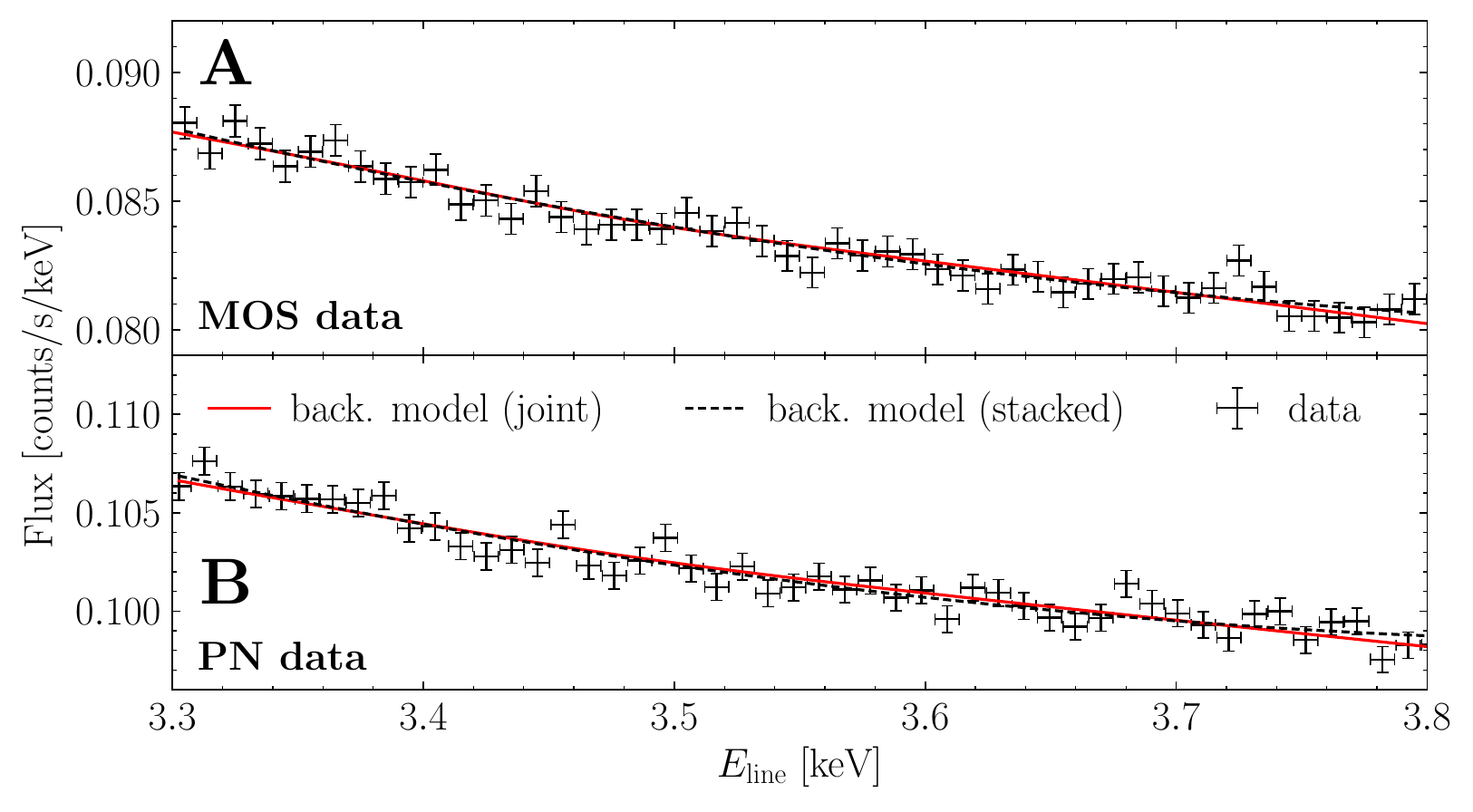}
\includegraphics[width = 0.39\textwidth]{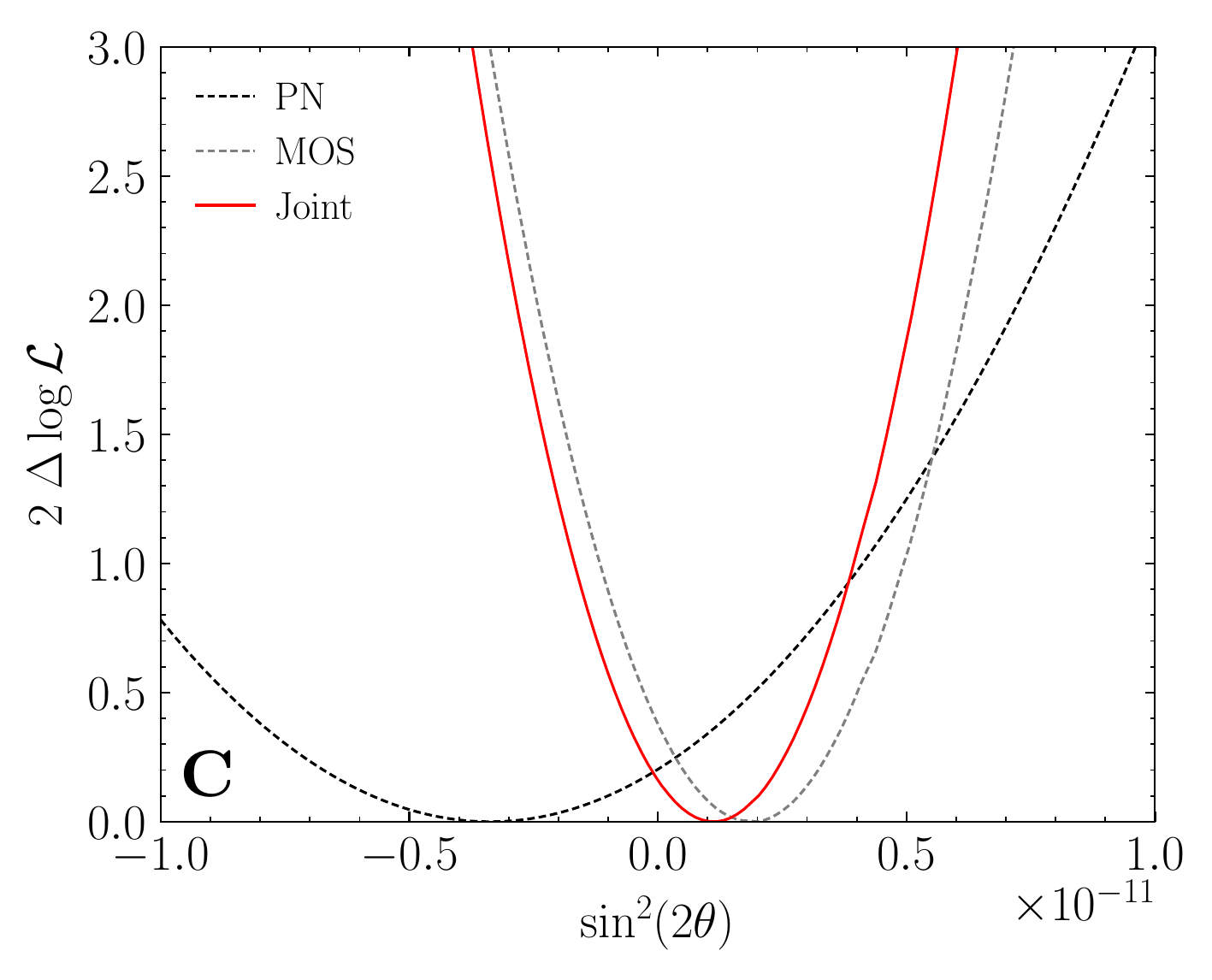}
\caption{\textbf{Analysis of stacked data.} The stacked MOS (A) and PN (B) data as presented in Fig.~\ref{fig: spectra}.  The red curves are the sums of the best-fit null-hypothesis models from the analyses of the individual exposures, as presented in Fig.~\ref{fig: spectra}, while the dashed black curves are the best-fit null-hypothesis models from fits of the quadratic background model to the stacked data. (C) The profile likelihoods as functions of the signal-strength parameter $\sin^2(2 \theta)$, including negative values, from analyses of the stacked data with the quadratic background model.  Results are shown for MOS and PN individually as well as combined.   }
\label{fig: stacked}
\end{figure}

The stacked MOS and PN datasets are shown in Fig.~\ref{fig: spectra} (see also Figs.~\ref{fig: stacked}A and B) from 3.3 keV to 3.8 keV.  These datasets are constructed by summing the individual spectra.  We then fit models to these stacked spectra that consist of a null hypothesis component to describe the smooth background emission and a signal component to model the contribution from a putative UXL at {\it e.g.} 3.55 keV.  We independently fit models to the MOS and PN data, since these instruments have different instrument responses, in order to construct profile likelihoods for the UXL flux as determined by the two cameras.  We then join these two profile likelihoods together in order to obtain our final constraint on the 3.55 keV UXL.  Note that since all of the energy bins shown in Fig.~\ref{fig: spectra}, for both the MOS and PN data, have over $10^3$ counts, we use a Gaussian likelihood instead of a Poisson likelihood.

The red curves in Fig.~\ref{fig: spectra} are the best sums of the best-fit null hypothesis models from the analyses of the individual exposures.  While each of the curves that goes into the sum is modeled as a combination of two power-law components, the combined spectrum does not need to follow a simple power-law.  However, as shown below, we find that a quadratic background model with three model parameters is able to describe the data at the level of statistical noise.
In particular, we take the background model for the stacked data analysis to be $dN/dE(E, {\bf \theta_{\rm nuis}}) = A + B \left(E/1 \, \, {\rm keV} \right) + C \left(E/1 \, \, {\rm keV} \right)^2$, where $dN/dE$ are the fluxes appearing in Fig.~\ref{fig: spectra} in units of counts s$^{-1}$keV$^{-1}$, and where ${\bf \theta_{\rm nuis}} = \{ A, B,C\}$ are the nuisance parameters.  The signal component is as shown in Fig.~\ref{fig: spectra} and is obtained by forward modeling the 3.55 keV UXL through the detector responses from each individual exposure and then summing the resulting spectra.  As seen in Fig. 2, the effect of the UXL is to produce a spectral feature with a width $\sim$200 eV, though that feature is narrower for the MOS data than for the PN data.   

Fitting the quadratic background plus signal model to the PN (MOS) data produces a fit with a $\chi^2 / {\rm DOF}$, for $50 - 4$ DOF, of 1.14 (0.81).  The expected 68\% containment interval for the $\chi^2 / {\rm DOF}$ with 46 DOF under the null hypothesis is $\sim$[0.79, 1.21]; this interval contains the values measured for both the PN and MOS data.  We then compute the profile likelihood for the signal-model flux parameter $\sin^2(2\theta)$ by profiling over the background nuisance parameters.  The profile likelihood is shown in Fig.~\ref{fig: stacked}C.  We show the profile likelihoods for the PN and MOS data analyzed separately and also for the combination.  The 95\% upper limit is found by taking the upper value for $\sin^2(2\theta)$ under the condition $2 \Delta \log {\mathcal L} \approx 2.71$; this leads to the upper limit $\sin^2(2 \theta) < 5 \times 10^{-12}$, which strongly constrains the DM interpretation of the 3.5 keV UXL.    

\begin{thebibliography}{10}

\bibitem{PhysRevD.25.766}
P.~B. Pal, L.~Wolfenstein, {\it Phys. Rev. D\/} {\bf 25}, 766 (1982).

\bibitem{Dodelson:1993je}
S.~Dodelson, L.~M. Widrow, {\it Phys. Rev. Lett.\/} {\bf 72}, 17 (1994).

\bibitem{Shi:1998km}
X.-D. Shi, G.~M. Fuller, {\it Phys. Rev. Lett.\/} {\bf 82}, 2832 (1999).

\bibitem{Kusenko:2006rh}
A.~Kusenko, {\it Phys. Rev. Lett.\/} {\bf 97}, 241301 (2006).

\bibitem{Bulbul:2014sua}
E.~Bulbul, {\it et~al.\/}, {\it Astrophys. J.\/} {\bf 789}, 13 (2014).

\bibitem{Boyarsky:2014jta}
A.~Boyarsky, O.~Ruchayskiy, D.~Iakubovskyi, J.~Franse, {\it Phys. Rev. Lett.\/}
  {\bf 113}, 251301 (2014).

\bibitem{Abazajian:2017tcc}
K.~N. Abazajian, {\it Phys. Rept.\/} {\bf 711-712}, 1 (2017).

\bibitem{Jeltema:2014qfa}
T.~E. Jeltema, S.~Profumo, {\it Mon. Not. Roy. Astron. Soc.\/} {\bf 450}, 2143
  (2015).

\bibitem{Gu:2015gqm}
L.~Gu, {\it et~al.\/}, {\it Astron. Astrophys.\/} {\bf 584}, L11 (2015).

\bibitem{Shah:2016efh}
C.~Shah, {\it et~al.\/}, {\it Astrophys. J.\/} {\bf 833}, 52 (2016).

\bibitem{Urban:2014yda}
O.~Urban, {\it et~al.\/}, {\it Mon. Not. Roy. Astron. Soc.\/} {\bf 451}, 2447
  (2015).

\bibitem{Boyarsky:2014ska}
A.~Boyarsky, J.~Franse, D.~Iakubovskyi, O.~Ruchayskiy, {\it Phys. Rev. Lett.\/}
  {\bf 115}, 161301 (2015).

\bibitem{Cappelluti:2017ywp}
N.~Cappelluti, {\it et~al.\/}, {\it Astrophys. J.\/} {\bf 854}, 179 (2018).

\bibitem{Horiuchi:2013noa}
S.~Horiuchi, {\it et~al.\/}, {\it Phys. Rev.\/} {\bf D89}, 025017 (2014).

\bibitem{Malyshev:2014xqa}
D.~Malyshev, A.~Neronov, D.~Eckert, {\it Phys. Rev.\/} {\bf D90}, 103506
  (2014).

\bibitem{Anderson:2014tza}
M.~E. Anderson, E.~Churazov, J.~N. Bregman, {\it Mon. Not. Roy. Astron. Soc.\/}
  {\bf 452}, 3905 (2015).

\bibitem{Tamura:2014mta}
T.~Tamura, R.~Iizuka, Y.~Maeda, K.~Mitsuda, N.~Y. Yamasaki, {\it Publ. Astron.
  Soc. Jap.\/} {\bf 67}, 23 (2015).

\bibitem{Aharonian:2016gzq}
F.~A. Aharonian, {\it et~al.\/}, {\it Astrophys. J.\/} {\bf 837}, L15 (2017).

\bibitem{SM}
Materials and methods are available as supplementary materials.

\bibitem{Navarro:1995iw}
J.~F. Navarro, C.~S. Frenk, S.~D.~M. White, {\it Astrophys. J.\/} {\bf 462},
  563 (1996).

\bibitem{Turner:2000jy}
M.~J.~L. Turner, {\it et~al.\/}, {\it Astron. Astrophys.\/} {\bf 365}, L27
  (2001).

\bibitem{Struder:2001bh}
L.~Struder, {\it et~al.\/}, {\it Astron. Astrophys.\/} {\bf 365}, L18 (2001).

\bibitem{Lumb:2002sw}
D.~H. Lumb, R.~S. Warwick, M.~Page, A.~De~Luca, {\it Astron. Astrophys.\/} {\bf
  389}, 93 (2002).

\bibitem{Moretti:2008hs}
A.~Moretti, {\it AIP Conf. Proc.\/} {\bf 1126}, 223 (2009).

\bibitem{Catena:2009mf}
R.~Catena, P.~Ullio, {\it J. Cosmol. Astropart. Phys.\/} {\bf 1008}, 004
  (2010).

\bibitem{Abuter:2018drb}
R.~Abuter, {\it et~al.\/}, {\it Astron. Astrophys.\/} {\bf 615}, L15 (2018).

\bibitem{Cowan:2010js}
G.~Cowan, K.~Cranmer, E.~Gross, O.~Vitells, {\it Eur. Phys. J.\/} {\bf C71},
  1554 (2011).

\bibitem{Cowan:2011an}
G.~Cowan, K.~Cranmer, E.~Gross, O.~Vitells, \href{arXiv:1007.1727}{https://arxiv.org/abs/1007.1727} (2011).

\bibitem{nick_rodd_2020_3669387}
C.~Dessert, N.~L. Rodd, B.~R. Safdi, nickrodd/XMM-DM: XMM-DM \href{https://doi.org/10.5281/zenodo.3669387}{https://doi.org/10.5281/zenodo.3669387} (2020).

\bibitem{XMM-SAS}
 "Users Guide to the XMM-Newton Science Analysis System", Issue 14.0, 2018
  (ESA: XMM-Newton SOC).

\bibitem{Cicoli:2014bfa}
M.~Cicoli, J.~P. Conlon, M.~C.~D. Marsh, M.~Rummel, {\it Phys. Rev.\/} {\bf
  D90}, 023540 (2014).

\bibitem{Conlon:2014xsa}
J.~P. Conlon, F.~V. Day, {\it J. Cosmol. Astropart. Phys.\/} {\bf 1411}, 033
  (2014).

\bibitem{Kuntz:2008}
{Kuntz, K. D.}, {Snowden, S. L.}, {\it Astron. Astrophys.\/} {\bf 478}, 575
  (2008).

\bibitem{Rolke:2004mj}
W.~A. Rolke, A.~M. Lopez, J.~Conrad, {\it Nucl. Instrum. Meth.\/} {\bf A551},
  493 (2005).

\bibitem{James:1975dr}
F.~James, M.~Roos, {\it Comput. Phys. Commun.\/} {\bf 10}, 343 (1975).

\bibitem{Burkert:1995yz}
A.~Burkert, {\it IAU Symp.\/} {\bf 171}, 175 (1996).

\bibitem{Salucci:2000ps}
P.~Salucci, A.~Burkert, {\it Astrophys. J.\/} {\bf 537}, L9 (2000).

\bibitem{Navarro:1996gj}
J.~F. Navarro, C.~S. Frenk, S.~D.~M. White, {\it Astrophys. J.\/} {\bf 490},
  493 (1997).

\bibitem{Hopkins:2017ycn}
P.~F. Hopkins, {\it et~al.\/}, {\it Mon. Not. Roy. Astron. Soc.\/} {\bf 480},
  800 (2018).

\bibitem{Nesti:2013uwa}
F.~Nesti, P.~Salucci, {\it J. Cosmol. Astropart. Phys.\/} {\bf 1307}, 016
  (2013).

\bibitem{Struder:2003}
L.~{Str{\"u}der}, {\it et~al.\/}, {\it Nuclear Instruments and Methods in
  Physics Research A\/} {\bf 512}, 386 (2003).

\bibitem{Ruchayskiy:2015onc}
O.~Ruchayskiy, {\it et~al.\/}, {\it Mon. Not. Roy. Astron. Soc.\/} {\bf 460},
  1390 (2016).

\bibitem{Boyarsky:2018ktr}
A.~Boyarsky, D.~Iakubovskyi, O.~Ruchayskiy, D.~Savchenko, \href{arXiv:1812.10488}{https://arxiv.org/abs/1812.10488}  (2018).
\end{thebibliography}
\end{document}